\documentclass[10pt]{article}
\usepackage[english]{babel}
\usepackage{sansmathfonts}
\usepackage[T1]{fontenc}
\usepackage{amsmath,amsfonts,amssymb,amstext,amsthm,amscd}
\usepackage{mathtools}
\usepackage[a4paper,top=2cm,bottom=2.5cm,left=2.5cm,right=2.5cm]{geometry}
\usepackage{breakcites}
\usepackage{authblk}
\usepackage{enumitem}
\usepackage{booktabs}
\usepackage{mathrsfs}
\usepackage{dsfont}
\usepackage{graphicx}
\usepackage{subcaption}
\usepackage{wrapfig}
\usepackage{indentfirst}
\usepackage{latexsym}
\usepackage{sidecap}
\usepackage{hyperref}

\usepackage{lipsum}
\usepackage{cleveref}
\usepackage[dvipsnames]{xcolor}
\usepackage[nodisplayskipstretch]{setspace}

\newcommand\myshade{85}
\colorlet{mylinkcolor}{violet}
\colorlet{mycitecolor}{YellowOrange}
\colorlet{myurlcolor}{RoyalBlue}

\hypersetup{
  colorlinks=true,
  linkcolor=myurlcolor,
  citecolor=mycitecolor!\myshade!black,
  filecolor=magenta,
  urlcolor=magenta,
}
\usepackage{textcomp}
\usepackage{braket}
\usepackage{colortbl}
\usepackage{bbm}
\usepackage{comment}
\usepackage[colorinlistoftodos]{todonotes}
\usepackage{tikz}
\usetikzlibrary{arrows.meta, positioning, shapes.geometric, fit, calc, shadows}
\usepackage{ragged2e}

\newcommand{\GSM}{GSM$^2\,$}
\usepackage{lineno}

\usepackage{algorithm}
\usepackage{algorithmic}

\usepackage{siunitx}
\DeclareSIUnit{\MeV}{MeV}
\DeclareSIUnit{\hour}{h}
\DeclareSIUnit{\sec}{s}

\definecolor{rust}{HTML}{D55E00}
\definecolor{bio}{HTML}{3E7CB1}
\definecolor{sched}{HTML}{6C6C6C}
\definecolor{outc}{HTML}{2B9F5A}

\definecolor{initcolor}{RGB}{235, 245, 255}
\definecolor{initdraw}{RGB}{100, 150, 200}
\definecolor{physcolor}{RGB}{255, 240, 240}
\definecolor{physdraw}{RGB}{200, 100, 100}
\definecolor{agentcolor}{RGB}{240, 255, 240}
\definecolor{agentdraw}{RGB}{100, 180, 100}
\definecolor{biocolor}{RGB}{255, 250, 230}
\definecolor{biodraw}{RGB}{200, 170, 50}

\title{A stochastic agent-based extension of the GSM$^2$ model for particle therapy: cell-cycle dynamics, dose-rate dependence, and fractionation effects}

\author[1, 2, 3, *]{Francesco G. Cordoni}
\author[2]{Marco Battestini}
\author[4]{Marta Missiaggia}

\affil[1]{Department of Civil and Environmental Engineering, Louisiana State University (LSU), 4905 S Quad Dr, Baton Rouge, LA 70808, USA}
\affil[2]{Trento Institute for Fundamental Physics and Application (TIFPA), via Sommarive 15, Trento, 38123, Italy}
\affil[3]{Department of Civil, Environmental and Mechanical Engineering, University of Trento, via Mesiano 77, Trento, 38123, Italy}
\affil[4]{Department of Physics \& Astronomy, Louisiana State University (LSU), 202 Nicholson Hall, Baton Rouge, LA 70803, USA}

\begin{document}

\maketitle

\begin{abstract}
Linking the stochastic, microscopic energy deposition of ionizing radiation to the response of a cell population requires explicit, consistent treatment of damage induction, repair, cell-cycle progression, and spatial organization. We present a stochastic agent-based extension of the Generalized Stochastic Microdosimetric Model (GSM$^2$) in which each cell is an agent whose DNA lesions, cell-cycle phase, and fate evolve as a continuous-time Markov process. Particle arrivals, repair, misrepair, cell-cycle transitions, division, delayed radiation-induced death, and migration are modeled as competing events by an event-driven algorithm whose per-event cost does not depend on population size, so that spheroids of $\sim 1.5\times10^5$ cells are irradiated with track-structure resolution and followed for three days in about 18 minutes on a single workstation in the most demanding case. In the limit of a static population, the simulated dose-rate dependence of survival agrees with the Lea--Catcheside prediction for $^{1}$H and $^{12}$C ions over four orders of magnitude in dose rate ($10^{-5}$ to $10^{-2}$~Gy\,s$^{-1}$), verifying the coupling between particle arrival and repair. Dose-rate sparing decreases with LET and vanishes for low-energy carbon ions, consistent with the small contribution of inter-track lesion interactions to cell killing at high LET. With only two global factors of the repair kinetics adjusted on a single survival curve, aerobic X-ray irradiation of V79 spheroids, the model predicts, without further fitting the survival in three independent conditions, X-rays under anoxia and carbon ions under aerobic and anoxic conditions, reproducing the measured oxygen enhancement ratios within 3\% and the relative biological effectiveness within 12\%. Applied to three-dimensional tumor spheroids, the framework resolves the spatiotemporal response to single doses, split doses, and fractionated schedules of equal total dose. It separates the contributions of repair, reassortment, and repopulation, distinguishes cells committed to death from surviving clonogens, and quantifies the variability between stochastic realizations, which is larger for carbon ions. The framework provides a mechanistic, single-cell basis for studying time-dependent effects and uncertainty in the response to particle irradiation.
\end{abstract}


\section{Introduction}
Particle therapy represents a major advance in radiation oncology, offering superior dose conformality and enhanced biological effectiveness compared with conventional photon radiotherapy. Exploiting the finite range and sharp Bragg peak of charged particles enables highly localized dose delivery while sparing surrounding healthy tissues, a benefit that is particularly critical for pediatric malignancies and tumors near radiosensitive structures such as the brain, spinal cord, and optic pathways \cite{Salem2024ProtonReview, Khong2024PediatricPBT}. Heavier ions such as carbon exhibit increased effectiveness against radioresistant and hypoxic tumors owing to their high linear energy transfer (LET), which induces complex, clustered DNA damage that is substantially less repairable than that produced by low-LET radiation \cite{Helm2023HighLET, Sokol2023CarbonHypoxia}. As of 2024, over 120 proton therapy and 14 carbon-ion therapy centers are operational worldwide, with more than 300,000 patients treated with protons and over 50,000 with carbon ions by the end of 2022, numbers that continue to grow \cite{Durante2024CurrentStatus, PTCOG}.

Despite this physical precision, the full clinical potential of particle therapy remains constrained by several sources of uncertainty, among which biological uncertainty is one of the most consequential \cite{Durante2024CurrentStatus, bellinzona2021linking}. The biological response to radiation emerges from a cascade of processes spanning multiple spatial and temporal scales, from femtosecond energy deposition and nanometric track structure, through radiation chemistry and DNA damage formation, to cellular repair, fate commitment, and ultimately tissue- and organ-level outcomes. These processes are tightly coupled; the stochasticity of energy deposition at the micrometer or sub-micrometer scale propagates through biological repair pathways and manifests as macroscopic variability in tumor control and normal tissue toxicity. Capturing this causal chain requires multiscale models that integrate physics, chemistry, and biology across many orders of magnitude in space and time \cite{cordoni2022multiple}.

The biological consequences arising from this multitude of radiation-induced effects have been successfully captured in conventional photon radiotherapy using the linear-quadratic (LQ) model, which remains the predominant framework for describing cell survival and guiding clinical dose prescription and fractionation \cite{mcmahon2018linear,fowler1989linear}. While the LQ model is well supported empirically and computationally convenient, it is phenomenological in nature. Its parameters $\alpha$ and $\beta$ are fitted to experimental survival data and do not derive from an explicit description of the underlying damage and repair processes. In particle therapy, the situation is further complicated by the strong dependence of biological effectiveness on particle type, energy, and LET, which the LQ model cannot account for without modification. RBE-weighted dose models, such as the Microdosimetric Kinetic Model (MKM) and its variants, are widely used in clinical practice to estimate cell survival and prescribe biologically equivalent doses \cite{Inaniwa2018SMK, bellinzona2021linking}. These models extend the LQ framework to account for the increased effectiveness of high-LET radiation, but they rely on population-averaged quantities and are not designed to capture the stochastic, cell-level dynamics that govern radiation response in heterogeneous tissues.

A central source of this heterogeneity is the variability in radiosensitivity across the cell population, arising from differences in cell-cycle phase, intrinsic repair capacity, and local microenvironmental conditions. Radiosensitivity varies markedly across the cell cycle, with cells in G$_2$/M among the most sensitive and cells in late S phase among the most resistant \cite{hall2006radiobiology}, and the phase distribution differs between the proliferating rim and the quiescent core of a tumor. Oxygenation is a further well-known determinant. In fact, molecular oxygen fixes radiation-induced radicals into stable DNA lesions, so that hypoxic cells, common in poorly vascularised tumor regions, are substantially more radioresistant than well-oxygenated ones, an effect quantified by the oxygen enhancement ratio (OER) that decreases with increasing LET \cite{chang2021oxygen, Scifoni2013, Sokol2023CarbonHypoxia}. Superimposed on these sources is the stochastic variability in the number and spatial distribution of radiation-induced lesions at the single-cell level, which is largest for densely ionizing particles. Capturing this multi-layered heterogeneity and its consequences for treatment outcome requires models that explicitly resolve the biological state of individual cells rather than population averages. A mechanistic framework for TCP and NTCP prediction that accounts for oxygenation heterogeneity and cell-to-cell variability in radiosensitivity is developed in \cite{battestini2026ntcp}; the present work focuses on the complementary aspect of cell dynamics.

The response to radiation also depends on time. When the same dose is protracted or split into fractions, sublethal damage is partly repaired between particle arrivals or between fractions, surviving cells redistribute among cell-cycle phases, hypoxic regions may reoxygenate, and the population regrows; together, these processes form the classical four Rs of radiobiology \cite{hall2006radiobiology}. In the LQ framework, the effect of dose protraction is described by the Lea--Catcheside factor \cite{LeaCatcheside1942}, which rescales the quadratic term to account for repair during irradiation, whereas redistribution and repopulation are added through separate, empirical corrections. Moreover, lethally damaged cells do not disappear during irradiation. In most solid tumor cell lines, cells die predominantly through mitotic catastrophe after one or more abortive divisions over a period of one to several days \cite{Chu2002, ErikssonStigbrand2010}. During this time, they still occupy space, alter the local environment, and contribute to the measured cell number or tumor volume. A description in which these processes arise from the same underlying cellular dynamics, rather than being added as independent corrections, requires an explicit representation of individual cells in time. Capturing these effects is essential because radiotherapy is delivered in fractions, and a mechanistic understanding of how the time structure of irradiation shapes the cellular response is a prerequisite for the biological optimization of treatment schedules.

Considerable progress has been made in modeling individual components of the radiation response cascade. Track-structure Monte Carlo codes such as Geant4-DNA provide detailed descriptions of physical and early chemical interactions \cite{Bernal2015Geant4DNA}, while mechanistic simulation tools link microscopic energy deposition to DNA damage yields and repair \cite{Kundrat2010MSA}. However, bridging the gap between these physical descriptions and cell- and tissue-level biological outcomes in a fully mechanistic and stochastic manner remains an open challenge.

The Generalized Stochastic Microdosimetric Model (\GSM{}) \cite{cordoni2021generalized, cordoni2022cell, missiaggia2024cell} was developed to address this gap. \GSM{} describes the formation, interaction, and repair of lethal and sublethal DNA lesions through a microdosimetric master equation, naturally producing non-Poissonian lesion statistics, multi-exponential repair kinetics, and dose-dependent transitions from linear-quadratic to linear survival behavior \cite{cordoni2022multiple}. Although originally developed in the context of particle therapy, \GSM{} provides a mechanistic generalization of the LQ framework applicable to any radiation quality, including conventional photon beams, thereby making it a unified model for both standard and particle radiotherapy. It has been validated over a broad range of LET values and particle species \cite{bordieri2024validation}, coupled with detailed track-structure simulations \cite{bordieri2024validation, bordieri2025integrating}, and extended to include the homogeneous chemical stage, enabling the modeling of time-resolved damage formation and ultra-high dose rate (FLASH) effects \cite{battestini2023across, battestini2025multiscale}.

However, \GSM{}, in its current form, remains limited in biological resolution. While recent extensions have enabled prediction of clinically relevant endpoints, such as tumor control probability (TCP) and normal tissue complication probability (NTCP), by incorporating tissue heterogeneity and stochastic cell killing \cite{battestini2026ntcp}, the underlying cellular dynamics remain implicit. In particular, \GSM{} does not explicitly model cell-cycle progression, proliferation, or redistribution, nor does it resolve how individual cells evolve through distinct biological states following irradiation. Cellular fate is effectively reduced to survival or death, without accounting for radiation-induced cell-cycle arrest, delayed mitotic death, or the temporal coupling between DNA repair and cell-cycle phase. As a consequence, key radiobiological mechanisms, such as redistribution, repopulation, and the dependence of the response on the time structure of irradiation, are not represented at the single-cell level. Bridging this gap requires extending \GSM{} beyond population-level survival descriptions towards an explicit, stochastic representation of individual cells and their internal state dynamics.

Agent-based models (ABMs) provide a natural framework for this purpose \cite{zhang2009multiscale, west2023agent, Cogno2024ABMReview}. By representing each cell as an autonomous agent with its own internal state, stochastic behavior, and local environment, ABMs can capture non-local energy deposition, cell-cycle dependence, heterogeneous repair capacity, and spatial organization, features that are particularly relevant in particle therapy, where the spatial scale of energy deposition is comparable to cell dimensions. In biology and oncology, ABMs have been widely applied to study tumor dynamics and interactions with the microenvironment \cite{Cogno2024ABMReview, Norton2019MultiscaleABM}. In radiotherapy, however, their application has largely focused on conventional photon treatments \cite{jalalimanesh2017simulation, kempf2015spatio, garcia2024topas, Kunz2024AMBER}, and only a limited number of studies have addressed particle therapy, typically involving simplified irradiation scenarios, coarse biological endpoints, or small cell populations \cite{Cogno2024LungFibrosis}.

Extending ABMs to particle therapy at relevant scales remains challenging, primarily because of the computational cost of coupling the microdosimetric variability of energy deposition and the stochastic biological response of each cell with large, spatially resolved populations. Models linking Monte Carlo simulations of energy deposition to the radiobiological effect at the multicellular level have begun to emerge \cite{garcia2024topas, Cogno2024LungFibrosis}, but they typically rely on simplified dose--response relations at the cell level, restrict the irradiation to a single time point, or address populations of limited size. None currently provides a unified treatment across particle types and energies within a stochastic, cell-resolved framework that also resolves the time structure of irradiation at an affordable computational cost.

In this work, we present an agent-based extension of \GSM{} that addresses these limitations by exploiting the intrinsic Markovian structure of the \GSM{} formalism. Cellular radiation response is reformulated as a next-event, continuous-time Markov process in which DNA damage induction, repair, misrepair, cell-cycle progression, division, delayed radiation-induced death, and migration are treated as competing stochastic events with well-defined transition rates. Each cell evolves through discrete events, such as lesion repair, cell-cycle transitions, or fate commitment, rather than through explicit time-stepping, and the global scheduler updates only the cells affected by each event, so that the cost per event does not depend on the size of the population while the full stochastic fidelity of \GSM{} is preserved.

A key feature of the framework is the explicit, particle-by-particle modeling of the irradiation process. The arrival time of each particle is sampled stochastically from the prescribed dose rate and coupled to repair via the shared event queue, so that residual damage from previous particles is explicitly carried forward and no dose-protraction factor is imposed. This makes the framework suited to dose-rate and fractionation studies. We show further that, in the limit of a static population, where the classical theory applies, the dose-rate dependence predicted by the model must reproduce the Lea--Catcheside factor, while cell-cycle progression, delayed death, and proliferation determine the response beyond this limit.

The main contributions of this work are the following:
\begin{enumerate}
\item an event-driven, single-cell formulation of GSM$^2$ coupled to cell-cycle progression, proliferation, contact inhibition, delayed radiation-induced death, and migration, whose cost per event is independent of population size, so that spheroids of $\sim \times10^5$ cells can be simulated with track-structure resolution on a single workstation;
\item a verification of the coupling between particle arrival and repair against the Lea--Catcheside factor for $^{1}$H and $^{12}$C ions over four orders of magnitude in dose rate, and an explanation of the loss of dose-rate sparing at high LET in terms of the contribution of inter-track lesion interactions to cell killing;
\item an application to three-dimensional tumor spheroids irradiated with protons and carbon ions, in which the cell-cycle response to single doses is resolved, the contributions of repair, reassortment, and repopulation are separated in split-dose experiments, fractionated schedules of equal total dose are compared, the response to fractionated irradiation is resolved into surviving and committed cells, and the variability between stochastic realizations is quantified;
\item a benchmark against the clonogenic survival of V79 spheroids irradiated with X-rays and carbon ions under different oxygen tensions \cite{staab2004response}, in which the model, calibrated on a single aerobic X-ray curve, predicts the remaining conditions without further adjustment.
\end{enumerate}
The model and its implementation are described in Section~\ref{SEC:methods}, the results are presented in Section~\ref{SEC:results}, and their implications and limitations are discussed in Section~\ref{SEC:discussion}.

\section{Material and Methods}
\label{SEC:methods}

The framework couples the \GSM{} description of DNA damage induction and repair with an agent-based representation of a cell population. Table~\ref{tab:new_vs_inherited} separates the components inherited from \GSM{} from those introduced in this work; the components are described in the following subsections, and their parameters are summarised in Table~\ref{tab:params} and Appendix~\ref{app:params}.

\begin{table}[htbp]
\centering
\caption{Components inherited from \GSM{} and components introduced in this work.}
\label{tab:new_vs_inherited}
\begin{tabular}{p{0.45\linewidth}p{0.45\linewidth}}
\hline
Inherited from \GSM{} \cite{cordoni2021generalized, cordoni2022cell, missiaggia2024cell, bordieri2024validation, battestini2025multiscale} & Introduced in this work \\
\hline
Domain-based nucleus geometry and amorphous-track energy deposition & Particle-by-particle arrival in time at a prescribed dose rate \\
Stochastic induction of sublethal ($X$) and lethal ($Y$) lesions, OER modulation & Event-driven coupling of particle arrivals, repair and cell dynamics \\
Repair, misrepair and pairwise interaction kinetics ($r$, $a$, $b$) & Phase-dependent kinetics with explicit cell-cycle progression and arrest \\
Single-cell survival probability & Fate commitment and delayed radiation-induced death \\
Oxygen gradient in 3D biological structures & Proliferation, contact inhibition (quiescence) and migration on a 3-D lattice \\
 & Priority-queue scheduler with local updates \\
\hline
\end{tabular}
\end{table}

\subsection{Cell population}

We consider a population of $N_{c}$ cells, each modeled as a sphere of radius $R_{\mathrm{cell}}$ placed on the sites of a regular cubic lattice with spacing $h = 2R_{\mathrm{cell}}$, each site hosting at most one cell. Each site has 26 neighbors, and periodic boundary conditions are applied; the half-size of the simulation box, $X_{\mathrm{box}}$, is chosen so that the spheroid and its growth during the simulated period remain well inside the box (Section~\ref{SEC:protocols}). A spheroid of radius $R_{\mathrm{sph}}$ is obtained by occupying all sites within that distance from the center. Each cell contains a cylindrical nucleus of radius $R_N$ and length $2R_N$, in accordance with the standard \GSM{} formulation \cite{battestini2025multiscale}. The nucleus is subdivided into $N_d$ domains of radius $r_d$, consistent with the multiscale MKM approach \cite{bellinzona2021linking}, and the single-cell response to ionizing radiation is computed using the \GSM{} framework \cite{cordoni2021generalized, cordoni2022cell}.

\subsection{Radiation field and particle arrival modeling}
\label{SSEC:AT}

The population is irradiated by a parallel beam of radius $R_{\mathrm{beam}}$ that propagates along the $z$-axis and covers the entire spheroid. The radiation field is characterized by the particle type, energy, LET, dose, and temporal structure. The particle fluence required to deliver a dose $D$ follows from the relation between dose, fluence and LET \cite{ICRU85}:
\begin{equation}
F\,[\mathrm{cm^{-2}}] = \frac{D\,[\mathrm{Gy}]}{1.602 \times 10^{-9}\,\mathrm{LET}\,[\mathrm{keV}\,\mu\mathrm{m}^{-1}]},
\label{eq:fluence}
\end{equation}
and the mean number of particles crossing the beam area is $N_{\mathrm{par}} = F\,\pi R_{\mathrm{beam}}^2$ (with $R_{\mathrm{beam}}$ in cm). For a dose rate $\dot D$, the irradiation lasts $T = D/\dot D$ and particles arrive as a homogeneous Poisson process of rate
\begin{equation}
\lambda_{\mathrm{p}} = \frac{\dot D\,N_{\mathrm{par}}}{D},
\label{eq:arrival_rate}
\end{equation}
so that inter-arrival times are exponentially distributed with mean $1/\lambda_{\mathrm{p}}$. Each arrival is a track with a uniformly sampled entry point within the beam, whose energy deposition is computed as described in Section~\ref{SEC:AT_model}. Acute irradiation corresponds to the limit $\dot D \to \infty$, in which all particles arrive at $t = 0$.

\subsection{Oxygenation}
\label{SEC:oxygen}

The oxygen distribution inside the spheroid is modeled as the quasi-steady-state solution of a reaction--diffusion equation. Assuming radial symmetry, the oxygen fraction $C(r)$ at distance $r$ from the spheroid centre satisfies \cite{GrimesFletcherPartridge2014}
\begin{equation}
  \frac{D_{\mathrm{O_2}}}{r^2}\frac{\mathrm{d}}{\mathrm{d}r}\!\left(r^2 \frac{\mathrm{d}C}{\mathrm{d}r}\right) = \rho_{\mathrm{cell}}(r)\, q_0,
  \label{eq:O2diffusion}
\end{equation}
where $D_{\mathrm{O_2}}$ is the oxygen diffusion coefficient in tissue, $\rho_{\mathrm{cell}}(r)$ is the local volume packing fraction of living cells, and $q_0$ is the zero-order volumetric oxygen consumption rate at unit packing density.

For a spatially uniform packing density $\rho$, with $A = \rho\, q_0$, integrating Equation~\eqref{eq:O2diffusion} twice in spherical symmetry and imposing the surface boundary condition $C(R) = O_{2,\mathrm{rim}}$ and, where the core becomes anoxic, the no-flux conditions $C(r_{\mathrm{n}}) = 0$ and $C'(r_{\mathrm{n}}) = 0$ at the necrotic boundary $r_{\mathrm{n}}$, yields the closed-form profile \cite{Greenspan1972, GrimesKellyPartridge2014}
\begin{equation}
  C(r) =
  \begin{cases}
    O_{2,\mathrm{core}}, & r \leq r_{\mathrm{n}}, \\[8pt]
    O_{2,\mathrm{rim}} - \dfrac{A}{6D_{\mathrm{O_2}}}\bigl(R^2 - r^2\bigr) - \dfrac{A\, r_{\mathrm{n}}^3}{3D_{\mathrm{O_2}}}\!\left(\dfrac{1}{R} - \dfrac{1}{r}\right), & r_{\mathrm{n}} < r \leq R,
  \end{cases}
  \label{eq:O2profile}
\end{equation}
where $O_{2,\mathrm{core}}$ is a small nonzero floor value reflecting chronic hypoxia rather than strict anoxia. The necrotic radius $r_{\mathrm{n}}$ is the root of
\begin{equation}
  O_{2,\mathrm{rim}} = \frac{A}{6D_{\mathrm{O_2}}}\left(R^2 - 3\,r_{\mathrm{n}}^2 + \frac{2\,r_{\mathrm{n}}^3}{R}\right),
  \label{eq:rnecrotic}
\end{equation}
solved numerically by bisection. When the central oxygen value $O_{2,\mathrm{rim}} - AR^2/6D_{\mathrm{O_2}}$ is non-negative, which holds for spheroids smaller than the characteristic radius
\begin{equation}
  R^* = \sqrt{\frac{6D_{\mathrm{O_2}}\,O_{2,\mathrm{rim}}}{A}},
  \label{eq:Rstar}
\end{equation}
the entire spheroid is viable and $r_{\mathrm{n}} = 0$. The local oxygen fraction $C(r_i)$ assigned to cell $i$ enters the OER of Section~\ref{SEC:DNA}, so that the lesion yield is modulated by the local oxygenation \cite{Scifoni2013}.

We estimate the consumption rate $q_0$ from the volumetric oxygen consumption rate $a\Omega \approx \SI{15}{mmHg\,s^{-1}}$ reported in \cite{MuellerKlieser1986} for spheroids under 20\% O$_2$, which already represents the reduced in-situ rate of large spheroids; \cite{Freyer1985} independently showed that this in-situ rate is approximately 2.5-fold lower than the single-cell rate at a spheroid diameter of \SI{1300}{\micro\metre}, reflecting chronic metabolic adaptation to nutrient limitation. Converting $a\Omega$ from mmHg\,s$^{-1}$ to \%\,O$_2$\,s$^{-1}$ and adopting unit packing density gives $A = \rho\,q_0 \approx 2.0\;\%\,\mathrm{s}^{-1}$. This predicts a viable rim of approximately 130--140~\si{\micro\metre} for spheroids of radius 250--500~\si{\micro\metre}, consistent with the experimentally observed diffusion-limited viable rim of 150--200~\si{\micro\metre} \cite{Freyer1985, GrimesKellyPartridge2014}, and a necrosis onset radius $R^* \approx \SI{206}{\micro\metre}$. The profile is computed from the current spheroid (its center, radius, and local cell density) immediately before each irradiation and kept fixed between irradiations; in fractionated schedules, the oxygenation therefore changes from one fraction to the next as dying cells are removed and the spheroid grows. The parameters are listed in Table~\ref{tab:O2params}. 

\begin{table}[htbp]
\centering
\caption{Oxygenation model parameters.}
\label{tab:O2params}
\begin{tabular}{llll}
\hline
Parameter & Symbol & Value & Reference \\
\hline
Surface O$_2$ fraction & $O_{2,\mathrm{rim}}$ & $7.0\%$ & \cite{McKeown2014, Carreau2011} \\
Core O$_2$ floor & $O_{2,\mathrm{core}}$ & $0.1\%$ & \cite{WoutersBrown1997} \\
Diffusion coefficient & $D_{\mathrm{O_2}}$ & $\SI{2.0e3}{\micro\metre\squared\per\second}$ & \cite{GrimesKellyPartridge2014, GrimesFletcherPartridge2014} \\
Volumetric consumption & $A = \rho\,q_0$ & $\approx\SI{2.0}{\percent\per\second}$ & \cite{MuellerKlieser1986} \\
&&&\cite{Freyer1985}\\
\hline
\end{tabular}
\end{table}

\subsection{Energy deposition model}
\label{SEC:AT_model}

We adopt the Kiefer--Chatterjee parameterization of the Amorphous Track (AT) model, as implemented in the MKM \cite{Kase2008} and described in detail in \cite{bellinzona2021linking}. The AT model assumes cylindrical symmetry of energy deposition around the particle trajectory. Energy is deposited uniformly within a core of radius $R_c$; beyond $R_c$ the radial dose decreases as $r^{-2}$ up to the maximum penumbra radius $R_p$:
\begin{equation}
D_{\mathrm{AT}}(r) =
\begin{cases}
\dfrac{1}{\pi R_c^2}\Bigl(\mathrm{LET}^\ast - 2\pi K_p \ln\!\dfrac{R_p}{R_c}\Bigr), & r \le R_c, \\[10pt]
\dfrac{K_p}{r^2}, & R_c < r \le R_p, \\[6pt]
0, & r > R_p,
\end{cases}
\label{eq:DarcKC}
\end{equation}
where $K_p$ is the penumbra scaling factor and $\mathrm{LET}^\ast$ denotes the LET expressed in dose units. The specific energy $z^{(j,d)}$ deposited by a track in domain $d$ of cell $j$ is obtained from Equation~\eqref{eq:DarcKC} at the distance between the track and the domain center. Within each nucleus, the track-segment condition is assumed, meaning that the LET is taken as constant along the nucleus length $2R_N$.

Across the target, the track-segment condition \cite{Kase2008} does not need to be assumed: each particle loses energy along its path, so that its energy, its LET, and the radial dose profile of its track change with depth, and cells at different depths are irradiated by different radiation qualities, as along a Bragg curve. The target is divided into layers, each perpendicular to the beam and of thickness $ h$, one for each lattice plane. For each ion, the continuous-slowing-down range in water,
\begin{equation}
\mathrm{R}(E) = \int_{E_{\min}}^{E} \frac{\mathrm{d}E'}{S(E')},
\label{eq:csda}
\end{equation}
is computed once from tabulated stopping powers $S(E)$ in water, and the kinetic energy at the entrance of layer $i+1$ is obtained by range inversion,
\begin{equation}
E_{i+1} = \mathrm{R}^{-1}\bigl(R(E_i) - h\bigr),
\label{eq:range_step}
\end{equation}
starting from the beam energy at the first layer that contains cells; layers without cells are treated as non-absorbing, so that depth is measured from the surface of the target. In each layer $i$, the LET is evaluated at the local energy, $\mathrm{LET}_i = S(E_i)$, and the amorphous-track parameters $R_c$, $R_p$ and $K_p$ of Equation~\eqref{eq:DarcKC} are recomputed from $E_i$, so that the radial dose profile and its normalization to the LET refer to the same energy. Tracks are straight and parallel to the beam axis, with lateral scattering neglected: each particle crosses all layers at the same transverse position, and the dose to the domains within a layer is computed using that layer's track parameters. A particle whose residual energy falls below the validity range of the track model is considered stopped and deposits no dose in deeper layers. The number of particles is computed from the entrance LET, Equation~\eqref{eq:fluence}, so that the prescribed dose refers to the entrance of the target. Because the layer thickness $h = \SI{30}{\micro\metre}$ exceeds the nucleus length, the LET is constant within each nucleus, as stated above, but changes from one layer of cells to the next. 

\subsection{DNA damage induction model}
\label{SEC:DNA}

Following the \GSM{} framework \cite{cordoni2021generalized}, two types of radiation-induced DNA lesions are considered: sublethal lesions $X$ and lethal lesions $Y$. Each energy deposition $z^{(j,d)}$ induces Poisson-distributed lesion counts in domain $d = 1,\dots,N_d$ of cell $j$:
\begin{equation}
Z_X^{(j,d)} \sim \mathrm{Po}\bigl(\kappa\, z^{(j,d)}\bigr), \qquad Z_Y^{(j,d)} \sim \mathrm{Po}\bigl(\lambda\, z^{(j,d)}\bigr).
\end{equation}
The sublethal-lesion yield per domain is $\kappa = \kappa_{\mathrm{cell}}/N_d$, where the yield per cell depends on LET, ion species, and oxygenation as
\begin{equation}
\kappa_{\mathrm{cell}}(\mathrm{LET},[\mathrm{O_2}]) = \frac{\bar{\kappa}_{\mathrm{cell}}(\mathrm{LET})}{\mathrm{OER}(\mathrm{LET},[\mathrm{O_2}])}\,,
\label{eq:kappa}
\end{equation}
with LET in keV\,\textmu m$^{-1}$ and $\bar{\kappa}_{\mathrm{cell}}$ the ion-specific yield under aerobic conditions, given by the analytical parameterization of track-structure simulations derived in \cite{kundrat2020analytical}, which saturates at high LET. The lethal-lesion yield is $\lambda = 10^{-3}\,\kappa$. The oxygen enhancement ratio is \cite{Scifoni2013, Epp1972}
\begin{equation}
\mathrm{OER}(\mathrm{LET},[\mathrm{O_2}]) = \frac{K_{O_2}\,\dfrac{K_{\mathrm{LET}}M+\mathrm{LET}^\gamma}{K_{\mathrm{LET}}+\mathrm{LET}^\gamma}+[\mathrm{O_2}]}{K_{O_2}+[\mathrm{O_2}]},
\label{eq:OER}
\end{equation}
with $M = 3.4$, $K_{O_2} = 0.41\%$, $K_{\mathrm{LET}} = 8.27\times10^{5}$ and $\gamma = 3$ \cite{Scifoni2013}, $[\mathrm{O_2}]$ expressed in \% O$_2$.

\subsection{Single-cell response model}
\label{SEC:CellR}

Given the lesions in each domain $d$ of cell $j$, damage evolution is governed by the \GSM{} reaction pathways \cite{cordoni2021generalized}:
\begin{equation}
\label{EQ:path_abr}
\begin{cases}
X \xlongrightarrow{r} \emptyset, & \text{repair of sublethal lesions,} \\
X \xlongrightarrow{a} Y, & \text{conversion of sublethal to lethal lesions,} \\
X + X \xlongrightarrow{b} Y, & \text{pairwise interaction of sublethal lesions,} \\
\emptyset \xlongrightarrow{\lambda_{\mathrm{p}}} \bigl(Z_X^{(j,d)}, Z_Y^{(j,d)}\bigr), & \text{damage induction by particle arrival.}
\end{cases}
\end{equation}
The rates $r$, $a$, $b$ depend on the cell-cycle phase (Section~\ref{SEC:params}). The domain-wise propensities are
\begin{equation}
r\,X[d], \qquad a\,X[d], \qquad b\,X[d]\,(X[d]-1),
\end{equation}
with total rate $a_0 = \sum_{d=1}^{N_d}\bigl(r\,X[d] + a\,X[d] + b\,X[d]\,(X[d]-1)\bigr)$. Between particle arrivals, the lesion state of each cell is evolved with Gillespie's Stochastic Simulation Algorithm (SSA) \cite{Gillespie1977}: the waiting time to the next reaction is $\Delta t = -\tau\ln(u)/a_0$, with $u \sim U(0,1)$, and the reaction and domain are chosen with probabilities proportional to their propensities. The factor $\tau$ rescales the time unit of the repair kinetics. Survival after acute irradiation depends only on the ratios $a/r$ and $b/r$, so the absolute time scale of repair is not determined by the calibration to acute survival data (Section~\ref{SEC:params}); we set $\tau = 4$, which gives an effective repair half-time $\tau\ln 2/(r+a) \approx 1$~h for G$_1$ cells, consistent with split-dose recovery times of the order of one hour \cite{hall2006radiobiology}. The cell survives the current damage when all sublethal lesions have been repaired; the first misrepair or interaction event, or the presence of a lethal lesion, commits the cell to death (Section~\ref{SEC:fate}). If a new particle arrives before either outcome, its lesions are added to the current state, and the SSA resumes; by the process's memoryless property, this is exact.

\subsection{Fate assignment and delayed death}
\label{SEC:fate}

After irradiation, each damaged cell is assigned one of three outcomes: (i) \emph{lethal commitment}, if a lethal lesion is present or forms before repair is complete; (ii) \emph{recovery}, at the time $t_R$ at which the last sublethal lesion is repaired; or (iii) \emph{pending}, if neither has occurred within the current repair window, in which case the residual lesions are carried forward. Cells that are not damaged, or that are already committed to death, are not modified. A cell committed to death becomes \emph{doomed}; it keeps its lattice site, does not cycle or divide, and is removed after a delay $T_{\mathrm{death}}$ measured from the commitment time $t_L$, the time of the first lethal event in the \GSM{} kinetics, so that the cell is removed at $t_L + T_{\mathrm{death}}$, with
\begin{equation}
T_{\mathrm{death}} \sim \mathrm{Gamma}\!\left(k_{\mathrm{d}},\, \frac{m_{\phi}\, T_c}{k_{\mathrm{d}}}\right),
\label{eq:death_delay}
\end{equation}
where $T_c$ is the mean cell-cycle length, $k_{\mathrm{d}} = 3$ (coefficient of variation $1/\sqrt{3}$), and $m_{\phi}$ depends on the phase $\phi$ at commitment: $m_{\mathrm{G_2}} = m_{\mathrm{M}} = 1$, $m_{\mathrm{S}} = 1.25$, $m_{\mathrm{G_1}} = 1.5$ and $m_{\mathrm{G_0}} = 2.5$. This reflects the observation that, in most solid tumor cell lines, lethally irradiated cells die predominantly through mitotic catastrophe after one or more abortive divisions, over one to several days \cite{Chu2002, ErikssonStigbrand2010}: cells committed close to mitosis die first, and quiescent cells, which have no mitosis to trigger death, persist longest. In addition, a cell in M phase whose mitosis ends before its damage is repaired is committed to death at the end of mitosis.

\subsection{Single-cell agent definition and cell cycle}
\label{SEC:cycle}

Every cell is characterized by its lattice position, cell-cycle phase, oxygenation, and lesion state. The cell cycle takes values in $\{\mathrm{G_1},\mathrm{S},\mathrm{G_2},\mathrm{M}\}$, with phase durations sampled from Gamma distributions, $\tau_{\phi} \sim \mathrm{Gamma}(k_{\phi},\theta_{\phi})$, $\mathbb{E}[\tau_{\phi}] = k_{\phi}\theta_{\phi}$, with the parameters in Table~\ref{tab:params}; the mean values (G$_1$ 11~h, S 8~h, G$_2$ 4~h, M 1~h; $T_c = 24$~h) reflect typical durations reported for mammalian cells, with a dominant G$_1$ phase, an intermediate S phase and comparatively short G$_2$ and M phases \cite{hall2006radiobiology, Alberts2015MBoC}. These are generic values for a mammalian tumor cell, as in previous agent-based spheroid models that use the same phase-specific radiosensitivity \cite{kempf2015spatio}. Upon completing M phase with at least one empty neighboring site, a cell divides: one daughter remains at the parent site, and the other occupies an empty neighboring site chosen uniformly at random; both daughters start in G$_1$ with newly sampled phase durations.

Proliferation is limited by contact inhibition, so that a cell with no empty neighboring site among its 26 neighbors enters quiescence (G$_0$), does not progress through the cycle, and re-enters G$_1$ with a newly sampled duration as soon as a neighboring site is freed by the death or migration of another cell. G$_0$ cells use the radiosensitivity parameters of G$_1$ cells.

After irradiation, the cell-cycle clock of a damaged cell is held until its sublethal lesions are repaired: a cell in G$_1$, S, or G$_2$ resumes progression at the recovery time $t_R$, while a cell in M dies if mitosis ends before repair is complete (Section~\ref{SEC:fate}).

\subsection{Cell migration}
\label{SEC:migration}

Cell migration is modeled as an event-driven random walk on the lattice. A cell at site $i$ with $n_{\mathrm{empty}}(i)$ empty neighbouring sites hops at total rate
\begin{equation}
    \Lambda_i = \frac{D_m}{h^2}\, n_{\mathrm{empty}}(i),
    \label{eq:mig_rate}
\end{equation}
where $D_m$ [\textmu m$^2$\,h$^{-1}$] is the random-motility coefficient and $h = 2R_{\mathrm{cell}}$ is the lattice spacing, so that the waiting time to the next hop is $t_{\mathrm{migrate}}^{(i)} \sim \mathrm{Exp}(\Lambda_i)$. If $n_{\mathrm{empty}}(i) = 0$, then $\Lambda_i = 0$. The destination is chosen uniformly among the empty neighbors. The complete cell state (phase, timers, lesions) is moved to the destination, the origin site is vacated, and the empty-neighbor counts, quiescence states, and migration rates of all cells neighboring either site are updated. Quiescent cells that gain an empty neighbor re-enter G$_1$ unless they are doomed. Migration was switched off in all simulations of this work ($D_m = 0$), so that the reported dynamics result from proliferation, quiescence, and death only.

\subsection{Event-driven simulation algorithm}
\label{SEC:algorithm}

System dynamics are simulated with a next-event algorithm for continuous-time Markov processes \cite{Gillespie1977, GibsonBruck2000}. Four event categories compete in a single global timeline: (i) particle arrivals; (ii) repair and lethal events of the lesion kinetics; (iii) cell-cycle transitions, divisions, and removal of doomed cells; and (iv) migration hops. For each cell, the times of its next cell-cycle event, its death (if doomed), and its next migration hop are stored as absolute times in a binary min-heap. At each iteration, the earliest event is extracted from the heap, the global clock is advanced to its time, and the event is executed. Because an event can modify only cells within two lattice sites of the cell where it occurs (the daughter and both neighborhoods for a division, the neighbors for a death, the destination and both neighborhoods for a migration hop), only the timers of these cells are updated and reinserted in the heap; outdated heap entries are discarded when they reach the top (lazy invalidation). The cost of an event is therefore $\mathcal{O}(\log N)$ and independent of the population size, and no fixed time step is introduced. Particle arrivals are processed on the same timeline: during continuous irradiation, the dose is delivered as a sequence of short pulses, each followed by the repair window until the next pulse and by cell dynamics over the same interval.

\subsection{Model parameters}
\label{SEC:params}

Of all the parameters of the model, only the \GSM{} rate constants, the repair rate $r$, the conversion rate $a$, and the pairwise interaction rate $b$ of each cell-cycle phase are fitted in this work. All other parameters, namely the geometry, the lesion yields, the oxygen effect, the cell-cycle durations and the oxygenation, are taken from previous work, and the remaining ones, the repair time scale, the quiescence rule and the death delay, are modeling choices described in the corresponding sections (Table~\ref{tab:params}). The rate constants are calibrated so that the \GSM{} survival curves reproduce the phase-specific LQ parameters $(\alpha, \beta)$ reported in the literature \cite{hall2006radiobiology, Sinclair1966}: G$_1$, $\alpha = 0.351$~Gy$^{-1}$, $\beta = 0.040$~Gy$^{-2}$; S, $\alpha = 0.124$~Gy$^{-1}$, $\beta = 0.029$~Gy$^{-2}$; G$_2$/M, $\alpha = 0.793$~Gy$^{-1}$, $\beta = 0$. These values reflect the established pattern of phase-dependent radiosensitivity, with S-phase cells the most resistant and G$_2$/M cells the most sensitive, with a purely linear response \cite{hall2006radiobiology}, and have been used as the representative radiosensitivity of a generic tumor cell in previous agent-based spheroid models \cite{kempf2015spatio}. The calibration is shown in Appendix~\ref{app:Fit}, and all parameters are listed in Table~\ref{tab:params} and Appendix~\ref{app:params}.

\begin{table}[htbp]
\centering
\caption{Model parameters. Only the \GSM{} rate constants (\emph{fitted}) are calibrated in this work; all other parameters are taken from the cited works or set as described in the indicated section.}
\label{tab:params}
\begin{tabular}{lll}
\toprule
Parameter & Value & Source \\
\midrule
\multicolumn{3}{l}{\textit{Geometry}} \\
\addlinespace[2pt]
Cell radius $R_{\mathrm{cell}}$, lattice spacing $h$ & \SI{15}{\micro\metre}, \SI{30}{\micro\metre} & Sec.~\ref{SEC:protocols} \\
Nucleus radius $R_N$, domain radius $r_d$ & \SI{7.2}{\micro\metre}, \SI{0.8}{\micro\metre} & \cite{battestini2025multiscale} \\
\addlinespace[6pt]
\multicolumn{3}{l}{\textit{Damage induction}} \\
\addlinespace[2pt]
Sublethal yield $\kappa_{\mathrm{cell}}$ & Equation~\eqref{eq:kappa} & \cite{kundrat2020analytical} \\
Lethal yield $\lambda$ & $10^{-3}\kappa$ & \cite{bordieri2024validation} \\
OER parameters $M$, $K_{O_2}$, $K_{\mathrm{LET}}$, $\gamma$ & 3.4, 0.41\%, $8.27\times10^{5}$, 3 & \cite{Scifoni2013} \\
\addlinespace[6pt]
\multicolumn{3}{l}{\textit{Repair kinetics} ($r$, $a$, $b$ in h$^{-1}$)} \\
\addlinespace[2pt]
G$_1$, G$_0$ & 2.780, 0.01287, 0.04030 & \textbf{fitted} \\
S & 5.840, 0.00589, 0.05794 & \textbf{fitted} \\
G$_2$, M & 1.772, 0.02431, $5.70\times10^{-5}$ & \textbf{fitted} \\
Time scaling $\tau$ & 4 & Sec.~\ref{SEC:CellR} \\
\addlinespace[6pt]
\multicolumn{3}{l}{\textit{Cell cycle} ($k_{\phi}$, $\theta_{\phi}$ [h])} \\
\addlinespace[2pt]
G$_1$, S, G$_2$, M & (5.5, 2), (4, 2), (2, 2), (0.5, 2) & \cite{hall2006radiobiology} \\
Quiescence & all 26 neighbours occupied & Sec.~\ref{SEC:cycle} \\
\addlinespace[6pt]
\multicolumn{3}{l}{\textit{Fate}} \\
\addlinespace[2pt]
Death delay shape $k_{\mathrm{d}}$ & 3 & Sec.~\ref{SEC:fate} \\
Mean delay $m_{\phi}T_c$ & $m = 1$ (G$_2$, M), 1.25 (S), 1.5 (G$_1$), 2.5 (G$_0$) & Sec.~\ref{SEC:fate} \\
\addlinespace[6pt]
\multicolumn{3}{l}{\textit{Oxygen}} \\
\addlinespace[2pt]
Oxygenation & Table~\ref{tab:O2params} & Sec.~\ref{SEC:oxygen} \\
\bottomrule
\end{tabular}
\end{table}

\subsection{Dose-rate dependence and the Lea--Catcheside factor}
\label{SEC:LC}

In the linear-quadratic (LQ) model, the effect of dose protraction is described by the Lea--Catcheside dose-protraction factor $G$, which multiplies the quadratic term \cite{LeaCatcheside1942}:
\begin{equation}
  S(D) = \exp\!\left(-\alpha D - G\,\beta D^{2}\right), \qquad G = \frac{2}{D^{2}} \int_{0}^{T} \dot D(t)\, \mathrm{d}t \int_{0}^{t} \dot D(t')\, e^{-\mu (t - t')}\, \mathrm{d}t',
  \label{eq:lc}
\end{equation}
where $T$ is the irradiation time and $\mu$ the repair rate of sublethal damage. For a constant dose rate $\dot D$, with $T = D/\dot D$, Equation~\eqref{eq:lc} reduces to
\begin{equation}
  G(x) = \frac{2\left(x - 1 + e^{-x}\right)}{x^{2}}, \qquad x = \mu T,
  \label{eq:lc_const}
\end{equation}
which tends to 1 for acute delivery ($x \to 0$) and to 0 for very protracted exposures ($x \to \infty$). The linear term is not affected, because single-track and directly lethal events do not depend on the time between particle arrivals.

In \GSM{}, sublethal lesions are removed by first-order repair (rate $r$) and by first-order conversion into lethal lesions (rate $a$), so the corresponding Lea--Catcheside repair rate is $\mu = (r + a)/\tau$, i.e.\ $\mu \approx 0.70$~h$^{-1}$ for G$_1$ cells.

The Lea--Catcheside factor is used only as an independent reference for the dose-rate response; it is not part of the agent-based simulation, in which the arrival time of each particle is sampled explicitly, and lesions are repaired continuously between arrivals. For each radiation quality, $\alpha_0$ and $\beta_0$ are obtained from a weighted LQ fit, constrained to $\alpha, \beta \geq 0$, to the simulated acute survival. The Lea--Catcheside survival $S_{\mathrm{LC}}(D, \dot D)$ is then computed from Equation~\eqref{eq:lc_const} at the same doses as the simulations. 

Two derived quantities summarise the dose-rate response. The quadratic share,
\begin{equation}
  q = \frac{\beta D_{\max}}{\alpha_0 + \beta D_{\max}},
\end{equation}
is the fraction of the logarithmic cell kill at the highest dose $D_{\max}$ due to the quadratic term, and therefore an upper limit on the achievable dose-rate effect. The dose-rate sparing, $\mathrm{SF}(\dot D, D_{\max})/\mathrm{SF}_{\mathrm{acute}}(D_{\max})$, is the gain in survival obtained by protracting the dose.

\subsection{Benchmark against V79 spheroid survival}
\label{SEC:staab_methods}

To test the predictive capability of the framework, we used the clonogenic survival of V79-379A spheroids (diameter $200 \pm 30$~\textmu m) irradiated in stirred medium with 250~kV X-rays and with 227~MeV/u carbon ions in the plateau region (LET 18~keV\,\textmu m$^{-1}$), at external pO$_2$ of 145 and 0~mmHg \cite{staab2004response}. Cells were dissociated and plated immediately after irradiation, so this endpoint tests the radiation-response component of the model independently of the growth dynamics. Spheroids of this size are fully oxygenated in a stirred medium, so all cells were assigned the oxygen level of the gas phase (145~mmHg $\simeq$ 19\% O$_2$, and 0.1\% for the nominally anoxic condition, which is not strictly anoxic). X-rays were represented by a low-LET field of 1.7~keV\,\textmu m$^{-1}$ under the track-segment condition. The LQ parameters of the reference radiation in this experiment ($\alpha = \SI{0.092}{\per\gray}$, $\beta = \SI{0.030}{\per\gray\squared}$ for aerobic X-rays \cite{staab2004response}) differ from those used to calibrate the phase-specific parameters of Section~\ref{SEC:params}, so the GSM$^2$ rates were recalibrated on the aerobic X-ray survival. Because the spheroid survival is an average over the cell-cycle phases and does not determine phase-specific parameters, we kept the relative radiosensitivity of the phases and multiplied the rates by two global factors, $s_a$ for the misrepair rate $a$ and $s_b$ for the interaction rate $b$, fitted by minimizing the squared difference of $\ln S$. Energy deposition and lesions were computed once per condition and dose, and survival was evaluated as the mean exact GSM$^2$ completion probability of the cells, so that the fit only re-evaluated the repair kinetics. Survival under anoxia and after carbon-ion irradiation was then predicted without further adjustment. OER and RBE were computed from LQ fits to the simulated curves, using dose ratios averaged over survival levels of 37\%, 10\%, 1\%, and 0.1\%, as in \cite{staab2004response}. Irradiation in the extended Bragg peak was not simulated because the LET spectrum of the mixed field was not reported. Results are reported in Appendix~\ref{app:staab}.
\subsection{Simulation protocols}
\label{SEC:protocols}

\textit{Continuous dose rate.} To verify the coupling between particle arrival and repair in isolation, a static population of G$_1$ cells (no cell-cycle progression, division or migration) was irradiated with $^{1}$H at \SI{100}{\MeV} (0.5--\SI{6}{\gray}), $^{12}$C at \SI{80}{\MeV}/u (0.25--\SI{3}{\gray}) and $^{12}$C at \SI{10}{\MeV}/u (0.1--\SI{1.5}{\gray}), at dose rates of $10^{-5}$, $5\times10^{-5}$, $10^{-4}$, $10^{-3}$ and \SI{e-2}{\gray\per\second}, and acutely, under uniform oxygenation (21\% O$_2$) and the track-segment condition. 

\textit{Single irradiation.} Spheroids of radius \SI{350}{\micro\metre} ($N_c \approx 6650$ cells) in a box of half-size \SI{550}{\micro\metre} were irradiated with \SI{2}{\gray} of $^{1}$H at \SI{100}{\MeV} or $^{12}$C at \SI{80}{\MeV}/u and followed for \SI{72}{\hour}, with the oxygen profile of Section~\ref{SEC:oxygen} and delayed death (Section~\ref{SEC:fate}); an unirradiated control used the same initial spheroid and random sequence. The cell configuration was recorded at 0, 6, 12, 24, 48, and \SI{72}{\hour}.

\textit{Single-dose response.} Spheroids of radius \SI{250}{\micro\metre} ($N_c \approx 2420$ cells) in a box of half-size \SI{550}{\micro\metre} were grown for \SI{24}{\hour} and irradiated with a single acute dose of 0, 2, 4 or \SI{6}{\gray} of $^{1}$H at \SI{100}{\MeV} or $^{12}$C at \SI{80}{\MeV}/u, with the oxygen profile of the spheroid at irradiation, and followed for \SI{168}{\hour}. Each condition was simulated in 3 realizations; in each realization, all conditions started from the same grown spheroid.

\textit{Split dose.} Spheroids of radius \SI{350}{\micro\metre} ($N_c \approx 6650$ cells) in a box of half-size \SI{550}{\micro\metre} were irradiated with two fractions of \SI{1.5}{\gray} separated by intervals of 0--\SI{30}{\hour}, with $^{1}$H at 2, 10 and \SI{100}{\MeV} and $^{12}$C at 15, 20 and \SI{80}{\MeV}/u. For $^{1}$H at \SI{100}{\MeV}, three configurations were compared: the full model; a model in which all cells have the radiosensitivity of G$_1$ cells; and the \GSM{} survival without cell dynamics between fractions.

\textit{Fractionation.} Spheroids of radius \SI{250}{\micro\metre} ($N_c \approx 2420$ cells) in a box of half-size \SI{450}{\micro\metre} were grown for \SI{24}{\hour} and irradiated with three fractions of \SI{2}{\gray}, \SI{24}{\hour} apart, followed by \SI{96}{\hour} of free evolution, with $^{1}$H at \SI{100}{\MeV}, $^{12}$C at \SI{80}{\MeV}/u, or no irradiation. At each fraction, the beam covered the current spheroid, the oxygen profile was recomputed, and the track-structure dose was computed on the current cell configuration; fates were assigned only to cells not already committed to death, so that cells doomed by an earlier fraction kept their scheduled death. Cell numbers were recorded every hour. Each condition was simulated in 5 independent realizations; realization $k$ used the same random sequence for all conditions up to the first fraction.

\textit{Fractionated schedules of equal total dose.} Spheroids of radius \SI{350}{\micro\metre} ($N_c \approx 6650$ cells) were grown for \SI{24}{\hour} and irradiated with a total dose of \SI{6}{\gray} of $^{1}$H at \SI{100}{\MeV} or $^{12}$C at \SI{80}{\MeV}/u, delivered as $1 \times \SI{6}{\gray}$, $3 \times \SI{2}{\gray}$ every \SI{24}{\hour}, or $2 \times \SI{3}{\gray}$ \SI{48}{\hour} apart, together with an unirradiated control, and followed for \SI{168}{\hour} after the first fraction. Before each fraction, the oxygen profile was recomputed on the current spheroid, including the local cell density, and the track-structure dose was computed on the current cell configuration. Each condition was simulated in 3 realizations starting from the same grown spheroid.

\textit{Regrowth after single irradiation.} Spheroids of radius \SI{150}{\micro\metre} ($N_c \approx 520$ cells) in a box of half-size \SI{400}{\micro\metre} were irradiated with single doses of 0--12~Gy of $^{1}$H at \SI{100}{\MeV} and 0--8~Gy of $^{12}$C at \SI{80}{\MeV}/u and followed for up to 14 days. Each dose was simulated in 20 realizations, using 4 independent track-structure dose realizations, with fresh lesions and cell dynamics in every run; a run was stopped once the clonogenic population exceeded four times its initial value.

\textit{Computational cost.} Spheroids of radius 100, 150, 200, 250, 300, 400 and \SI{500}{\micro\metre} (from $\approx 150$ to $\approx 19\,400$ cells; box half-size: radius $+$ \SI{150}{\micro\metre}, rounded to the lattice spacing) were irradiated with \SI{2}{\gray} of $^{1}$H (100 and \SI{30}{\MeV}) or $^{12}$C (80 and \SI{10}{\MeV}/u) and followed for \SI{72}{\hour} with delayed death; each configuration was repeated three times after a discarded warm-up run that excludes compilation. As a stress test, spheroids of radius 600, 800 and \SI{1000}{\micro\metre} (up to $\approx 155\,000$ cells) were simulated once for $^{1}$H at \SI{100}{\MeV}, the most demanding case, and $^{12}$C at \SI{80}{\MeV}/u. The dose deposition, lesion sampling, fate assignment, and agent-based simulation were timed separately, using 30 threads of the workstation.

All protocol parameters are summarised in Tables~\ref{tab:app_geometry} and~\ref{tab:app_irradiation}.

\subsection{Model outputs and endpoints}

The model records, for every cell, its lesion history, phase, and fate, and for the population, the time course of cell numbers per phase. Three populations are distinguished throughout: \emph{clonogenic} cells, which are alive and not committed to death; \emph{doomed} cells, which are committed to death but not yet removed; and the \emph{total} number of cells, their sum, which corresponds to what a cell count or a spheroid volume measures. The \emph{surviving fraction} is the fraction of irradiated cells that are clonogenic immediately after fate assignment; the cell number at later times includes regrowth and doomed cells and is reported separately. Population-level endpoints, such as tumor control probability, can be derived by aggregating single-cell outcomes \cite{battestini2026ntcp}.

All simulations were performed on a workstation with an Intel\textsuperscript{\textregistered} Core\texttrademark{} i9-13900K processor (32 logical cores, maximum clock speed \SI{5.8}{\giga\hertz}) and \SI{62}{\gibi\byte} of RAM, running Ubuntu 24.04.3 LTS. The framework is implemented in Julia 1.12.1 \cite{Julia2017}. The source code, the scripts that produce all figures, and the simulation outputs are available at \url{https://github.com/francescogcordoni/GSM2_Julia}.

\begin{figure}[htbp]
\centering
\resizebox{\textwidth}{!}{
\begin{tikzpicture}[
    node distance=1.5cm and 2cm,
    every node/.style={font=\sffamily\small, align=center},
    block/.style={rectangle, draw, minimum width=3.8cm, minimum height=1.3cm, rounded corners, drop shadow={opacity=0.2}, fill=white},
    init/.style={block, fill=initcolor, draw=initdraw, line width=0.8pt},
    phys/.style={block, fill=physcolor, draw=physdraw, line width=0.8pt},
    agent/.style={block, fill=agentcolor, draw=agentdraw, line width=0.8pt},
    bio/.style={block, fill=biocolor, draw=biodraw, line width=0.8pt},
    decision/.style={diamond, draw, aspect=2, minimum width=3.2cm, inner sep=0pt, fill=white, drop shadow={opacity=0.2}},
    arrow/.style={-{Stealth[scale=1.2]}, line width=1pt, draw=gray!80},
    gsm/.style={label={[anchor=south east, font=\scriptsize\bfseries, text=black!70, fill=white, inner sep=1pt, rounded corners=1pt]north east:GSM$^2$}},
    new/.style={label={[anchor=south east, font=\scriptsize\bfseries, text=red!70!black, fill=white, inner sep=1pt, rounded corners=1pt]north east:new}}
]
    \node[init] (start) {3D Regular Lattice\\(Target Volume)};
    \node[init, new, below=of start] (agents) {Initialise Agents\\(position, phase)};
    \node[init, new, below=of agents] (params) {Assign Parameters\\($O_2$, cycle, GSM$^2$ rates)};
    \node[decision, new, right=4.5cm of agents] (scheduler) {Global Scheduler\\$\min(\tau_{all\_cells}, t_k)$};
    \node[phys, new, above right=1.2cm and 2.5cm of scheduler] (radfield) {Radiation Field Model\\(Particle Arrival $t_k$)};
    \node[phys, gsm, right=of radfield] (damage) {Energy Deposition (AT)\\$\Delta X, \Delta Y$ Induction};
    \node[bio, gsm, below right=1cm and 1cm of scheduler] (ssa) {Gillespie SSA (GSM$^2$)\\$\tau_{death}$ or $\tau_{repair}$};
    \node[decision, below=1.2cm of ssa] (fate) {SSA Event?};
    \node[bio, new, below=2.5cm of fate] (death) {\textbf{Commitment /}\\\textbf{Delayed Removal}\\Lethal ($Y\ge1$)};
    \node[agent, new, left=2.2cm of death] (cycle) {\textbf{Next Cycle Event}\\(clock held while $X>0$)};
    \node[agent, new, left=1cm of cycle] (migration) {\textbf{Cell Migration}\\(Lattice Displacement)};
    \node[bio, gsm, right=2.2cm of death] (repair) {\textbf{Repair Event}\\Resume Cycle Clock};

    \draw[arrow] (start) -- (agents);
    \draw[arrow] (agents) -- (params);
    \draw[arrow] (params.east) -- node[above=0.5cm, font=\scriptsize] {Start $t=0$} (scheduler.west);
    \draw[arrow] (scheduler.north) |- node[below, pos=0.8] {$t_k$} (radfield.west);
    \draw[arrow] (scheduler.south) |- node[above, pos=0.8] {$\tau_{cell\_ssa}$} (ssa.west);
    \draw[arrow] (scheduler.west) -- ++(-1.15,0) |- node[left,pos=0.6,xshift=-6pt, yshift=6pt] {$\tau_{cycle}$} (cycle.north);
    \draw[arrow] (scheduler.west) -- ++(-1.15,0) -- ++(0,-4.8) -| node[left, pos=0.15, yshift=5pt] {$\tau_{mig}$} (migration.north);
    \draw[arrow] (radfield) -- (damage);
    \draw[arrow] (damage.south) -- ++(0,-0.8) -| (ssa.north);
    \draw[arrow] (ssa) -- (fate);
    \draw[arrow] (fate.south) -- node[left, pos=0.4] {Lethal} (death.north);
    \draw[arrow] (fate.east) -| node[above, pos=0.4] {Repair} (repair.north);
    \draw[arrow, dashed, draw=red!60, line width=1.1pt] (cycle.east) -- node[above, font=\scriptsize, text=red!80] {If $X>0$ in M} (death.west);
    \draw[arrow, dashed] (repair.south) -- ++(0,-1.0) -- ++(4.0,0) |- (scheduler.east);
    \draw[arrow, dashed] (cycle.north) -- ++(0,1.0) |- (scheduler.south);
    \draw[arrow, dashed] (migration.north) -- ++(0,1.4) -- ++(2.2,0) |- ([xshift=-1.0cm]scheduler.south);

    \node[above=0.4cm of start, font=\bfseries\large, text=initdraw] {I. Initialisation};
    \node[above=0.4cm of radfield, font=\bfseries\large, text=physdraw] {III. Radiation Physics};
    \node[left=-0.5cm of scheduler, font=\bfseries\large, yshift=0.8cm] (looplabel) {Global Event Loop};
    \node[below=1.5cm of cycle, font=\bfseries\large, text=agentdraw, xshift=-1cm] {II. Biological State};
    \node[right=0.5cm of ssa, font=\bfseries\large, text=biodraw] {IV. GSM$^2$ Biological Endpoints};
\end{tikzpicture}
}
\caption{Overview of the simulation architecture. A global event loop extracts the next event from a shared priority queue and dispatches it to radiation physics (III), \GSM{} lesion kinetics (IV), or cell-level dynamics (II: cell-cycle progression, division, delayed death, and migration). Dashed arrows indicate feedback loops that reschedule events after a state change. Labels mark components inherited from \GSM{} (\textbf{GSM$^2$}) and components introduced in this work (\textbf{new}).}
\label{fig:architecture}
\end{figure}

\begin{figure}[hb!]
    \centering
    \includegraphics[width=0.9\linewidth]{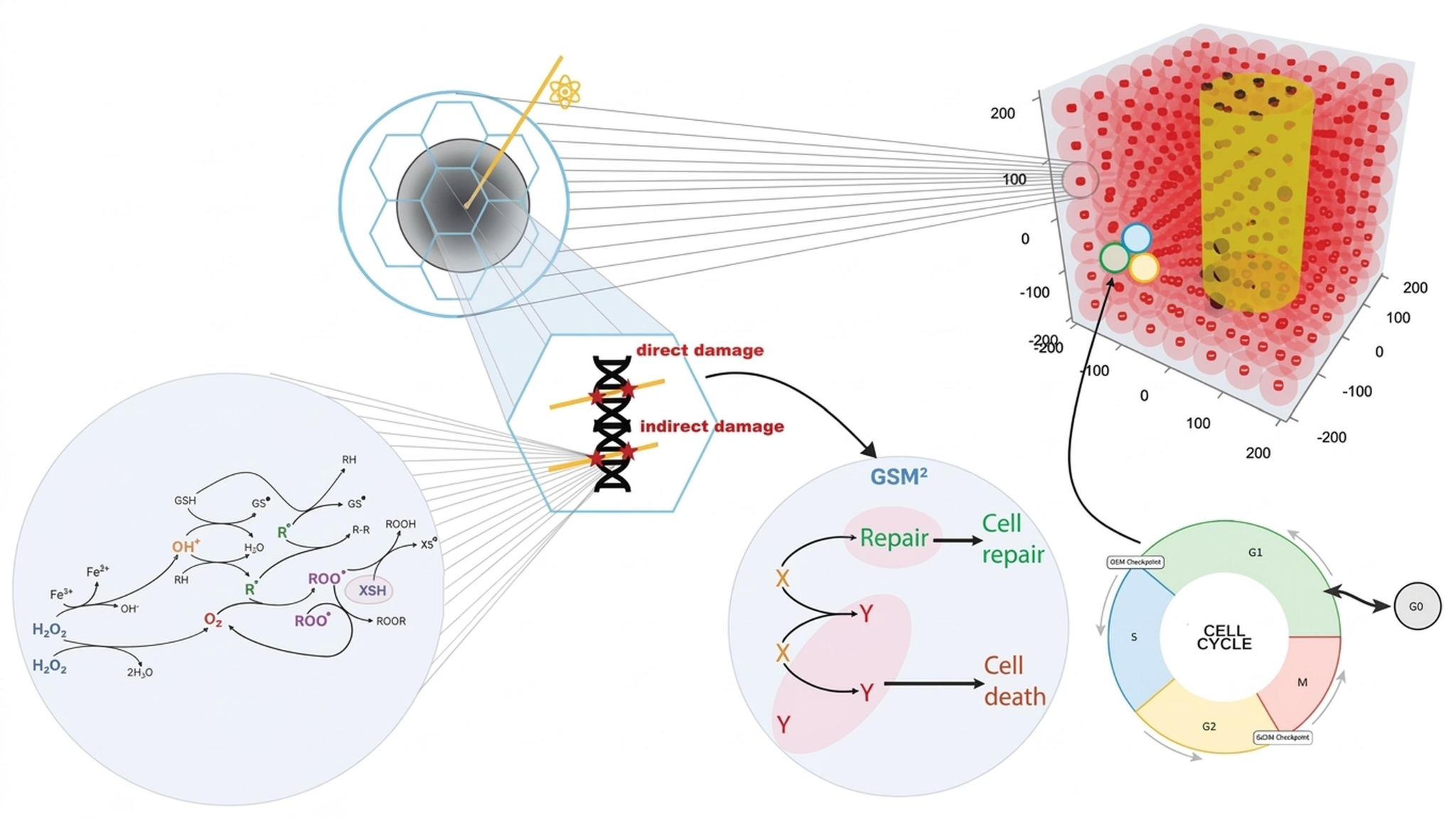}
    \caption{Schematic representation of the GSM$^2$-ABM coupling. Each cell agent hosts an independent GSM$^2$ instance that processes incoming energy deposition events and evolves the lesion state stochastically via SSA.}
    \label{fig:scheme}
\end{figure}

\section{Results}\label{SEC:results}

\subsection{Verification against the Lea--Catcheside factor}
\label{SEC:res_doserate}

\begin{figure}[htbp]
    \centering
    \begin{subfigure}[b]{0.49\linewidth}
        \centering
        \includegraphics[width=\linewidth]{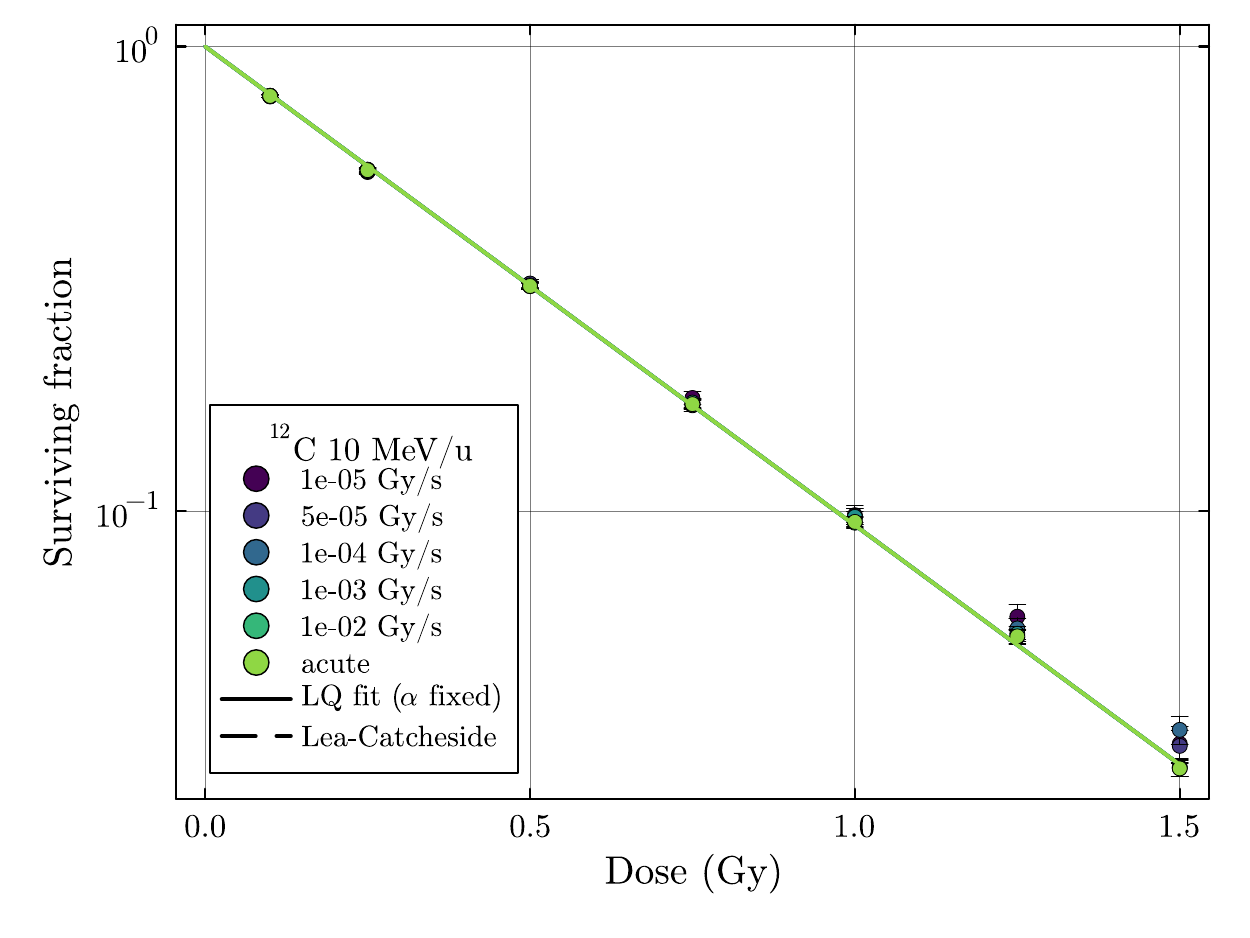}
        \caption{}
        \label{fig:sf_12C_10}
    \end{subfigure}
    \hfill
    \begin{subfigure}[b]{0.49\linewidth}
        \centering
        \includegraphics[width=\linewidth]{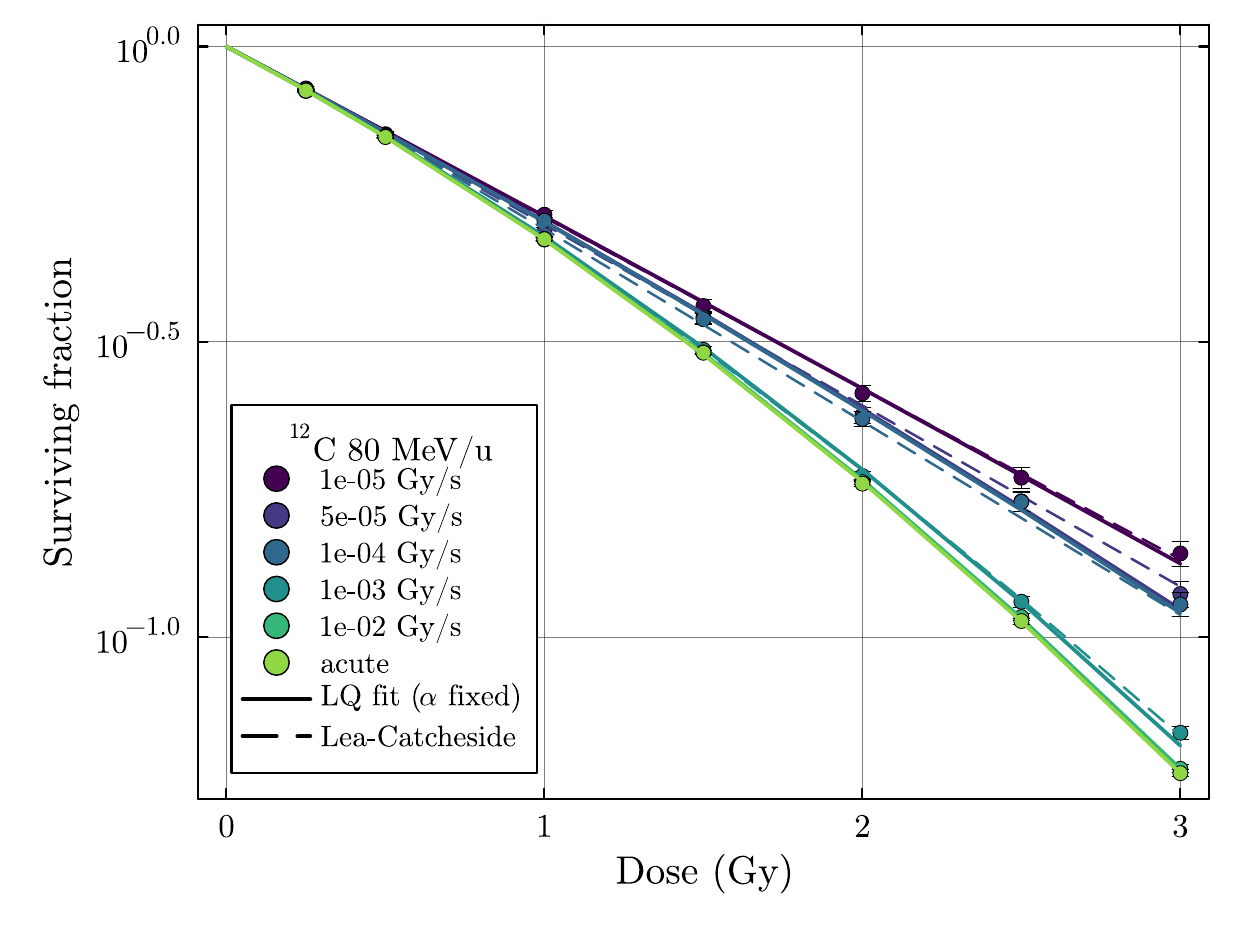}
        \caption{}
        \label{fig:sf_12C_80}
    \end{subfigure}
    \begin{subfigure}[b]{0.49\linewidth}
        \centering
        \includegraphics[width=\linewidth]{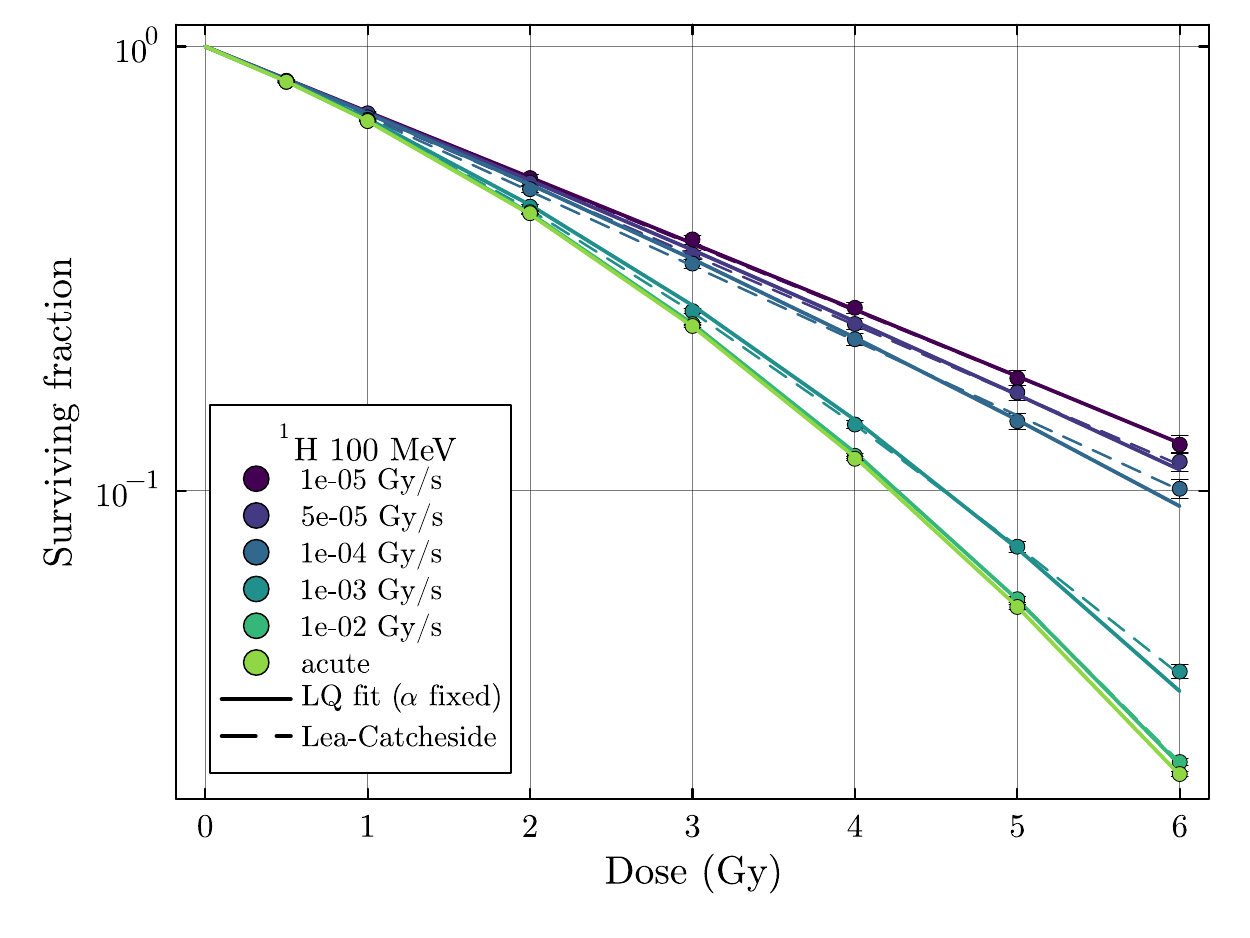}
        \caption{}
        \label{fig:sf_1H_100}
    \end{subfigure}
    \caption{Surviving fraction as a function of dose for a static G$_1$ population irradiated with (a) $^{12}$C at \SI{10}{\MeV\per u}, (b) $^{12}$C at \SI{80}{\MeV\per u} and (c) $^{1}$H at \SI{100}{\MeV}. Colors: dose rate from \SI{e-5}{\gray\per\sec} to \SI{e-2}{\gray\per\sec} and acute delivery. Symbols: simulations ($\pm$ one standard error); solid lines: LQ fit with $\alpha$ fixed to its acute value; dashed lines: Lea--Catcheside prediction.}
    \label{fig:survival_curves}
\end{figure}

\begin{figure*}[htbp]
    \centering
    \begin{subfigure}[b]{0.48\linewidth}
        \centering
        \includegraphics[width=\linewidth]{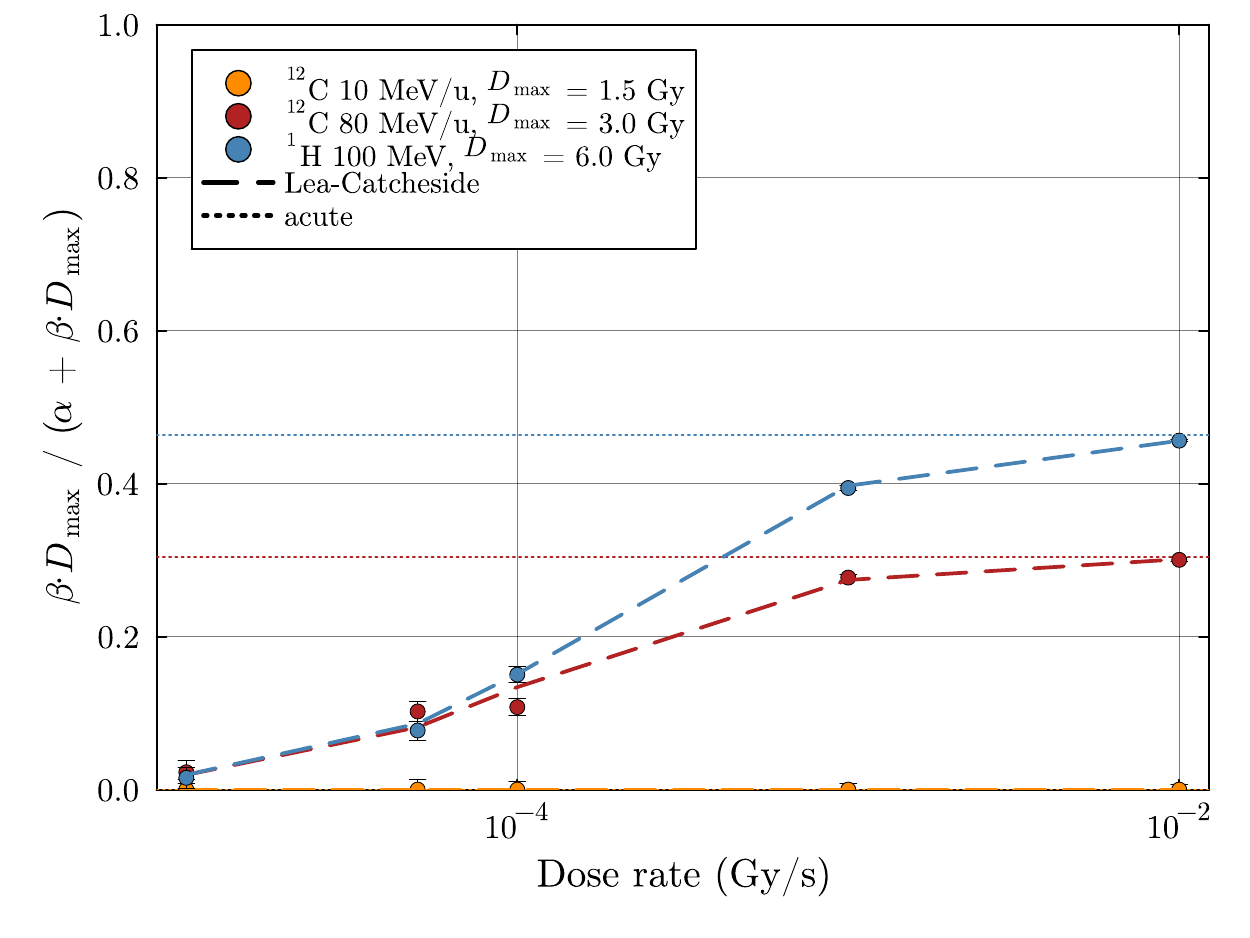}
        \caption{}
        \label{fig:quadratic_share}
    \end{subfigure}
    \hfill
    \begin{subfigure}[b]{0.48\linewidth}
        \centering
        \includegraphics[width=\linewidth]{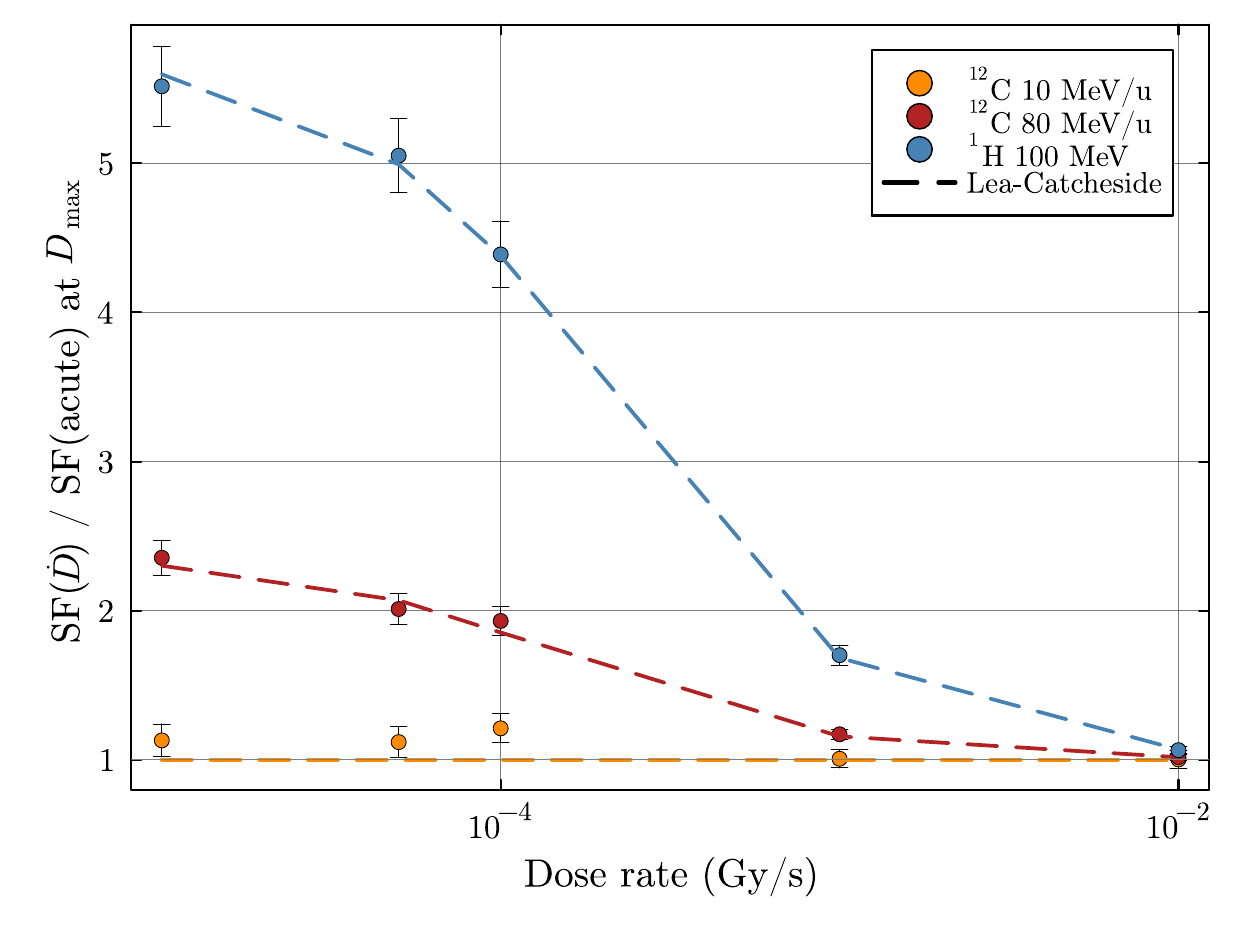}
        \caption{}
        \label{fig:sparing}
    \end{subfigure}
    \   \caption{Dose-rate response of a static G$_1$ population for $^{1}$H at 100~MeV, $^{12}$C at 80~MeV/u and $^{12}$C at 10~MeV/u ($D_{\max} = 6$, 3 and 1.5~Gy). (a) Quadratic share of the cell kill, $q = \beta D_{\max}/(\alpha + \beta D_{\max})$, with acute values as dotted lines; (b) dose-rate sparing at $D_{\max}$, $\mathrm{SF}(\dot D)/\mathrm{SF}_{\mathrm{acute}}$. Symbols: simulations ($\pm$ one standard error); dashed lines: Lea--Catcheside prediction.}
    \label{fig:lq_params}
\end{figure*}

To verify the coupling between stochastic particle arrival and repair, we irradiated a static population of G$_1$ cells, without cell-cycle progression or division, at dose rates from $10^{-5}$ to $10^{-2}$~Gy\,s$^{-1}$, and compared the results with acute delivery and with the Lea--Catcheside prediction (Section~\ref{SEC:LC}). In this static limit the Lea--Catcheside factor is the expected answer, so the comparison tests the event-driven treatment of particle arrivals and repair on its own.

Figure~\ref{fig:survival_curves} shows the surviving fraction as a function of dose for $^{12}$C at 10~MeV/u (a), $^{12}$C at 80~MeV/u (b) and $^{1}$H at 100~MeV (c). Each color corresponds to a dose rate; symbols are the simulations, solid lines LQ fits with $\alpha$ fixed to its acute value, and dashed lines the Lea--Catcheside prediction computed from the acute survival curve. For protons, lowering the dose rate clearly raises the survival curves; the effect is much weaker for carbon ions at 80~MeV/u and absent at 10~MeV/u. In all cases, lowering the dose rate reduced $\beta$ as predicted by the Lea--Catcheside factor, while $\alpha$ did not change within its uncertainty.

Figure~\ref{fig:lq_params} summarizes these results as a function of dose rate, with one color per radiation quality. Panel (a) shows the quadratic share $q$, the fraction of the cell kill at the highest dose $D_{\max}$ that comes from the quadratic term, that is, from interactions between lesions produced by different tracks; panel (b) shows the dose-rate sparing, the survival at $D_{\max}$ divided by the survival after acute delivery. Symbols are the simulations, dashed lines the Lea--Catcheside prediction, and dotted lines in (a) the acute values. The size of the dose-rate effect depends strongly on LET. For $^{1}$H at 100~MeV, the quadratic term accounts for about 46\% of the cell kill at 6~Gy under acute delivery; protracting the dose to $10^{-5}$~Gy\,s$^{-1}$ removes almost all of this contribution and increases survival about 5.5-fold. For $^{12}$C at 80~MeV/u, both the quadratic share and the sparing are smaller. For $^{12}$C at 10~MeV/u, cell kill is dominated by single-track events: the fitted $\beta$ is zero, the quadratic share is below 1\% at 1.5~Gy, and survival does not depend on dose rate within statistical uncertainty (sparing $1.1 \pm 0.1$ at $10^{-5}$~Gy\,s$^{-1}$). In all cases, the simulations agree with the Lea--Catcheside prediction within statistical uncertainty, confirming that the event-driven coupling between particle arrivals and repair reproduces the expected behavior in the absence of cell-cycle dynamics.

\subsection{Spatiotemporal response of a spheroid}
\label{SEC:res_single}

\begin{figure}[htbp]
    \centering
    \includegraphics[width=0.65\linewidth]{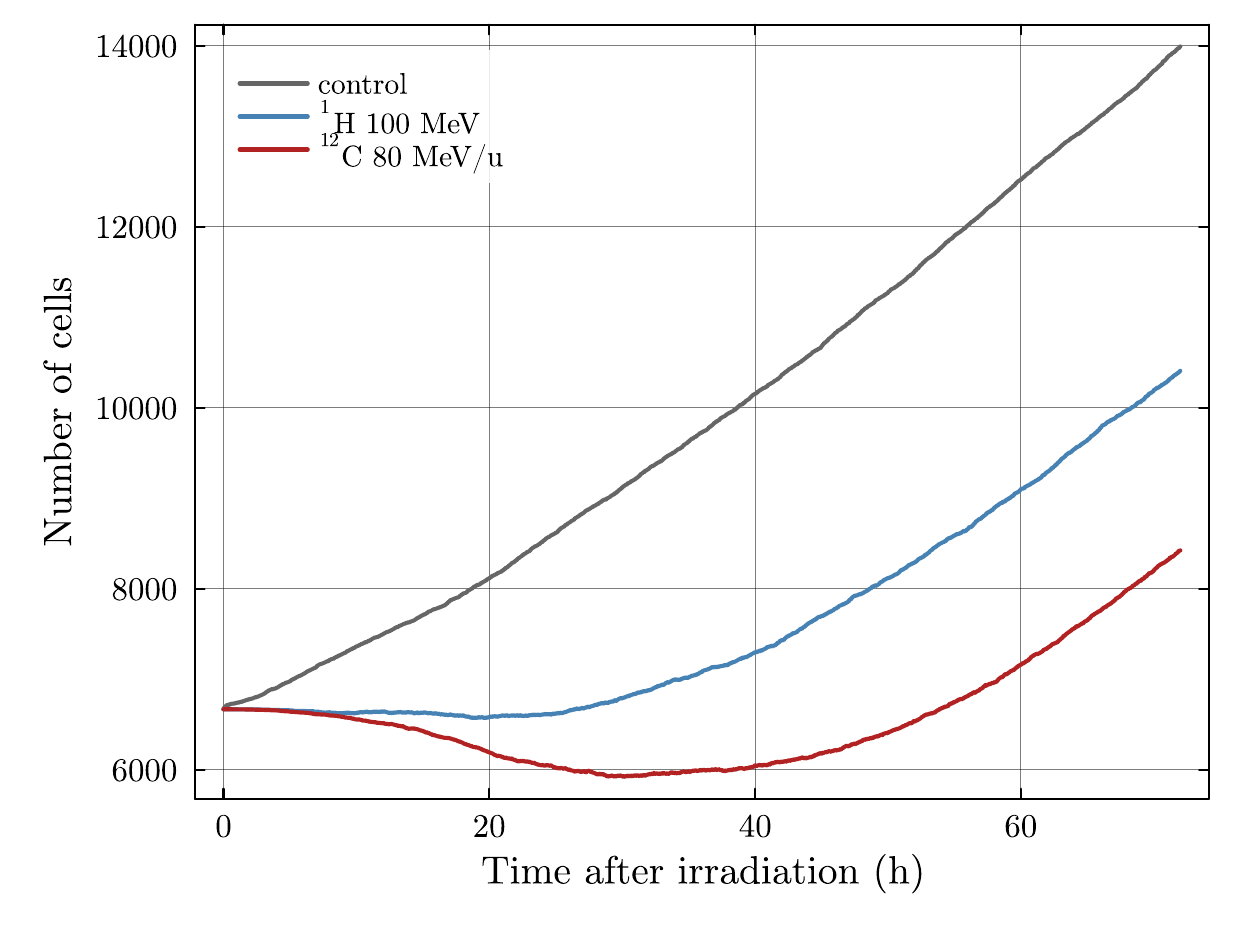}
    \caption{Total number of cells (including doomed cells, committed to death but not yet removed) as a function of time after 2~Gy, for no irradiation (gray), $^{1}$H at \SI{100}{\MeV} (blue) and $^{12}$C at \SI{80}{\MeV\per u} (red).}
    \label{fig:total_cells}
\end{figure}

Figure~\ref{fig:total_cells} shows the total number of cells after 2~Gy of $^{1}$H (blue) at 100~MeV and $^{12}$C at 80~MeV/u (red), compared with the unirradiated control. The total includes doomed cells, which remain on the lattice until their delayed death: after $^{1}$H irradiation the cell number stays almost constant, reaching a minimum of $\approx 6550$ cells ($\approx 98\%$ of the initial 6650) at about 20~h, and increases to $\approx 10\,300$ cells at 72~h; after $^{12}$C irradiation it decreases to $\approx 5980$ cells ($\approx 90\%$) at 30--40~h and reaches $\approx 8400$ cells at 72~h, compared with $\approx 14\,000$ cells for the control. The total cell number, therefore, responds to irradiation with a delay of about one day and underestimates the killing, which is visible only in the clonogenic population.

\begin{figure}[htbp]
    \centering
    \includegraphics[width=\linewidth]{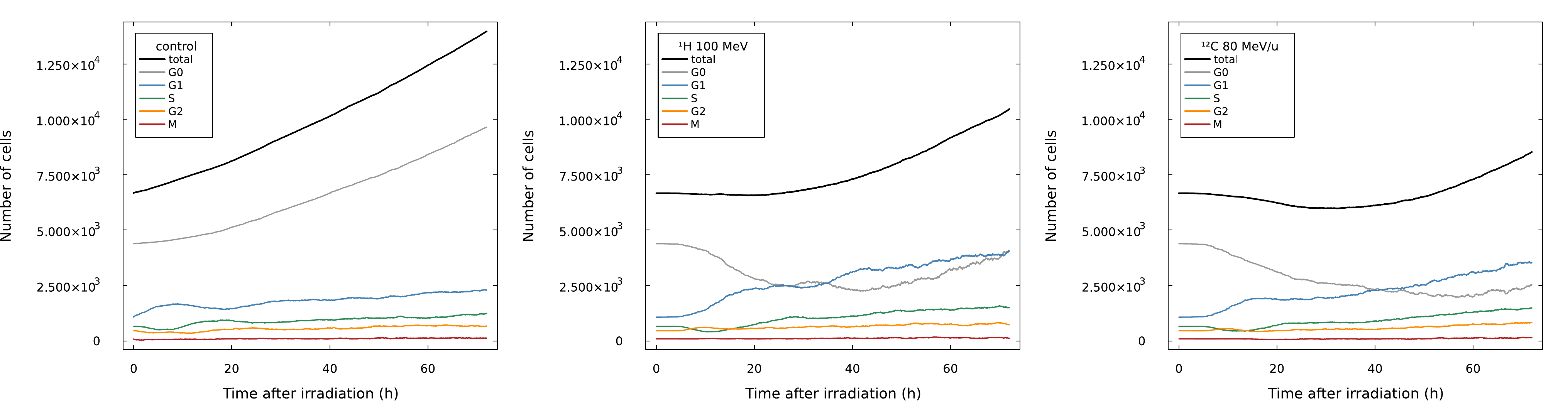}
    \caption{Temporal evolution of cell-cycle phase populations after no irradiation (left), 2~Gy of $^{1}$H at \SI{100}{\MeV} (center) and 2~Gy of $^{12}$C at \SI{80}{\MeV\per u} (right). Total cell count (black), $\text{G}_{1}$ (blue), S (green), $\text{G}_{2}$ (orange), M (red), and $\text{G}_{0}$ (dashed grey).}
    \label{fig:phase_breakdown}
\end{figure}

The phase-resolved population dynamics are shown in Figure~\ref{fig:phase_breakdown} for the unirradiated case (left panel) and both ion species, $^{1}$H at \SI{100}{\MeV} (center) and $^{12}$C at \SI{80}{\MeV\per u} (right). After irradiation, the quiescent G$_0$ population drops sharply, as the death of neighboring cells frees space and allows quiescent cells to re-enter the cycle, which produces a transient increase in the G$_1$ population. The cycling populations then evolve as doomed cells are removed and the spheroid regrows.

\begin{figure*}[t]
  \centering
  \begin{subfigure}[b]{0.32\linewidth}
    \centering
    \includegraphics[width=\linewidth]{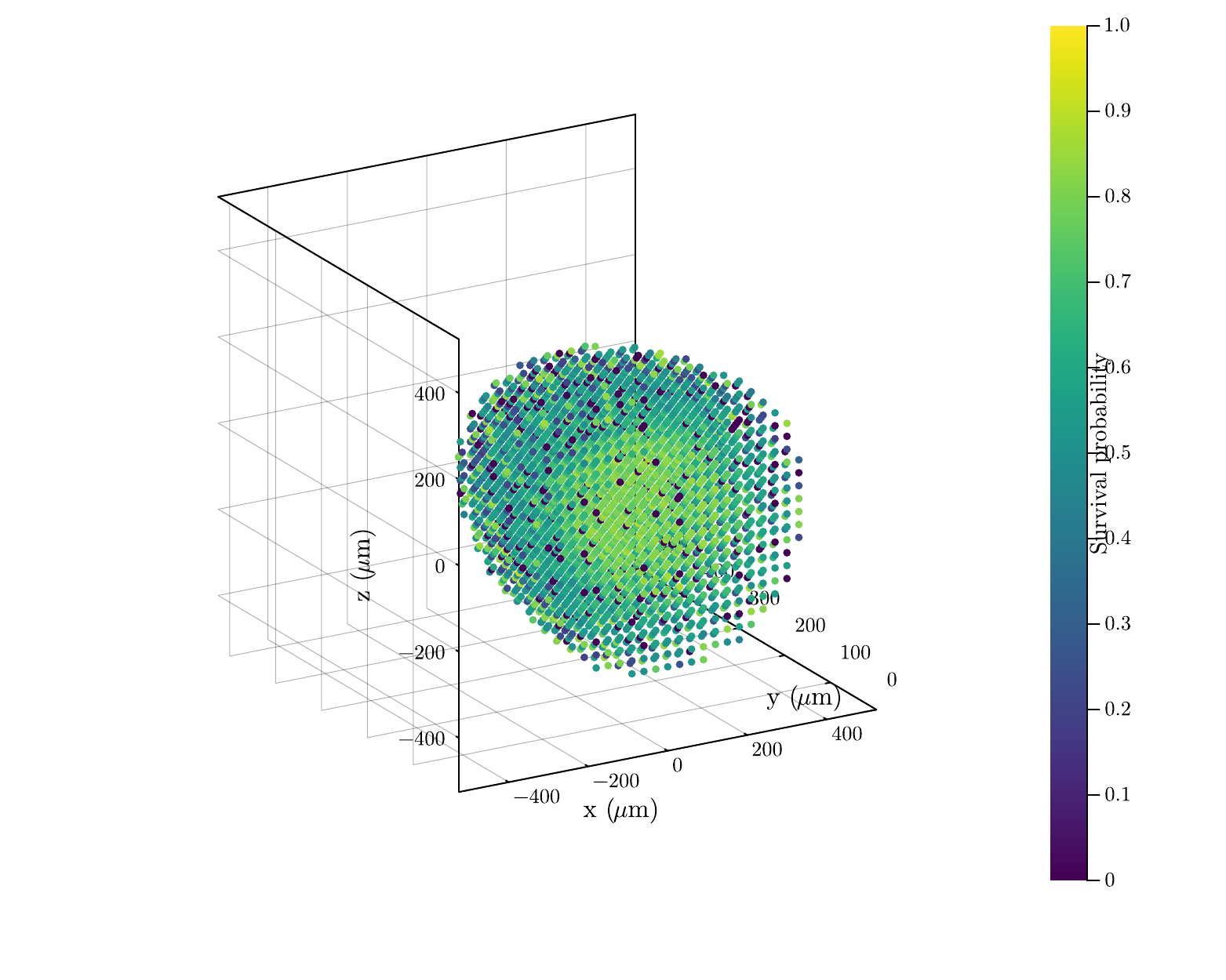}
    \caption{}
    \label{fig:sph_H_sp}
  \end{subfigure}\hfill
  \begin{subfigure}[b]{0.32\linewidth}
    \centering
    \includegraphics[width=\linewidth]{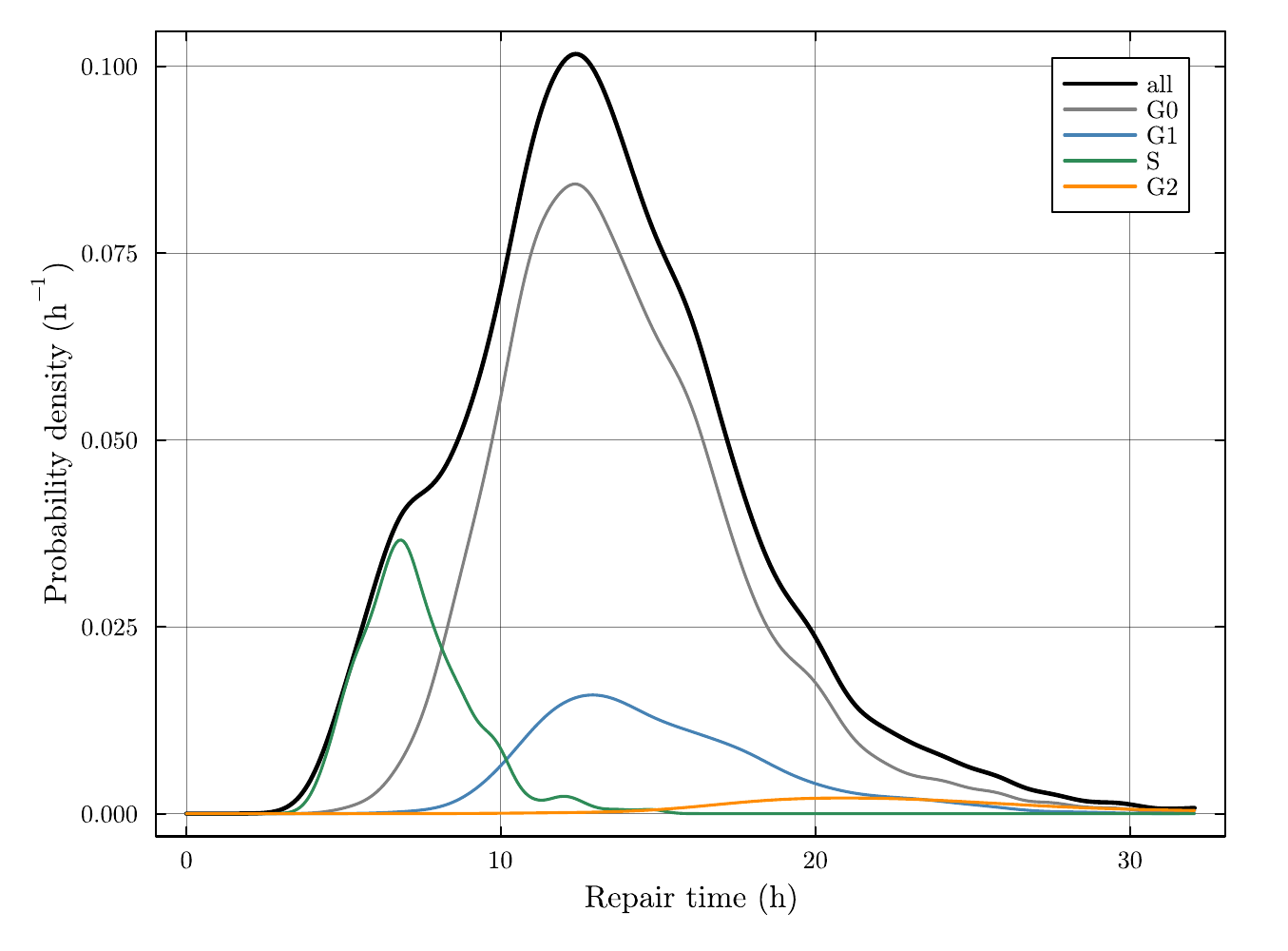}
    \caption{}
    \label{fig:sph_H_repair}
  \end{subfigure}\hfill
  \begin{subfigure}[b]{0.32\linewidth}
    \centering
    \includegraphics[width=\linewidth]{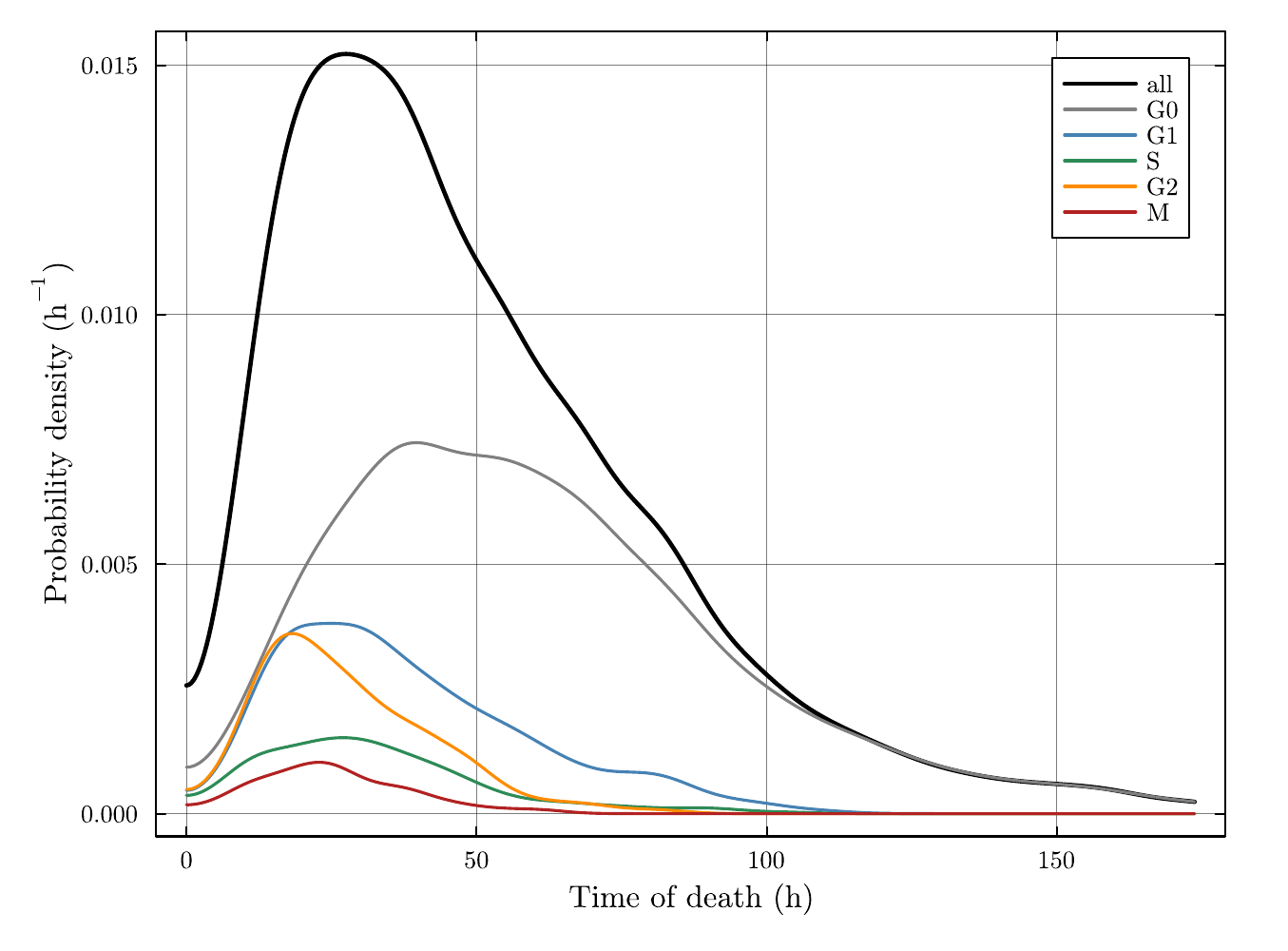}
    \caption{}
    \label{fig:sph_H_death}
  \end{subfigure}

  \vspace{0.5em}
  \begin{subfigure}[b]{0.32\linewidth}
    \centering
    \includegraphics[width=\linewidth]{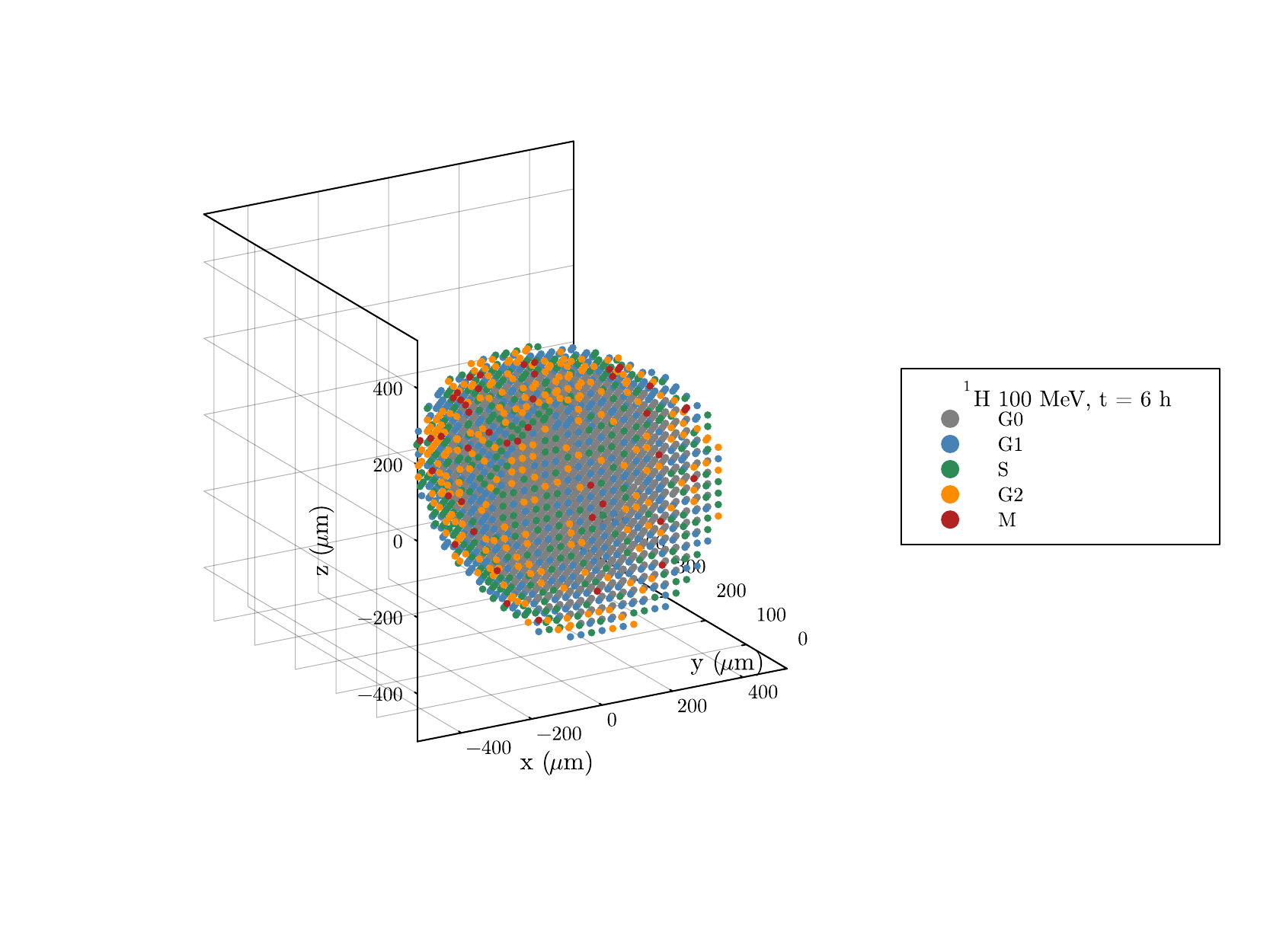}
    \caption{6~h}
    \label{fig:sph_H_t6}
  \end{subfigure}\hfill
  \begin{subfigure}[b]{0.32\linewidth}
    \centering
    \includegraphics[width=\linewidth]{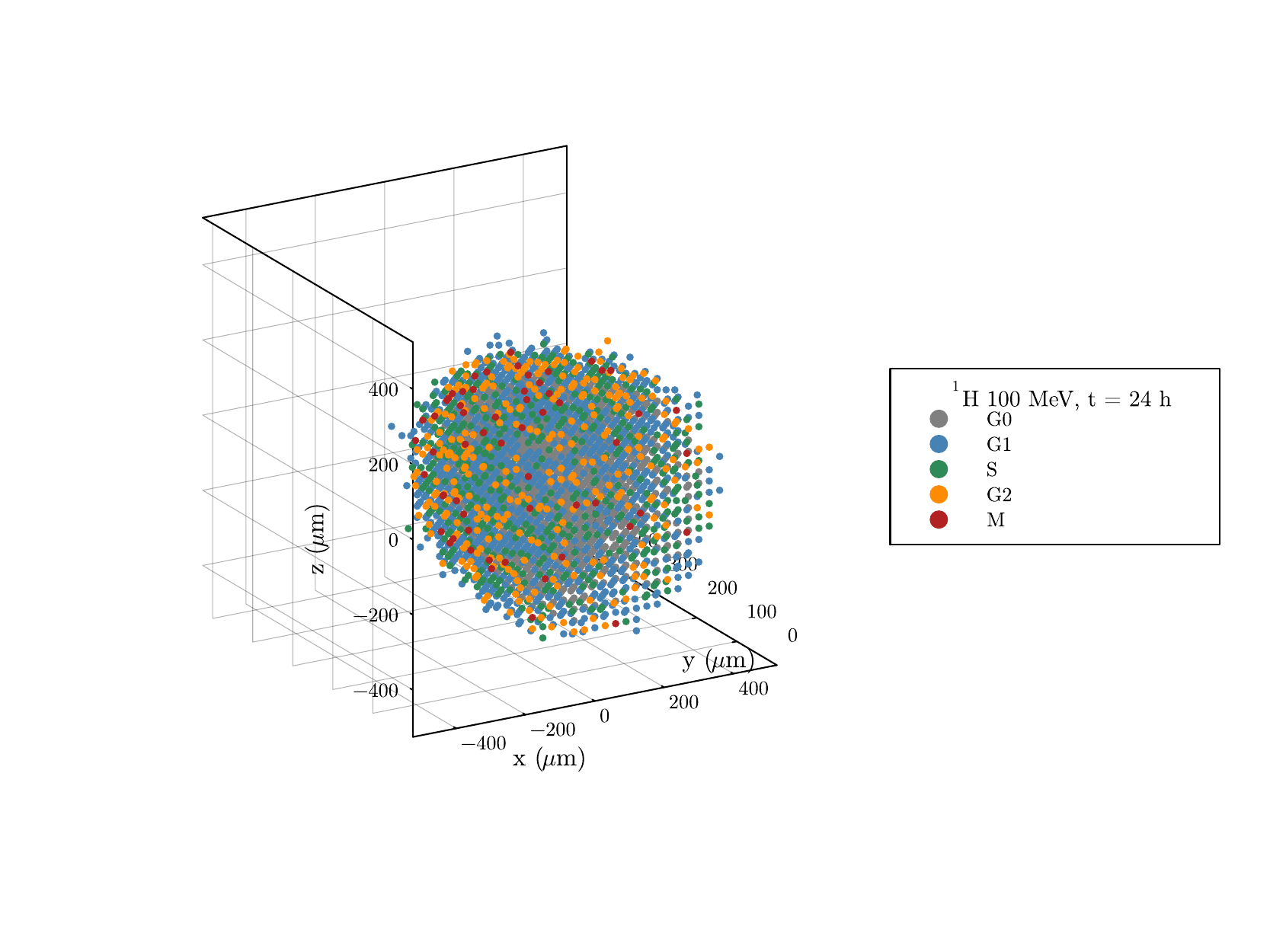}
    \caption{24~h}
    \label{fig:sph_H_t24}
  \end{subfigure}\hfill
  \begin{subfigure}[b]{0.32\linewidth}
    \centering
    \includegraphics[width=\linewidth]{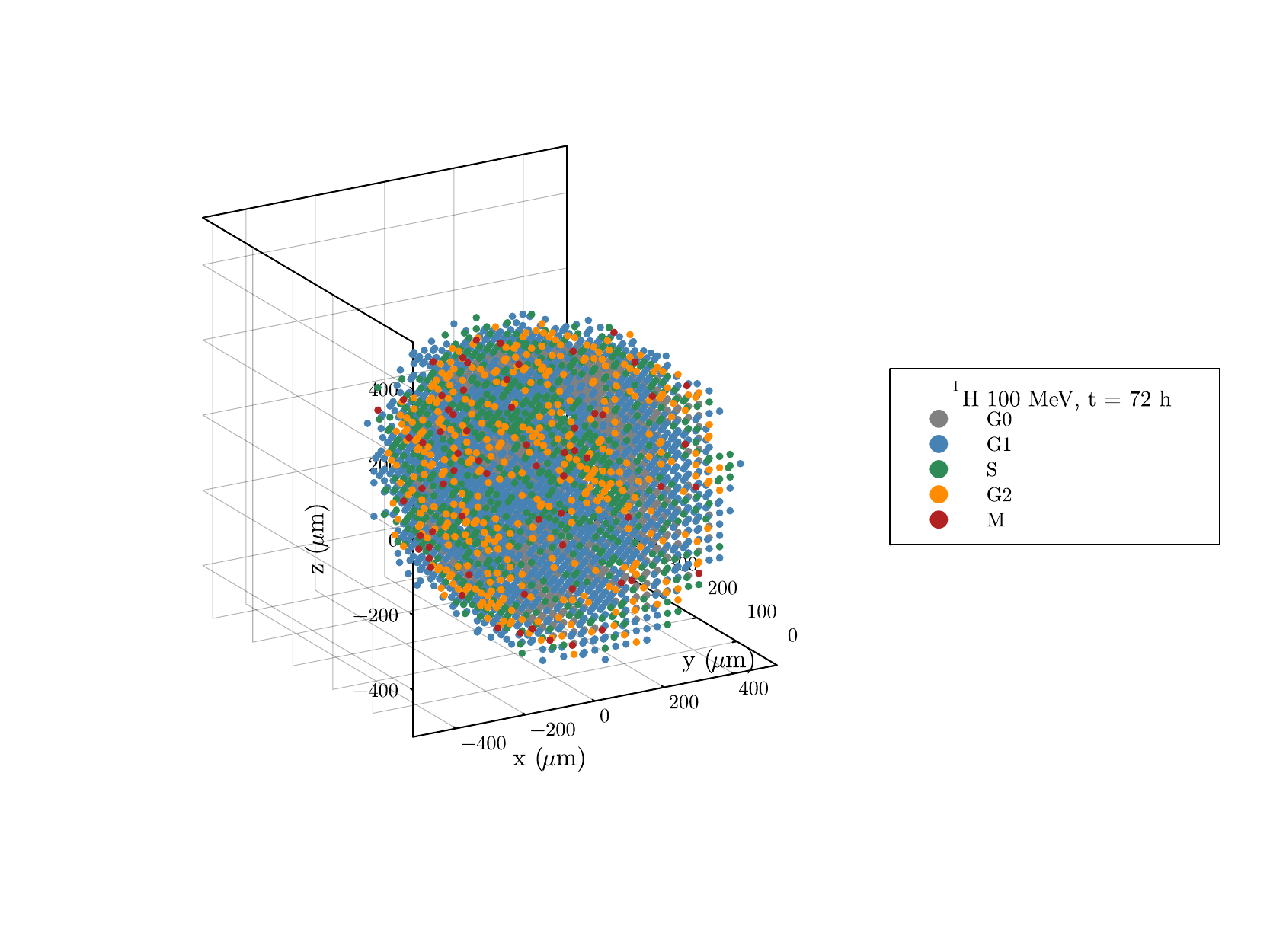}
    \caption{72~h}
    \label{fig:sph_H_t72}
  \end{subfigure}
  \caption{Response of a spheroid to 2~Gy of $^{1}$H at 100~MeV. (a) GSM$^2$ survival probability of each cell right after irradiation (half spheroid, cut face $y = 0$ in front); (b) distribution of the time needed to complete repair and (c) of the time of radiation-induced death, for all cells (black) and by cell-cycle phase at irradiation (colors; curves weighted by the fraction of cells in each phase); (d)--(f) cell-cycle phase at 6, 24 and 72~h.}
  \label{fig:spheroid3d_H}
\end{figure*}

\begin{figure*}[t]
  \centering
  \begin{subfigure}[b]{0.32\linewidth}
    \centering
    \includegraphics[width=\linewidth]{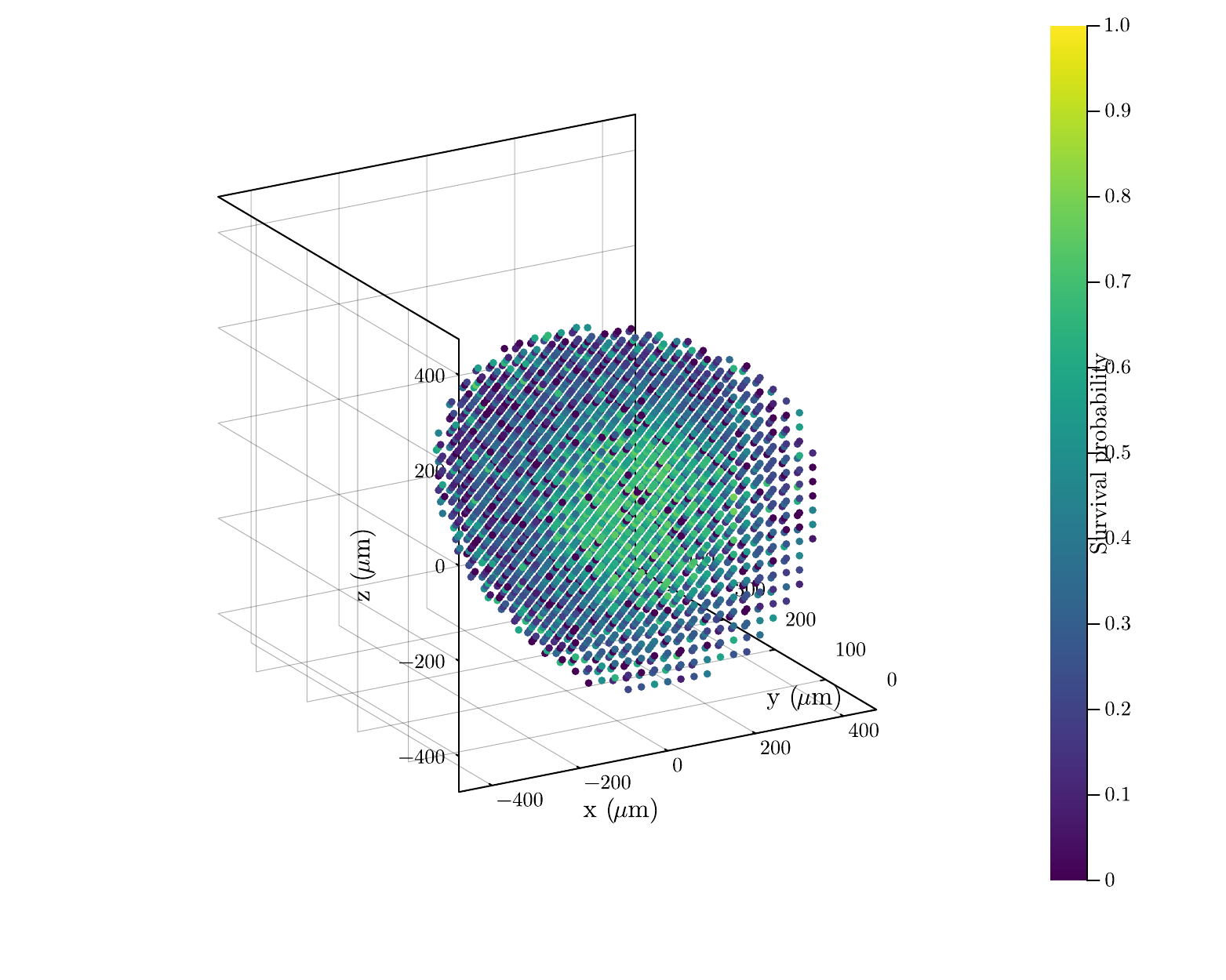}
    \caption{}
    \label{fig:sph_C_sp}
  \end{subfigure}\hfill
  \begin{subfigure}[b]{0.32\linewidth}
    \centering
    \includegraphics[width=\linewidth]{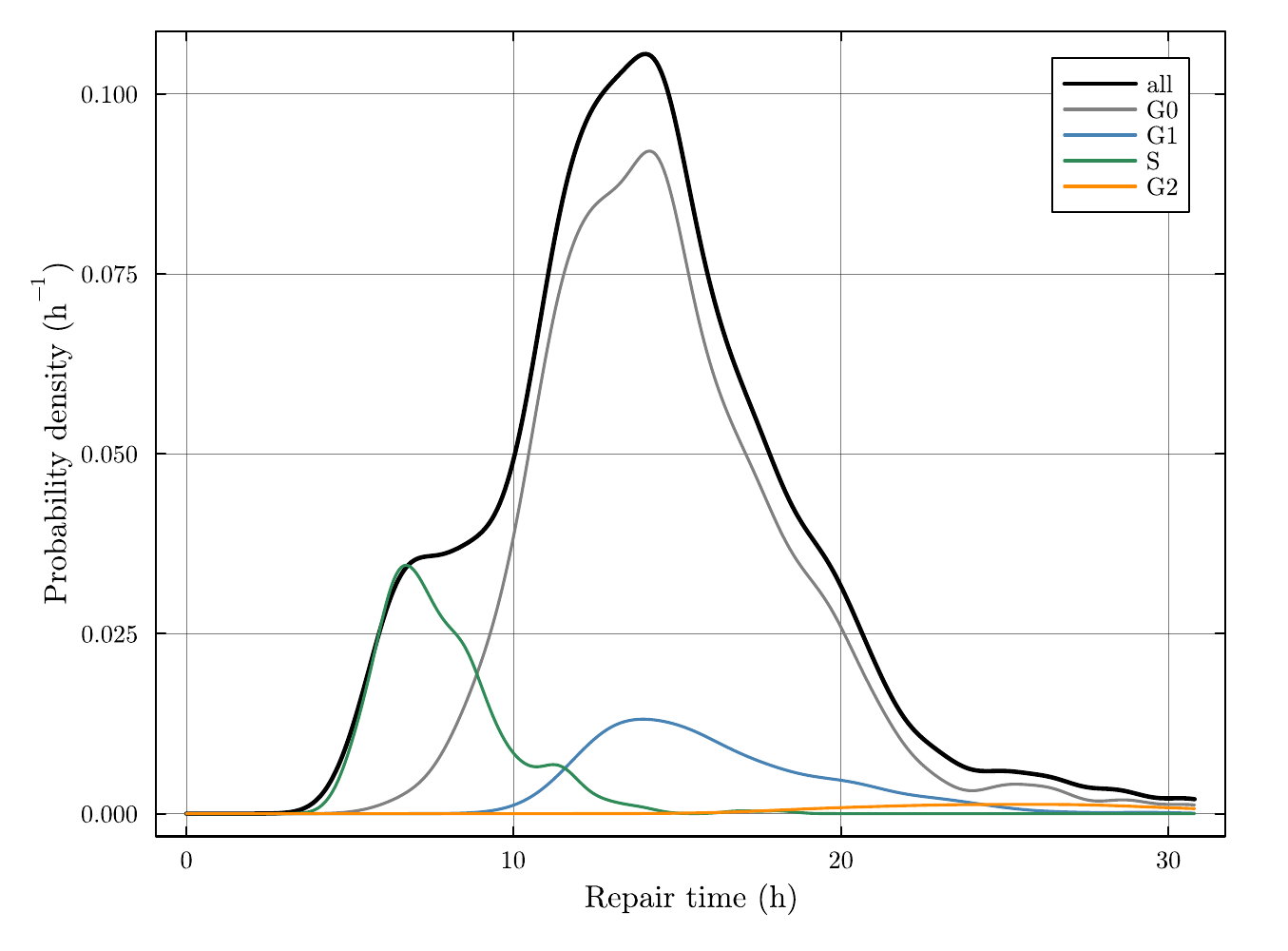}
    \caption{}
    \label{fig:sph_C_repair}
  \end{subfigure}\hfill
  \begin{subfigure}[b]{0.32\linewidth}
    \centering
    \includegraphics[width=\linewidth]{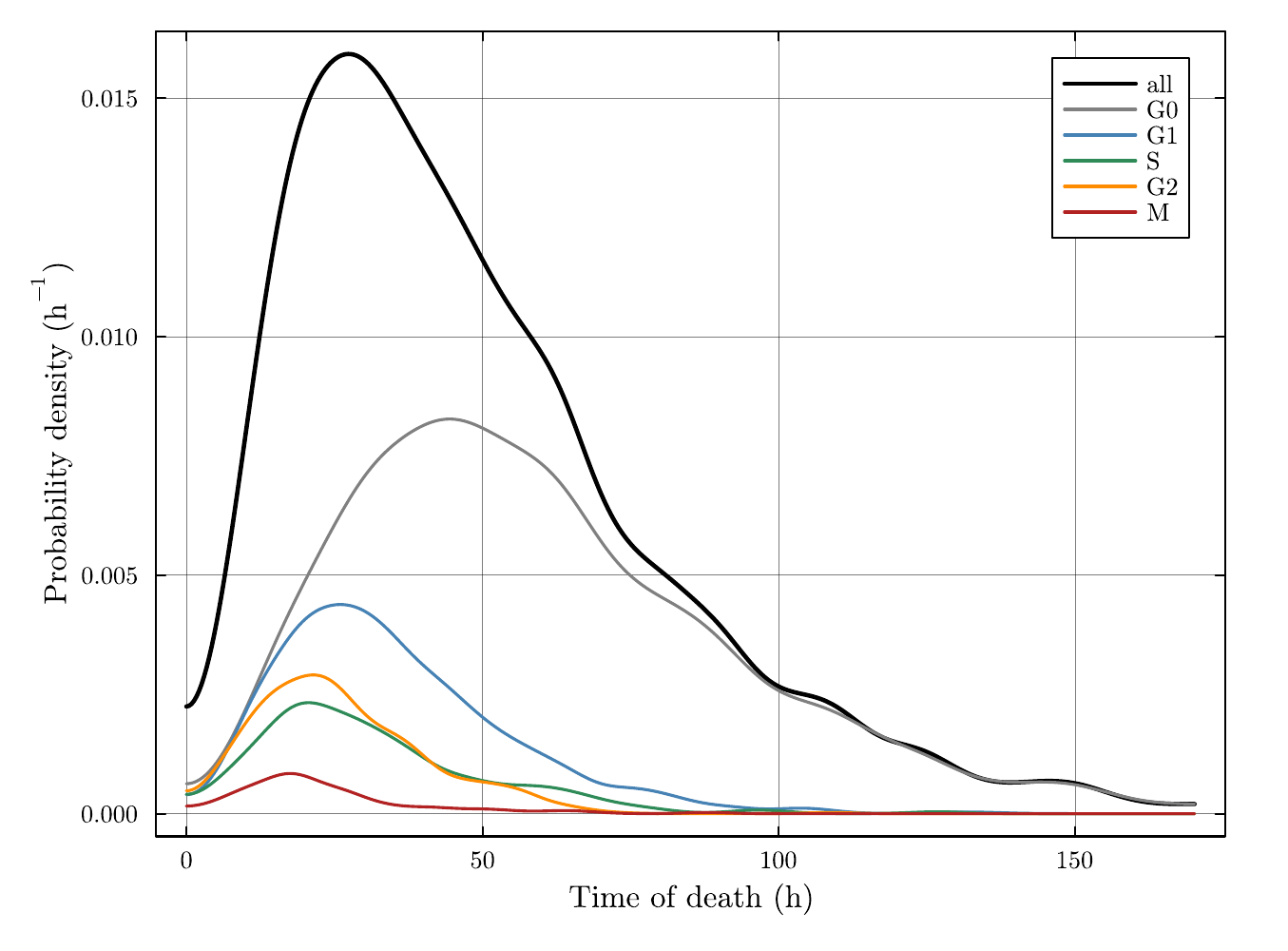}
    \caption{}
    \label{fig:sph_C_death}
  \end{subfigure}

  \vspace{0.5em}
  \begin{subfigure}[b]{0.32\linewidth}
    \centering
    \includegraphics[width=\linewidth]{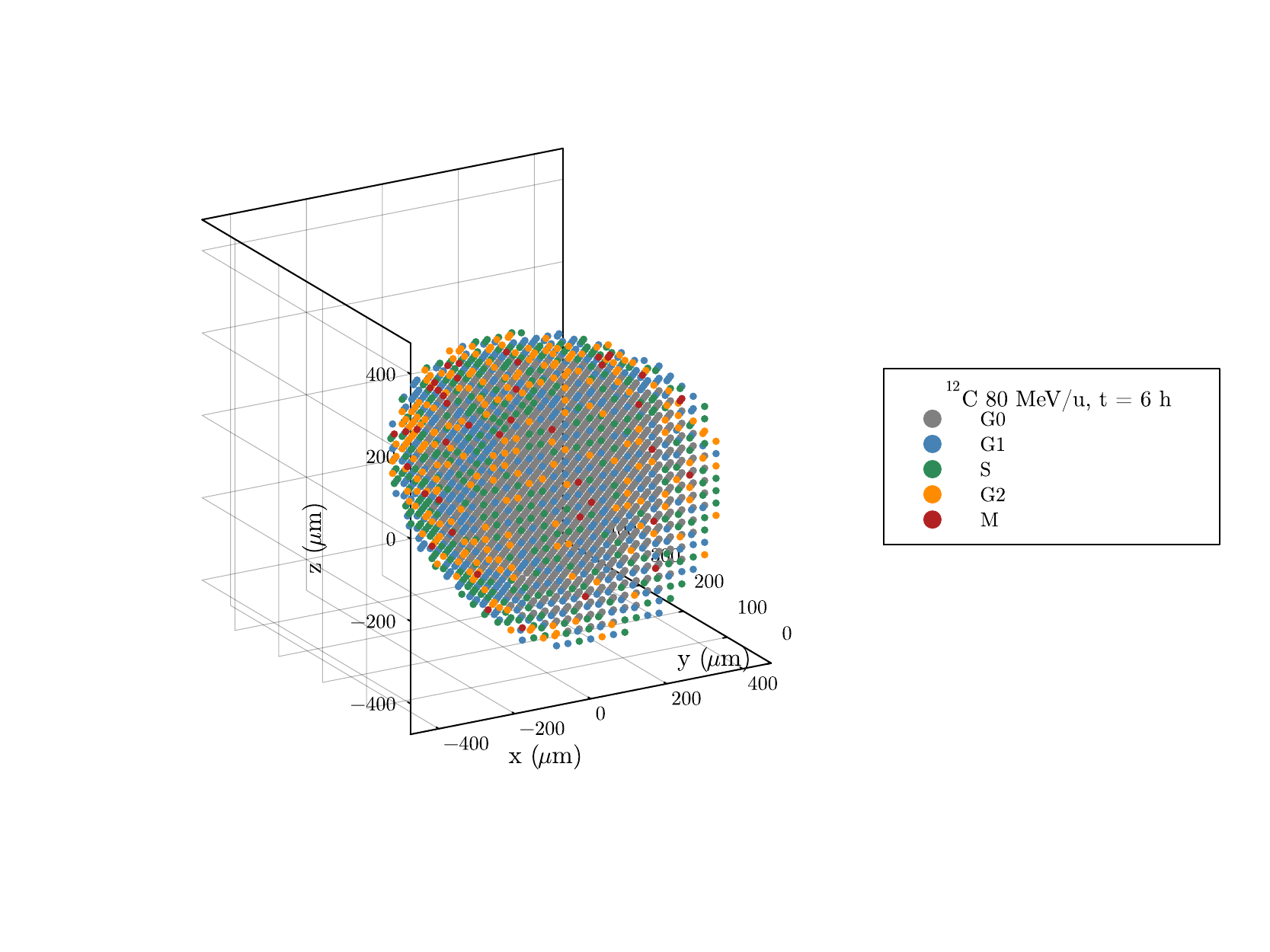}
    \caption{6~h}
    \label{fig:sph_C_t6}
  \end{subfigure}\hfill
  \begin{subfigure}[b]{0.32\linewidth}
    \centering
    \includegraphics[width=\linewidth]{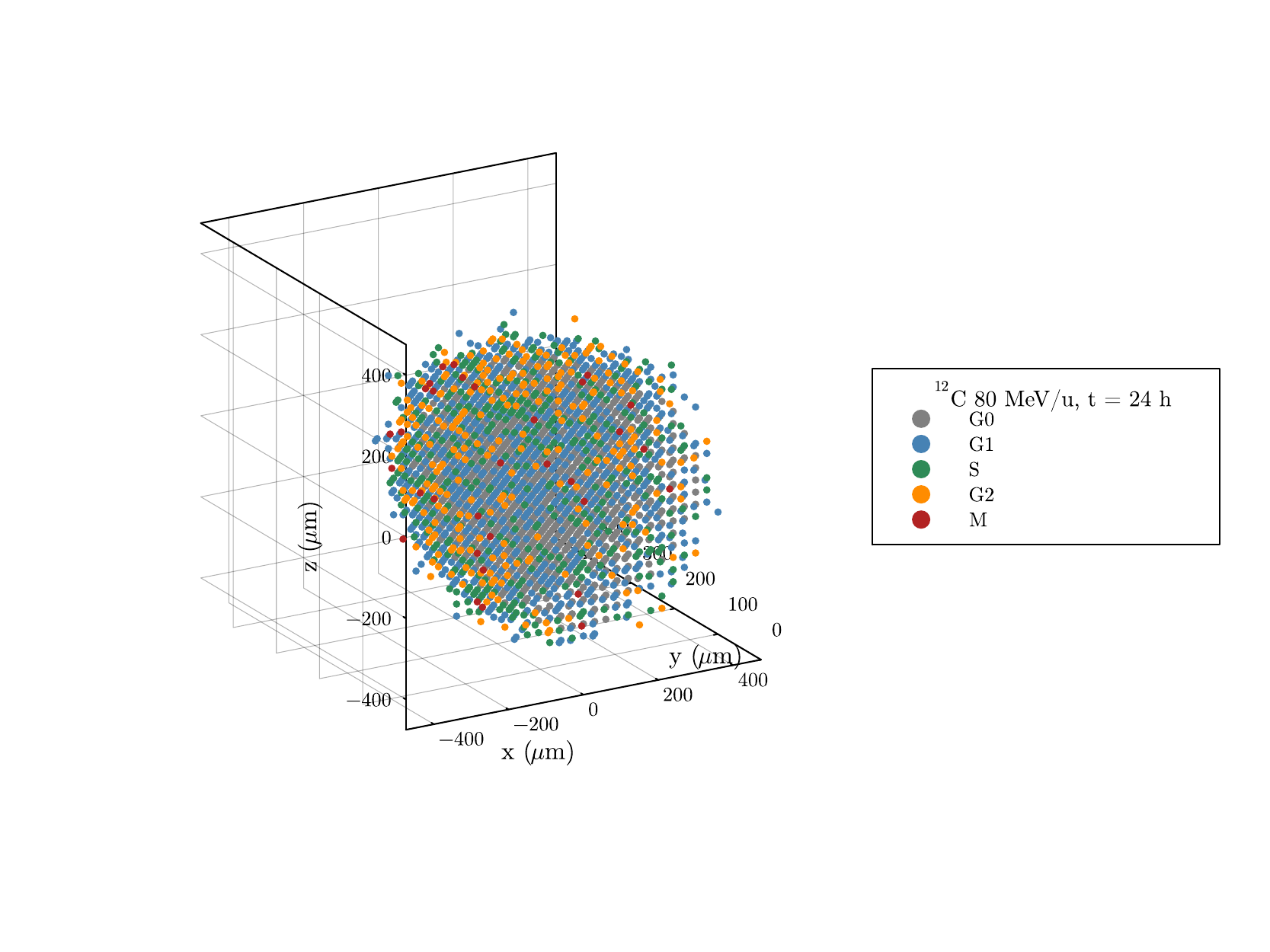}
    \caption{24~h}
    \label{fig:sph_C_t24}
  \end{subfigure}\hfill
  \begin{subfigure}[b]{0.32\linewidth}
    \centering
    \includegraphics[width=\linewidth]{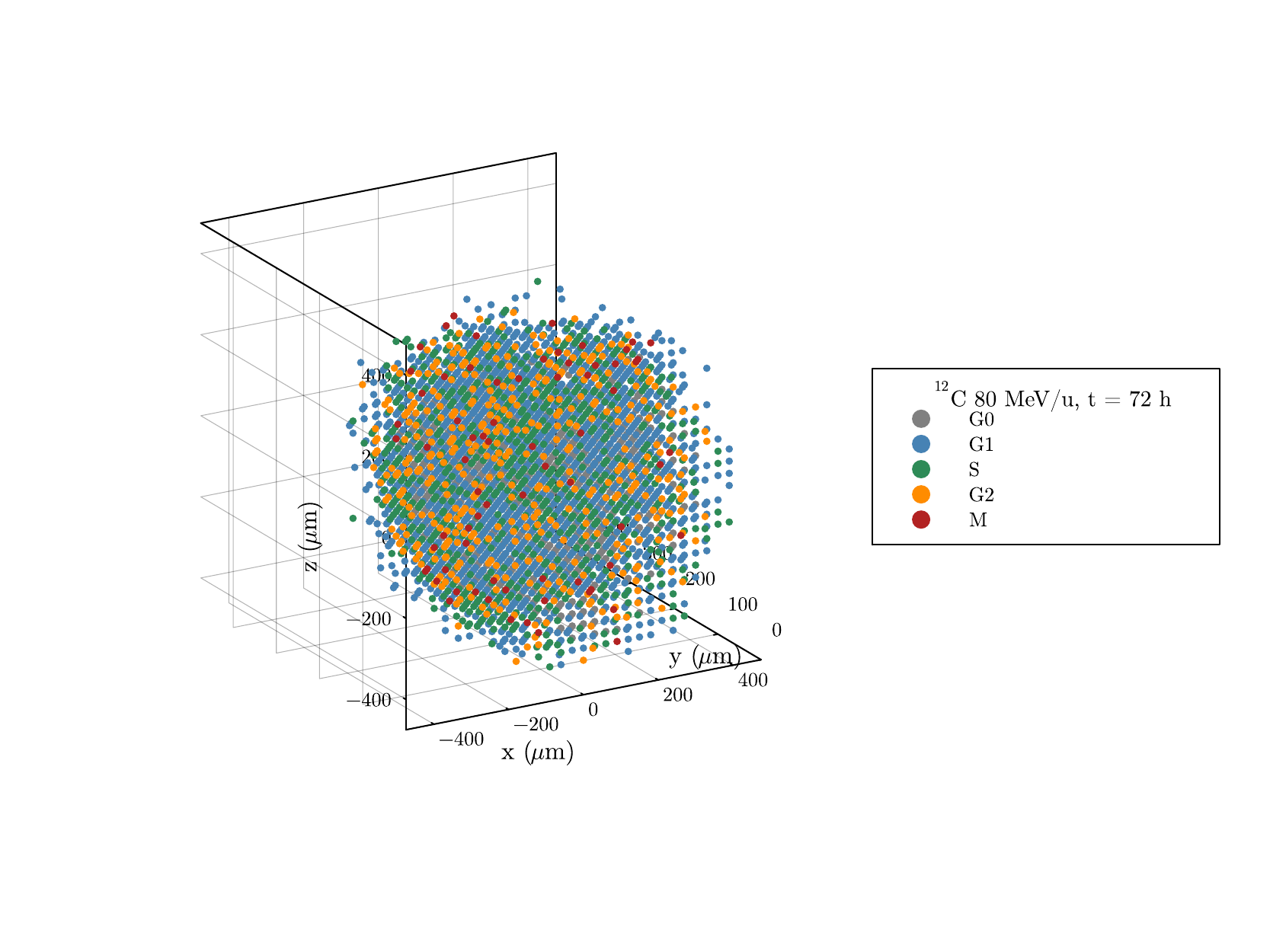}
    \caption{72~h}
    \label{fig:sph_C_t72}
  \end{subfigure}
  \caption{As Fig.~\ref{fig:spheroid3d_H}, for 2~Gy of $^{12}$C at 80~MeV/u.}
  \label{fig:spheroid3d_C}
\end{figure*}

Figures~\ref{fig:spheroid3d_H} and~\ref{fig:spheroid3d_C} show the response of individual cells after 2~Gy of $^{1}$H at 100~MeV and $^{12}$C at 80~MeV/u, respectively. For protons, the survival probability immediately after irradiation varies only moderately among cells (panel a), primarily with cell-cycle phase, with the lowest values occurring in G$_2$/M cells; the hypoxic core is visible as a region of higher survival. For carbon ions, the cell dose is much more broadly distributed: cells crossed by a track receive a high dose, while those almost unaffected receive a low dose; correspondingly, survival probability is heterogeneous, with many cells in the normoxic outer region having a low survival probability. The higher survival of the hypoxic core is again evident because, at 80~MeV/u, the LET of carbon ions remains low enough to produce a substantial oxygen effect. The repair of sublethal lesions (panel b) is completed within about 20~h in most damaged cells (median 12~h for $^{1}$H and 13~h for $^{12}$C), with S-phase cells, which have the highest repair rate, repairing first, so that the fate of almost every cell is decided within the first day after irradiation. Radiation-induced death, in contrast, extends over one to five days (panels c): cells committed in G$_2$ or M die first, around 20~h, followed by cells in S and G$_1$, while quiescent G$_0$ cells, which have no mitosis to trigger their death, die last, with a median of about 45~h; the distribution of death times is therefore set by the phase composition of the spheroid at irradiation rather than by the radiation quality. The bottom rows show the consequences for the spheroid (panels d--f). At 6~h the doomed cells are distributed throughout the volume, both in the core and in the rim; the spheroid has not changed size, but its composition has. At 24~h doomed cells are being removed, and the sites they free allow quiescent cells in the former core to re-enter the cycle. At 72~h the proton-irradiated spheroid has recovered a compact structure with a cycling rim, whereas the carbon-irradiated one remains depleted. The spatial coexistence of doomed and surviving cells, the temporal separation between fate decision and death, and the recruitment of quiescent cells into the space left by dying cells are direct outputs of the agent-based description and are not accessible to population-level survival models.

\begin{figure}[htbp]
    \centering
    \begin{subfigure}[b]{\linewidth}
        \centering
        \includegraphics[width=\linewidth]{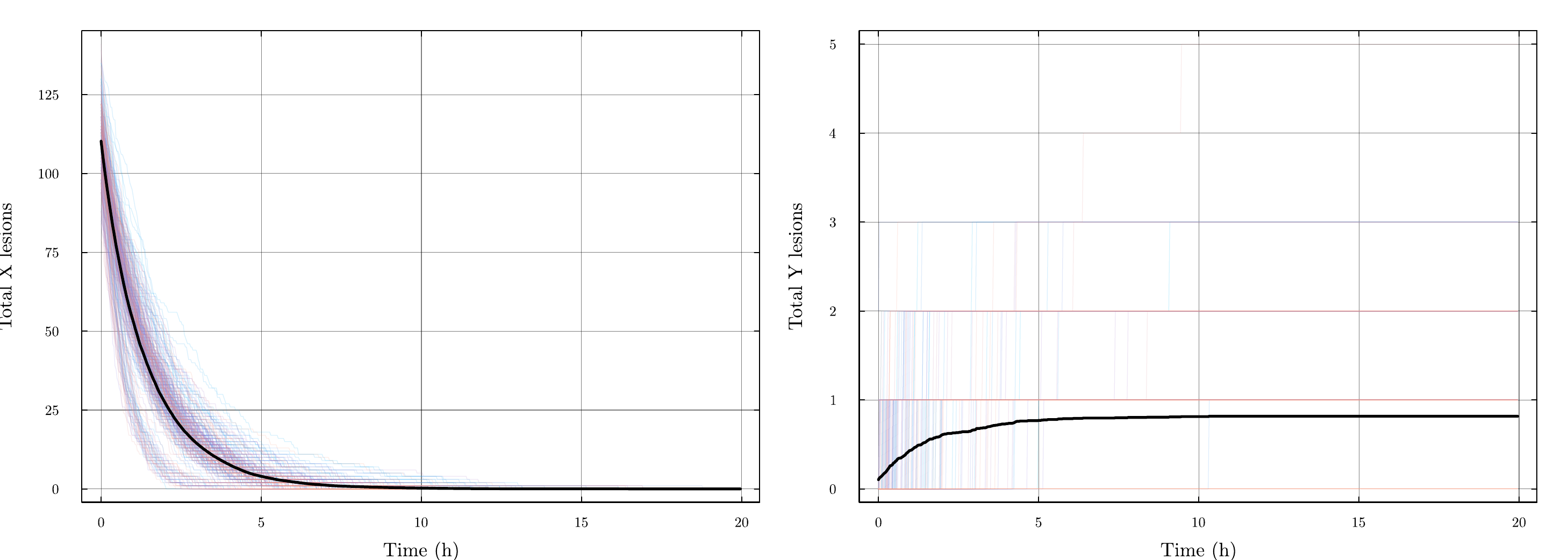}
        \caption{$^{1}$H ions at \SI{80}{\MeV}.}
        \label{fig:damage_1H_80MeV}
    \end{subfigure}
    \vspace{0.5em}
    \begin{subfigure}[b]{\linewidth}
        \centering
        \includegraphics[width=\linewidth]{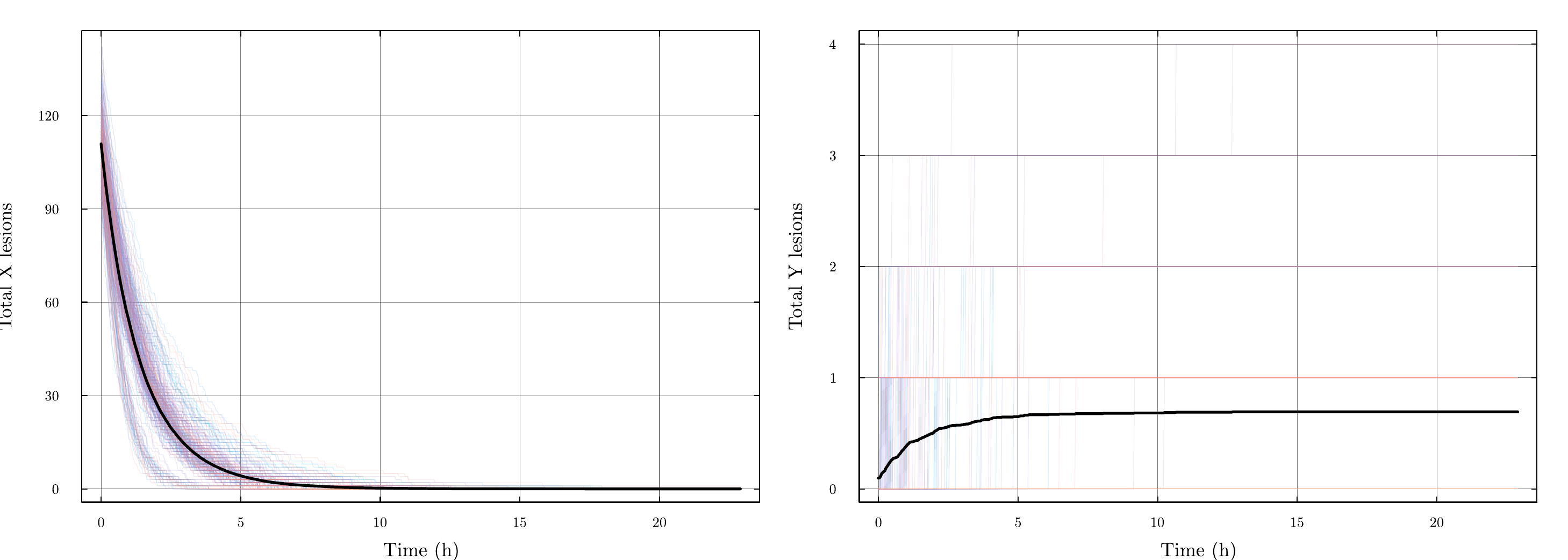}
        \caption{$^{1}$H ions at \SI{30}{\MeV}.}
        \label{fig:damage_1H_30MeV}
    \end{subfigure}
    \vspace{0.5em}
    \begin{subfigure}[b]{\linewidth}
        \centering
        \includegraphics[width=\linewidth]{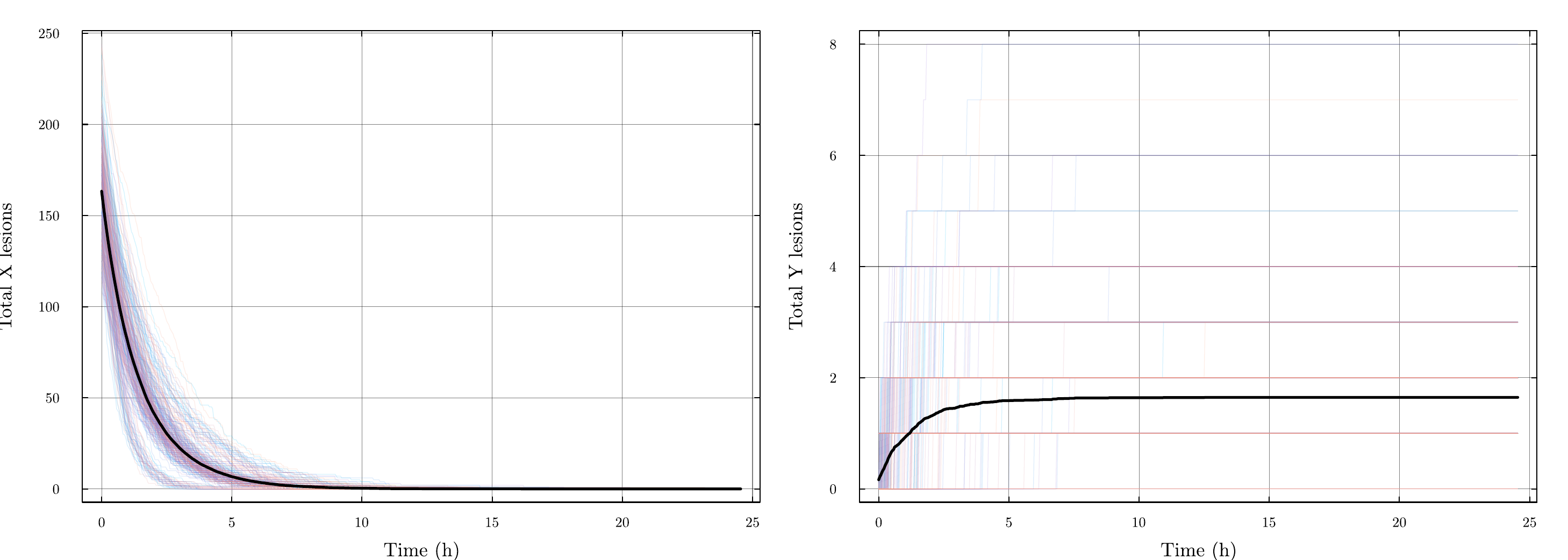}
        \caption{$^{12}$C ions at \SI{80}{\MeV\per u}.}
        \label{fig:damage_12C_80MeVu}
    \end{subfigure}
    \caption{Temporal evolution of sublethal lesions (left) and lethal lesions (right) per cell for $^{1}$H ions at \SI{80}{\MeV} (a), $^{1}$H ions at \SI{30}{\MeV} (b), and $^{12}$C ions at \SI{80}{\MeV\per u} (c), chosen to span increasing LET. The ensemble mean is shown in black; each shaded trajectory corresponds to an individual cell realization.}
    \label{fig:damage_evolution}
\end{figure}

Figure~\ref{fig:damage_evolution} shows the stochastic temporal evolution of DNA damage at the single-cell level for three radiation qualities of increasing LET. In each panel, the left plot shows the number of sublethal lesions per cell, and the right plot shows the number of lethal lesions per cell, both as a function of time after irradiation. Each shaded trajectory corresponds to an individual cell, illustrating the cell-to-cell variability inherent to the stochastic damage model, while the black curve represents the ensemble mean. The spread of individual trajectories is broadest immediately after irradiation and narrows over time as repair progresses. The mean lethal lesion count increases from protons at \SI{80}{\MeV} (Figure~\ref{fig:damage_1H_80MeV}) to protons at \SI{30}{\MeV} (Figure~\ref{fig:damage_1H_30MeV}) and carbon ions at \SI{80}{\MeV\per u} (Figure~\ref{fig:damage_12C_80MeVu}), consistent with the increasing LET.

\subsection{Dose dependence of the response}
\label{SEC:res_dose}

\begin{figure}[htbp]
\begin{subfigure}[b]{0.24\linewidth} \includegraphics[width=\linewidth]{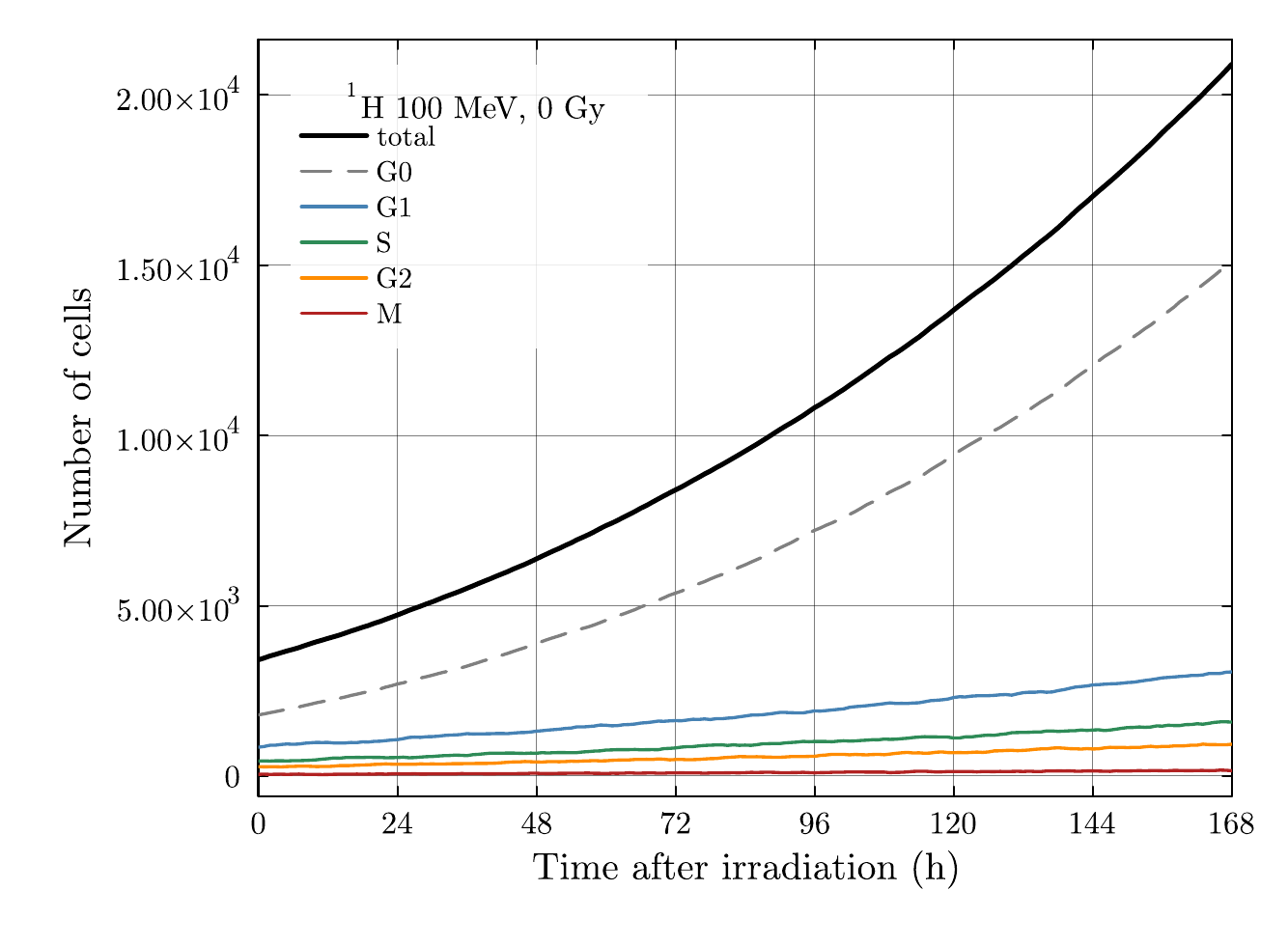} \end{subfigure}
\begin{subfigure}[b]{0.24\linewidth} \includegraphics[width=\linewidth]{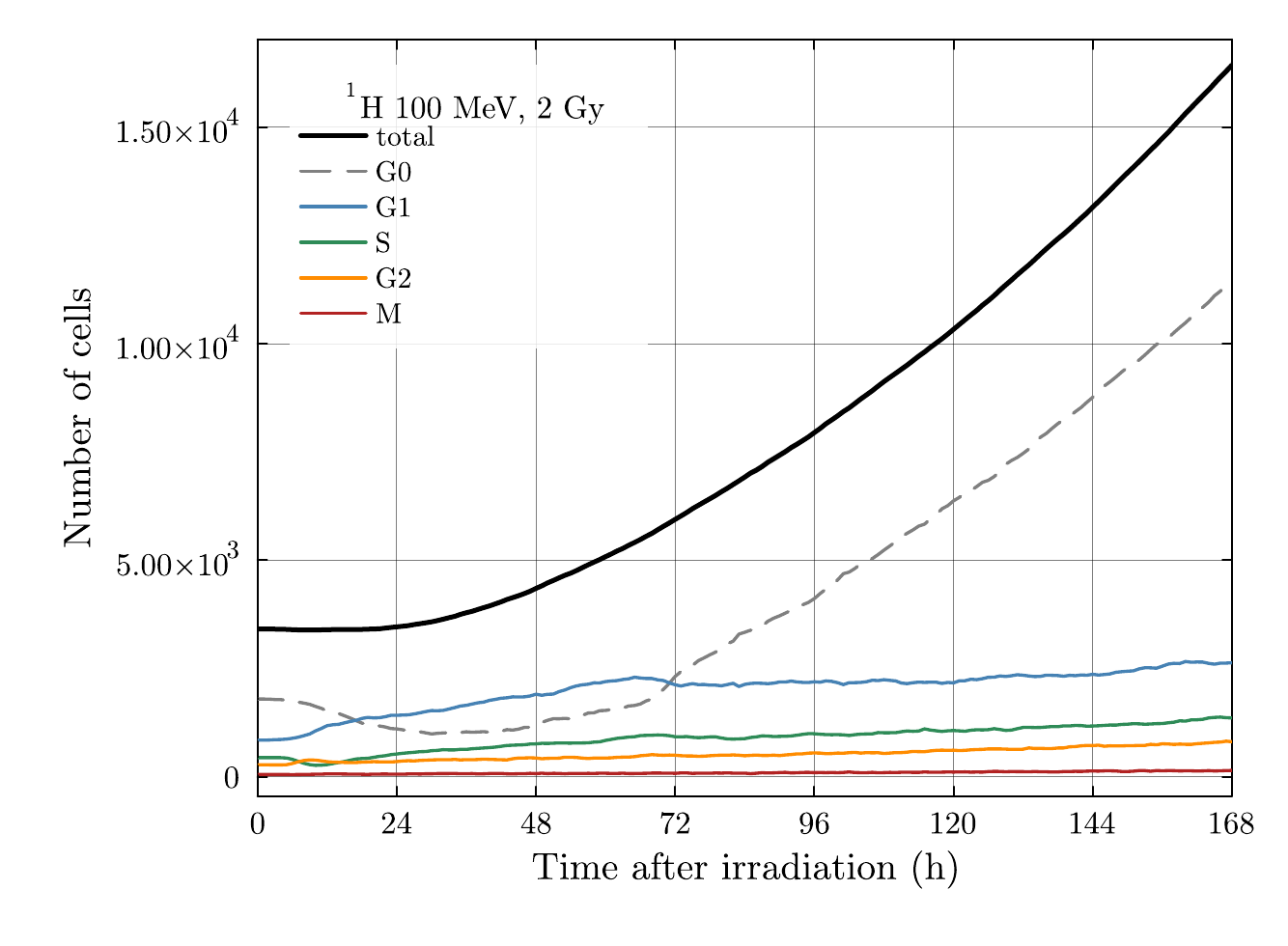} \end{subfigure}
\begin{subfigure}[b]{0.24\linewidth} \includegraphics[width=\linewidth]{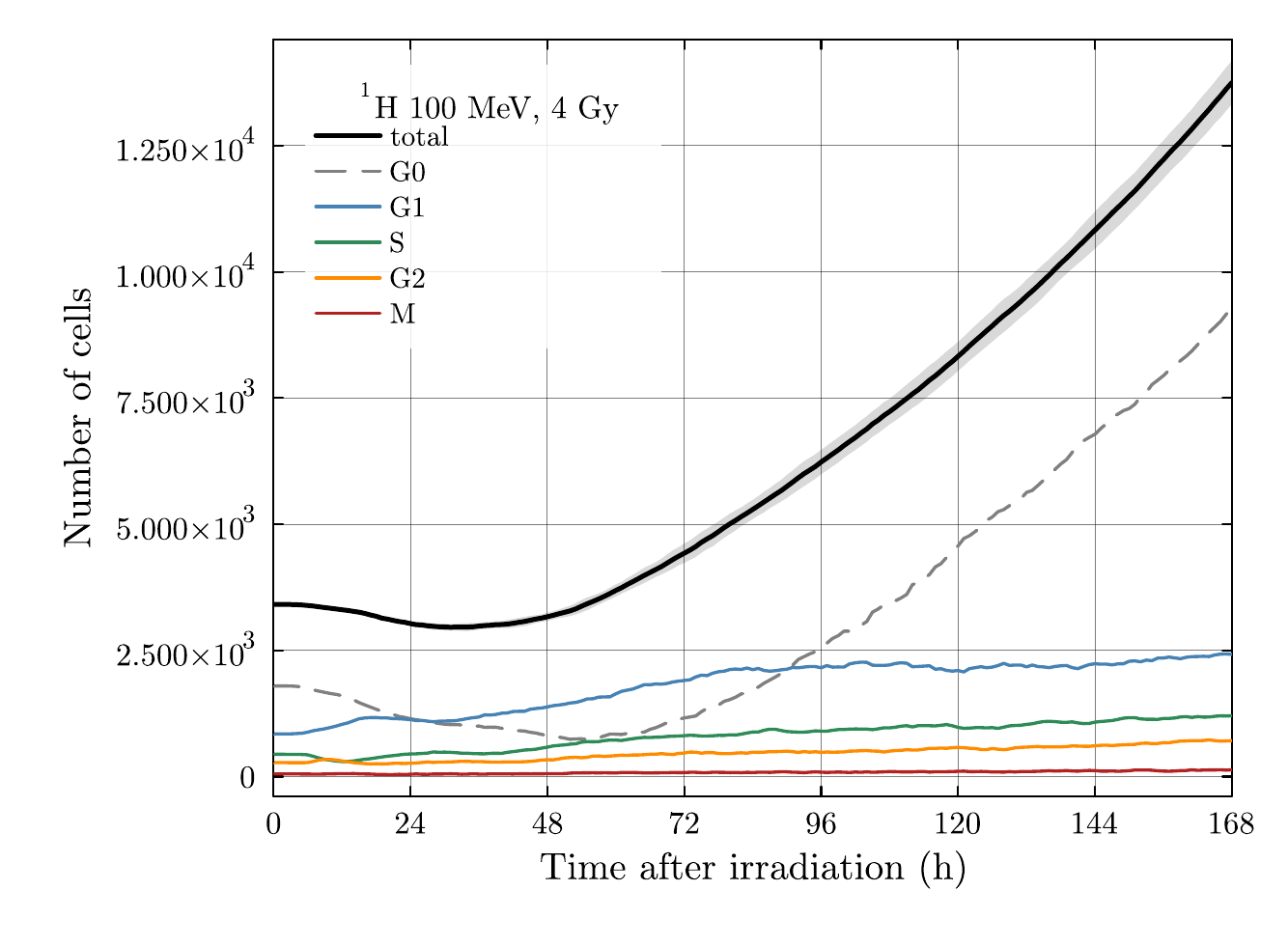} \end{subfigure}
\begin{subfigure}[b]{0.24\linewidth} \includegraphics[width=\linewidth]{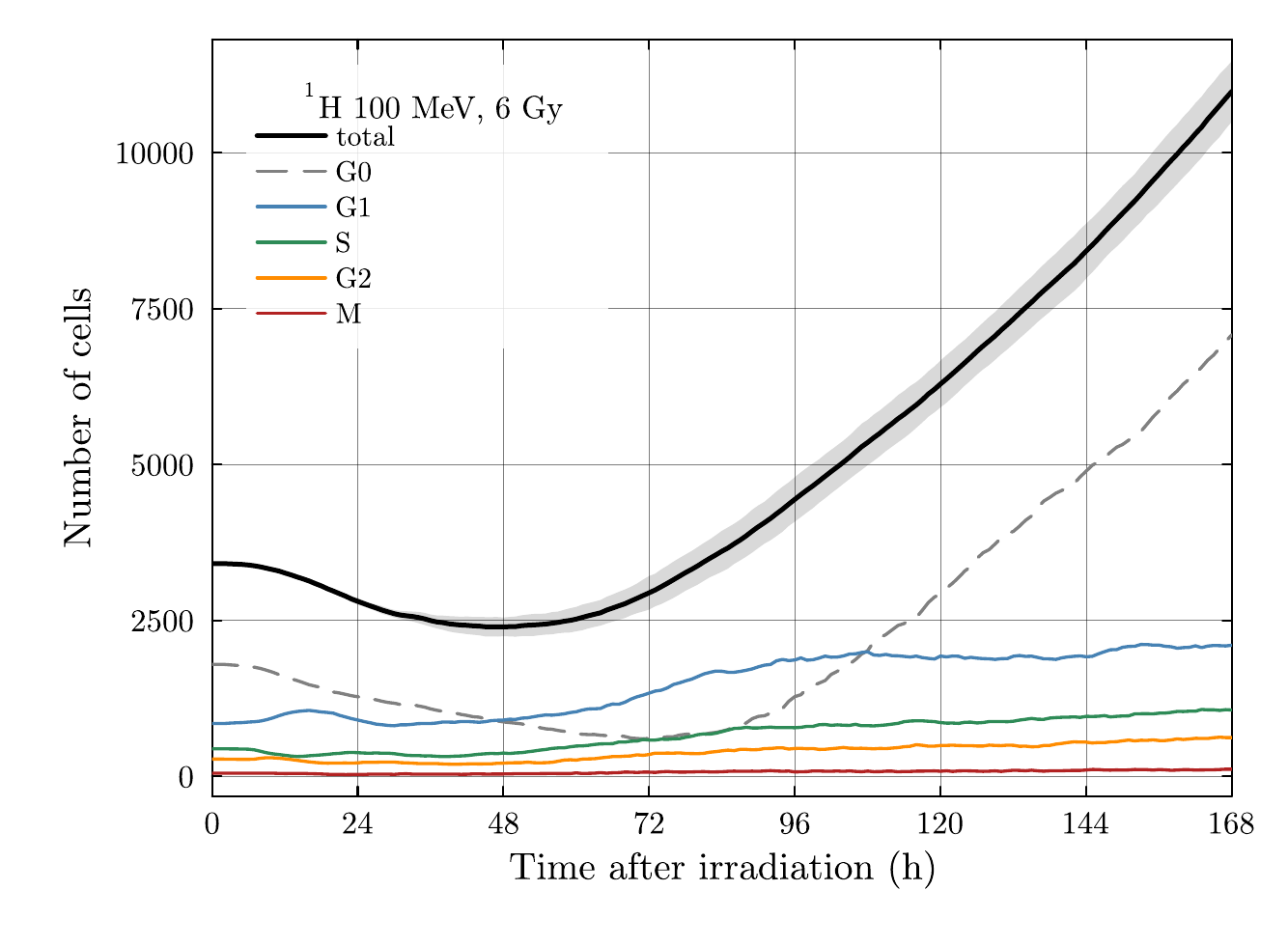} \end{subfigure} \vspace{0.5em} 
\begin{subfigure}[b]{0.24\linewidth} \includegraphics[width=\linewidth]{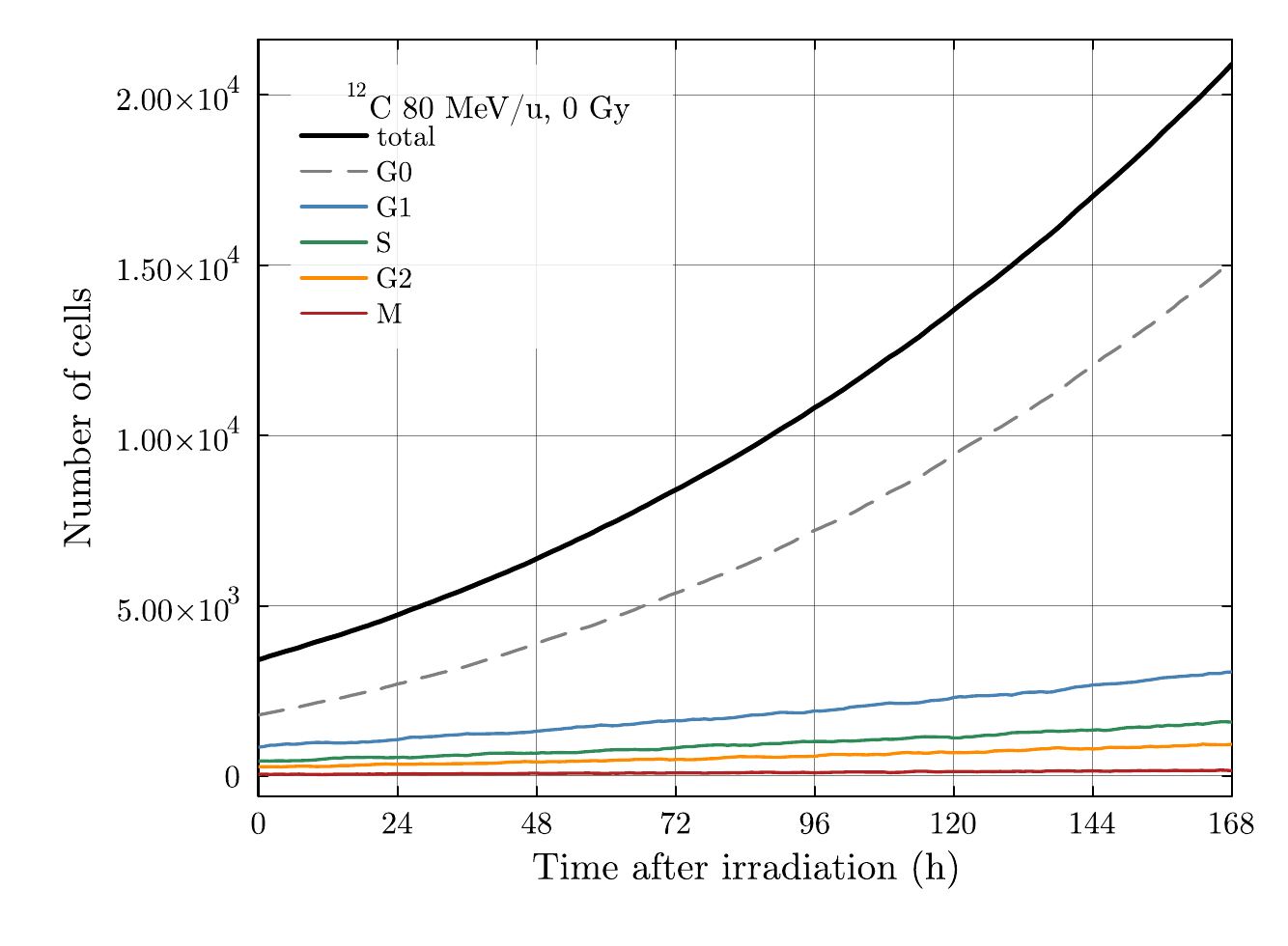} \end{subfigure}
\begin{subfigure}[b]{0.24\linewidth} \includegraphics[width=\linewidth]{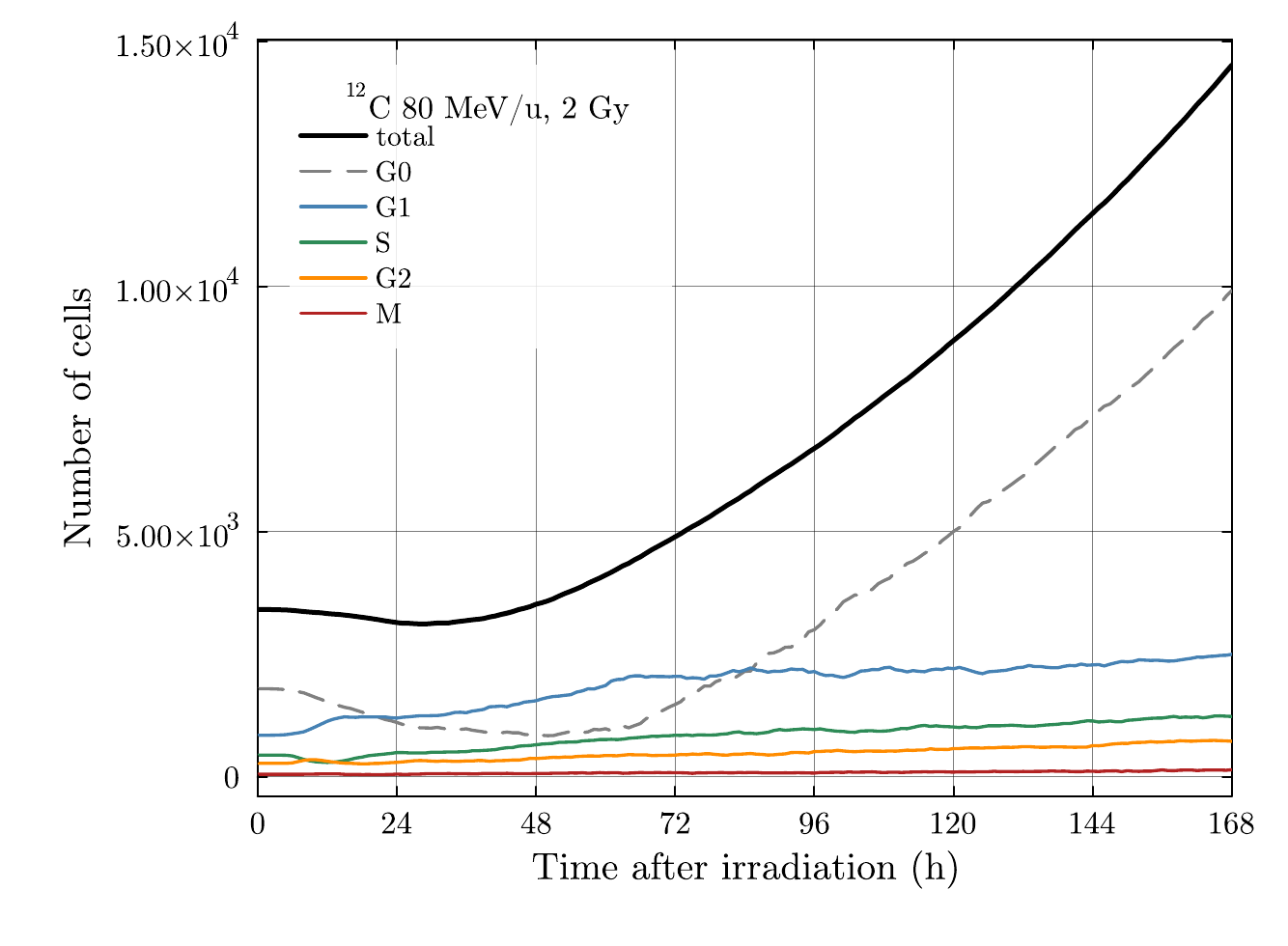} \end{subfigure}
\begin{subfigure}[b]{0.24\linewidth} \includegraphics[width=\linewidth]{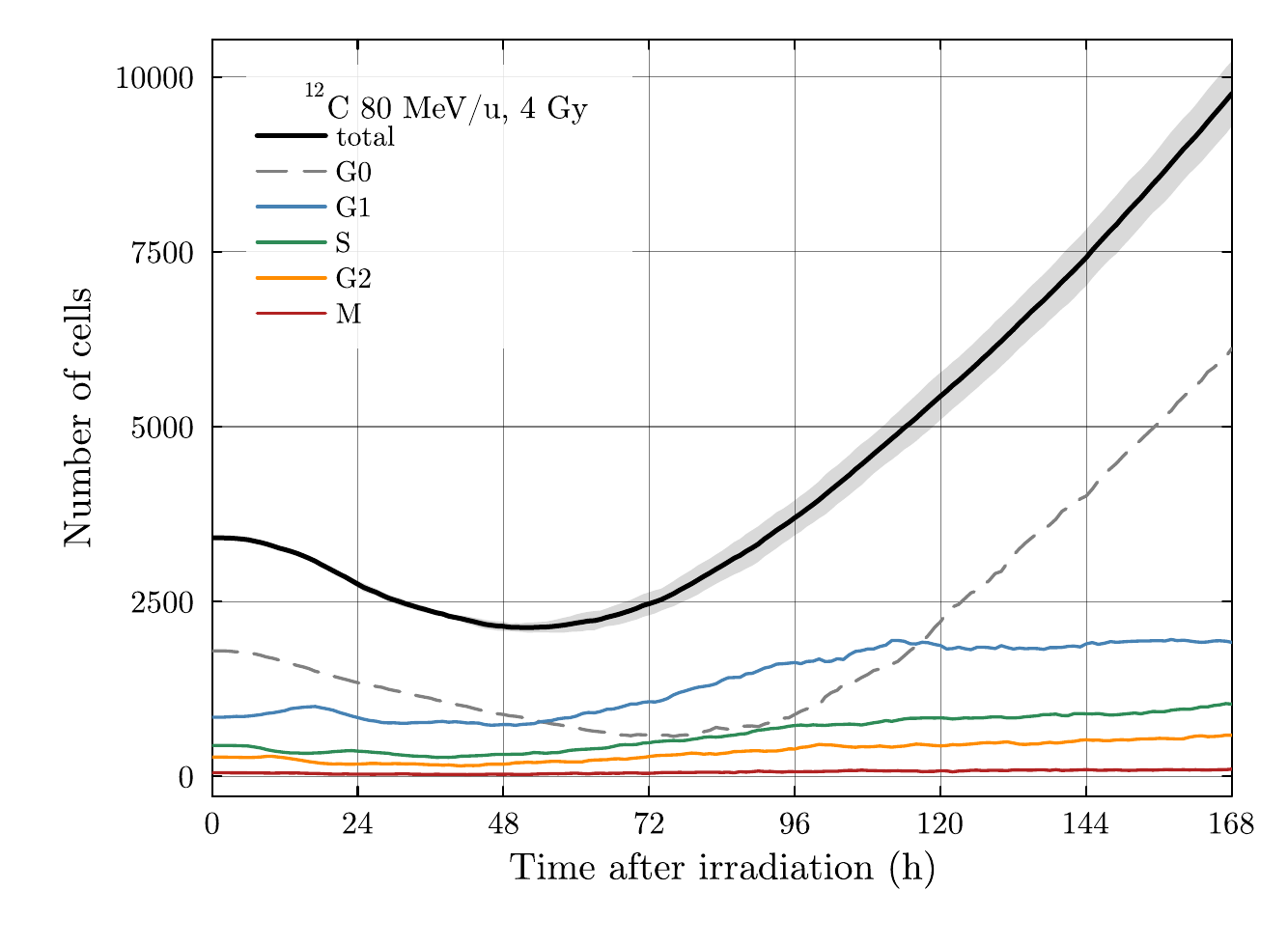} \end{subfigure}
\begin{subfigure}[b]{0.24\linewidth} \includegraphics[width=\linewidth]{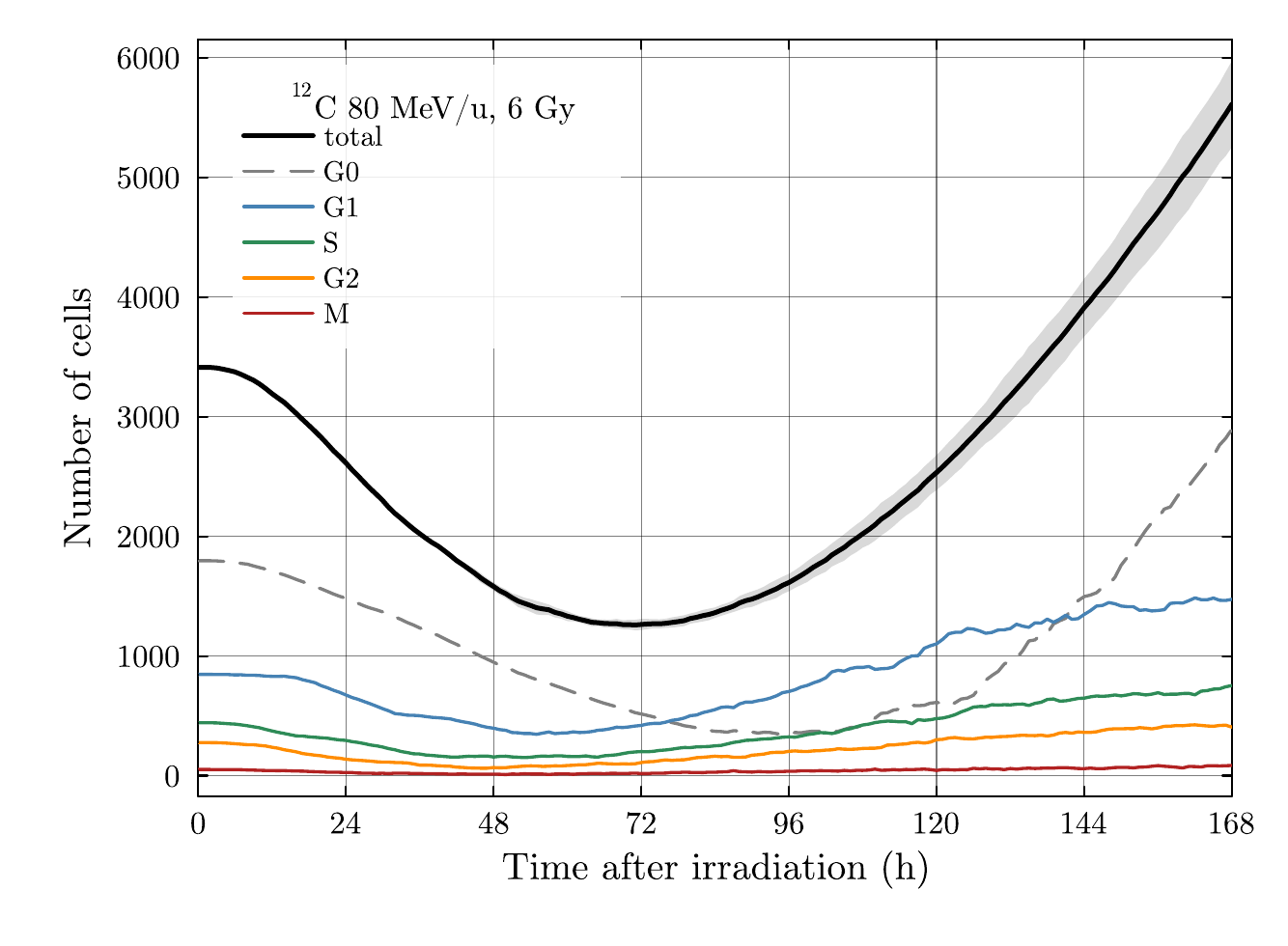} \end{subfigure}
  \caption{Total cells (black) and cells in G$_0$ (dashed grey), G$_1$ (blue), S (green), G$_2$ (orange) and M (red), including doomed cells, after a single acute dose of 0, 2, 4 and 6~Gy (columns) of $^{1}$H at 100~MeV (top row) and $^{12}$C at 80~MeV/u (bottom row), for spheroids of radius \SI{250}{\micro\metre}; mean $\pm$ SD over 3 realizations.}
  \label{fig:dose_phases}
\end{figure}

Figure~\ref{fig:dose_phases} shows the number of cells as a function of time after a single dose of 0, 2, 4 and 6~Gy (columns) of $^{1}$H at 100~MeV (top row) and $^{12}$C at 80~MeV/u (bottom row): the black line is the total number of cells, and the colored lines are the cells in each phase, G$_0$ (dashed grey), G$_1$ (blue), S (green), G$_2$ (orange) and M (red), including cells committed to death. With increasing dose, the drop of the quiescent G$_0$ population after irradiation becomes deeper and longer-lasting, because more cells in the core are committed to death and the sites they free allow more quiescent cells to re-enter the cycle; the resulting wave of G$_1$ cells, followed by S, G$_2$ and M, is delayed and broadened by the arrest of damaged cells until repair is complete. The total cell number reaches its minimum 9--71~h after irradiation, at 70\% of the initial value after 6~Gy of $^{1}$H and 37\% after 6~Gy of $^{12}$C, and the time needed to recover the initial cell number increases from 20~h after 2~Gy to 81~h after 6~Gy of $^{1}$H, and from 46~h to 137~h for $^{12}$C. At equal dose, carbon ions produce a deeper and longer-lasting depletion of all cycling phases, consistent with their lower surviving fraction. The cell-cycle composition of the spheroid therefore changes with dose, altering its radiosensitivity to subsequent irradiation, which underlies the schedule dependence discussed below.

\begin{figure}[htbp]
    \centering
    \includegraphics[width=0.45\linewidth]{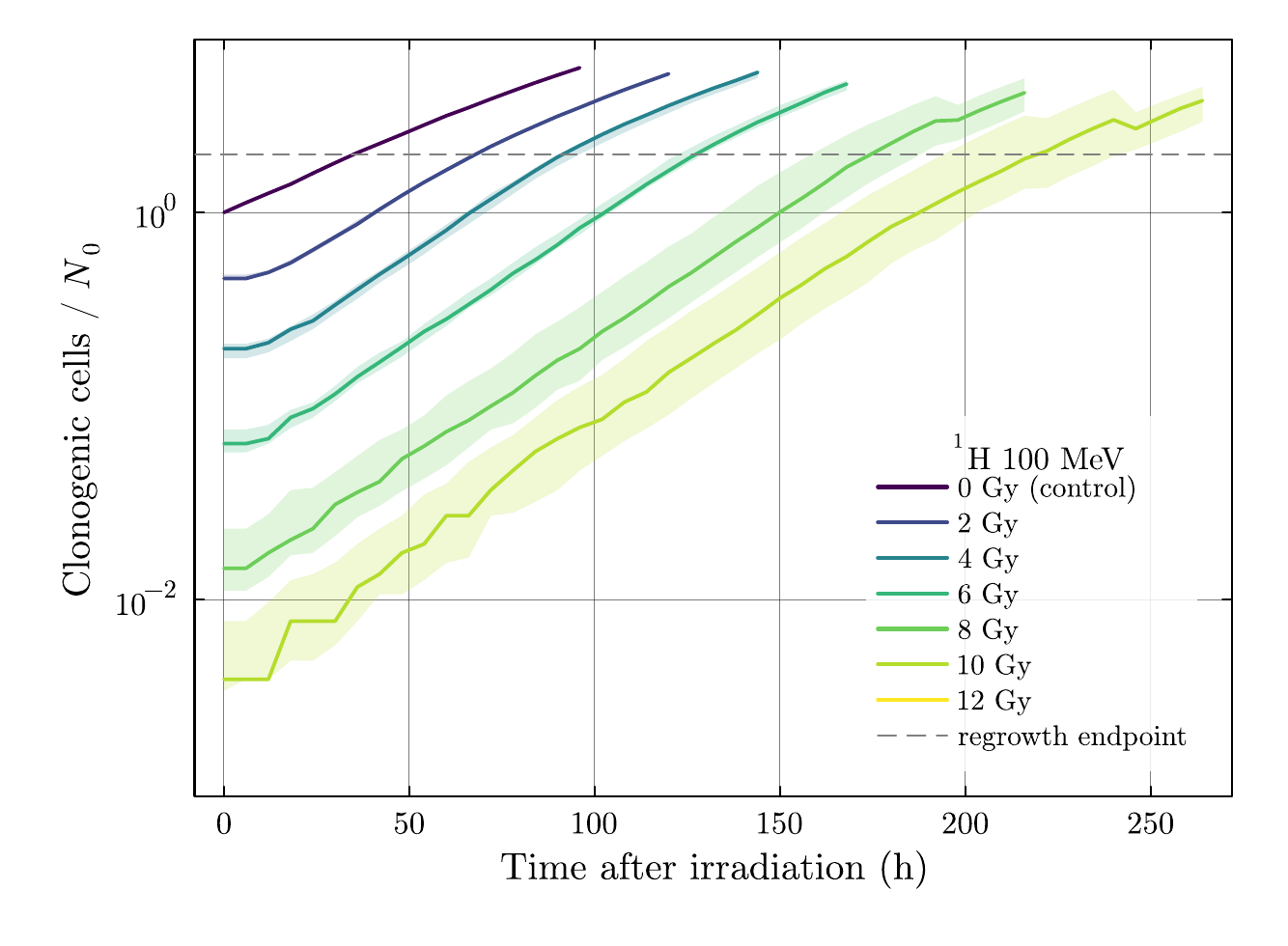}
    \includegraphics[width=0.45\linewidth]{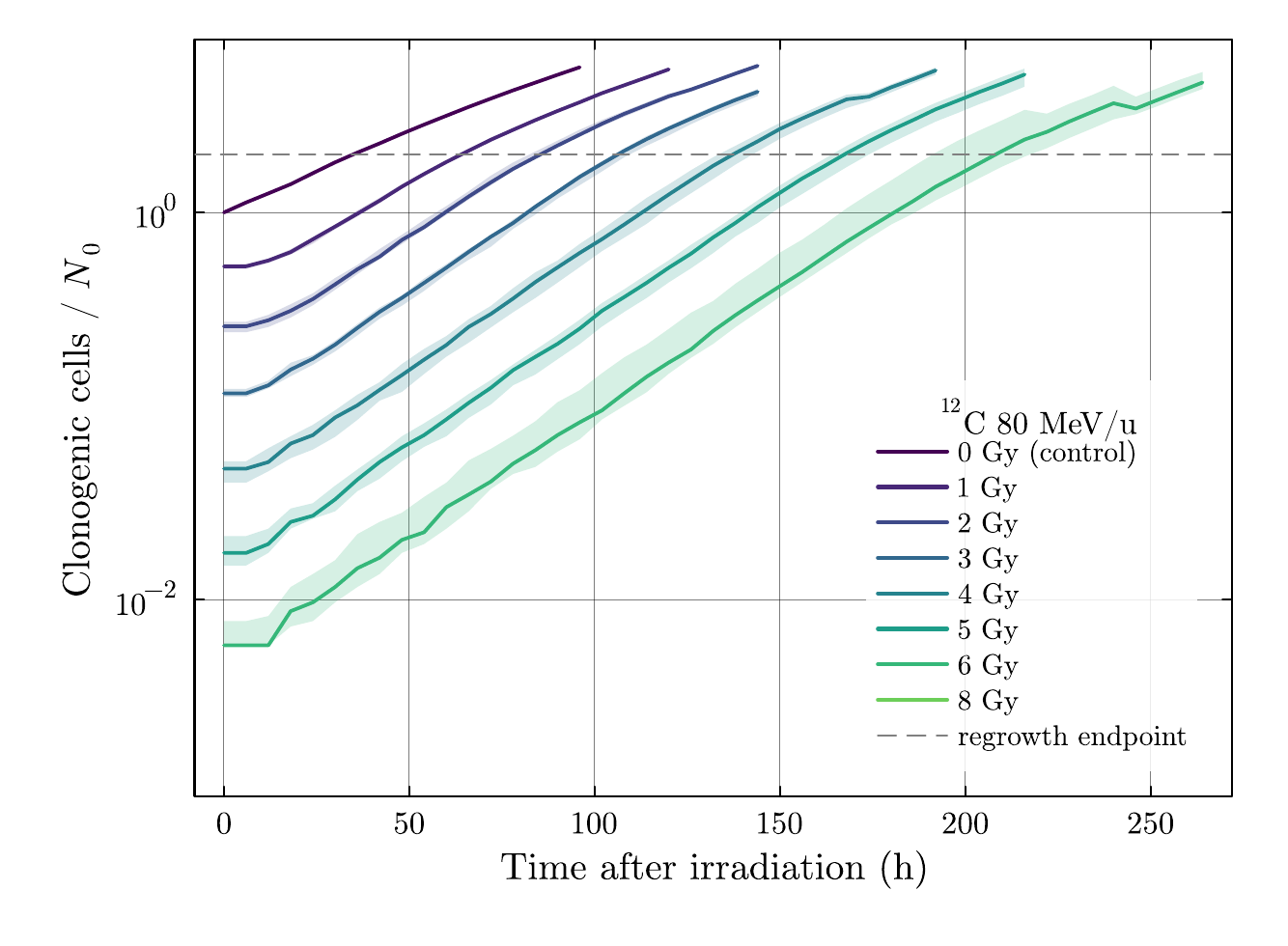}
    \caption{Clonogenic cells normalized to the initial number of cells as a function of time after irradiation, for protons at 100~MeV (left) and carbon ions at 80~MeV/u (right), for spheroids of radius \SI{150}{\micro\metre}. Colors: dose. Median over 20 realizations; bands: interquartile range; dashed line: regrowth endpoint (twice the initial number).}
    \label{fig:growth}
\end{figure}

Figure~\ref{fig:growth} shows the regrowth of the clonogenic population after single doses, for $^{1}$H at 100~MeV (left) and $^{12}$C at 80~MeV/u (right). Regrowth is defined as the time after irradiation at which the number of clonogenic cells reaches twice the initial number. The clonogenic number drops immediately after irradiation to the surviving fraction and then grows from the surviving clonogens; the time needed to reach twice the initial number increases with dose, from 36~h for the control to 177~h after 8~Gy of $^{1}$H and 246~h after 8~Gy of $^{12}$C. The interquartile range widens with dose because fewer clonogens survive; at the highest doses, a fraction of the spheroids are controlled, leaving no clonogenic cells.

\subsection{Split-dose irradiation}
\label{SEC:res_split}

\begin{figure}[htbp]
    \centering
    \includegraphics[width=0.65\linewidth]{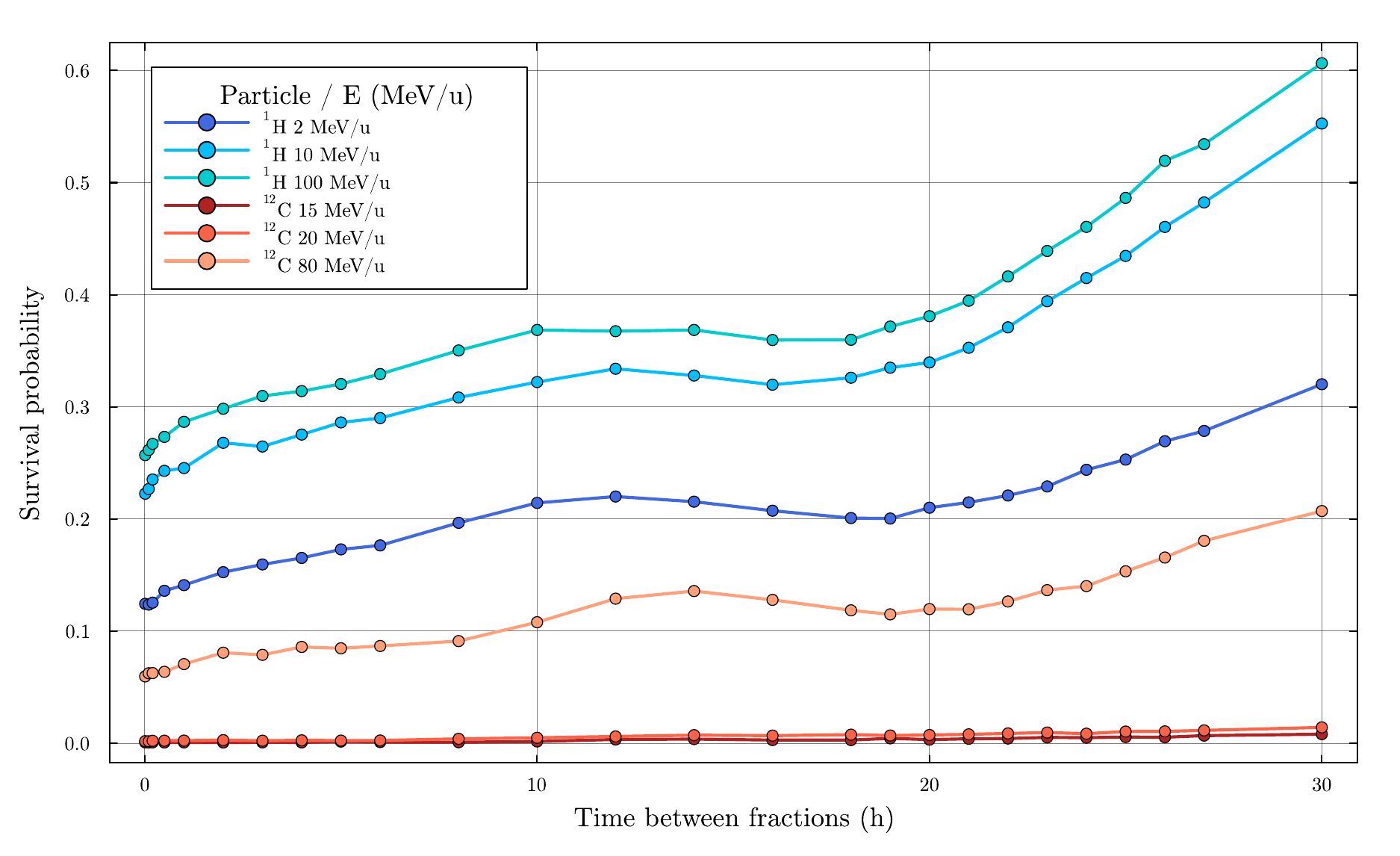}
    \caption{Survival probability as a function of the time of delivery of the second fraction for a fractionated irradiation scheme of \SI{1.5}{\gray} $+$ \SI{1.5}{\gray}, for $^{1}$H at 2, 10 and 100~MeV and $^{12}$C at 15, 20 and 80~MeV/u.}
    \label{fig:surv_prob_vs_time}
\end{figure}

Figure~\ref{fig:surv_prob_vs_time} shows, for a two-fraction scheme of \SI{1.5}{\gray} $+$ \SI{1.5}{\gray}, the surviving fraction after the second fraction (vertical axis) as a function of the time between the two fractions (horizontal axis), with one curve for each radiation quality. The survival probability varies with inter-fraction time, reflecting the competing contributions of sublethal damage repair and cell-cycle redistribution in the interval between fractions. Survival increases with the interval across all radiation qualities, but the increase is largest for low-LET protons and is almost absent for carbon ions at the lowest energies, where sublethal damage that can be repaired between fractions contributes little to cell killing.

\begin{figure}[htbp]
    \centering
    \begin{subfigure}[t]{0.49\linewidth}
        \centering
        \includegraphics[width=\linewidth]{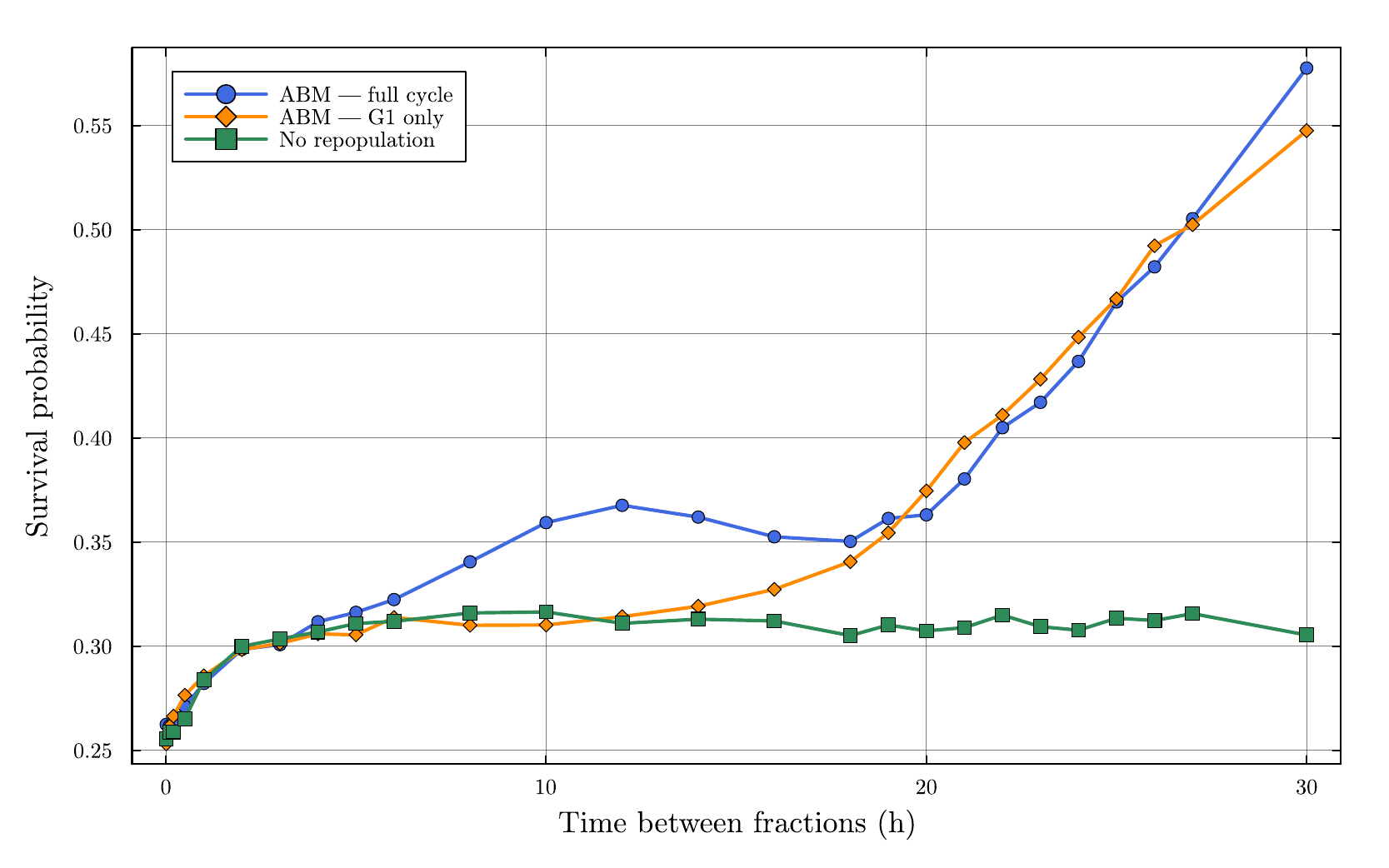}
        \caption{}
        \label{fig:surv_vs_time}
    \end{subfigure}
    \hfill
    \begin{subfigure}[t]{0.49\linewidth}
        \centering
        \includegraphics[width=\linewidth]{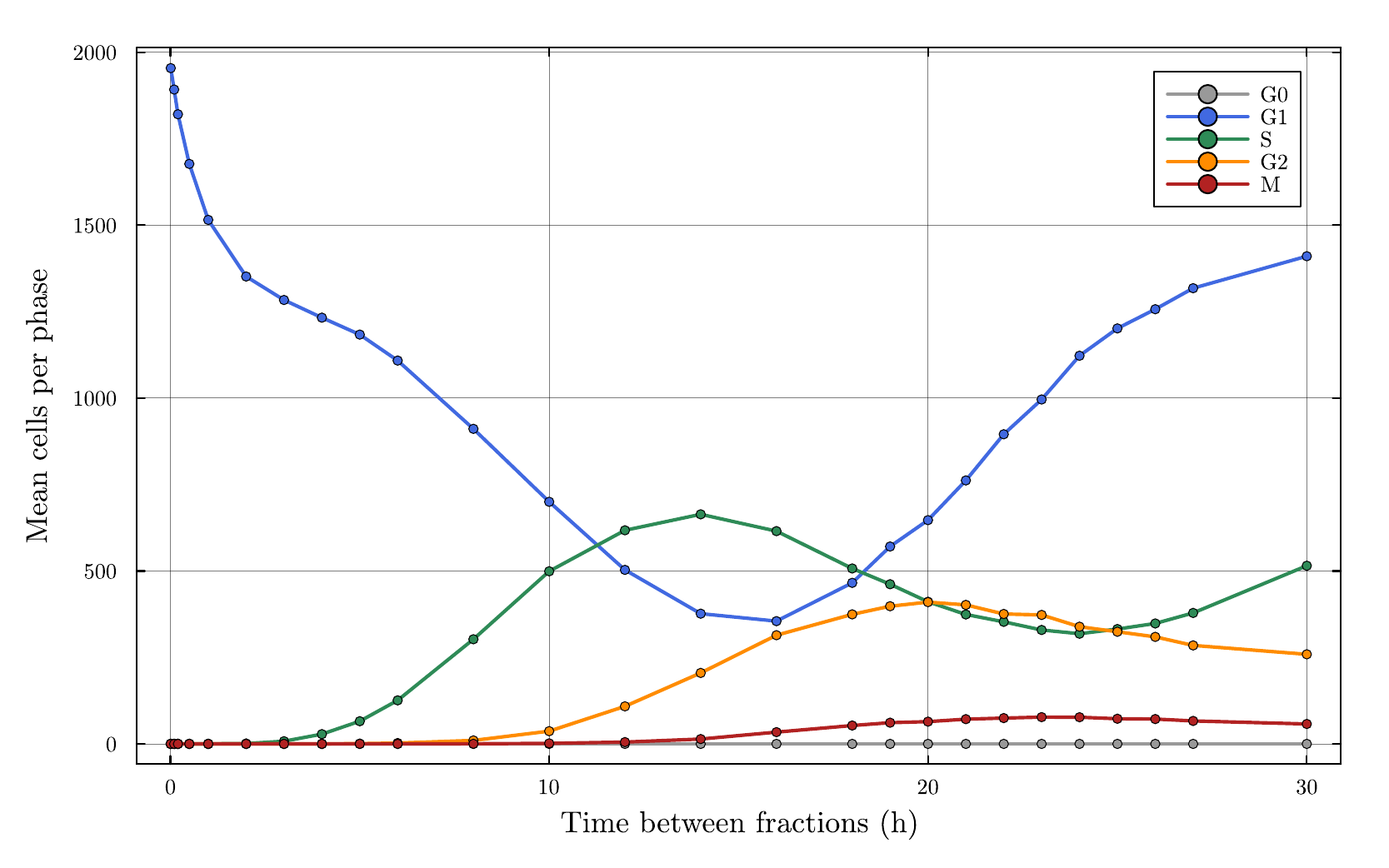}
        \caption{}
        \label{fig:phases_vs_time}
    \end{subfigure}
    \caption{Survival probability as a function of inter-fraction time (a) and cell-cycle phase composition at the time of the second fraction (b) for $^{1}$H at \SI{100}{\MeV} and a scheme of \SI{1.5}{\gray} $+$ \SI{1.5}{\gray}. (a) Three model configurations: the full model, a model in which all cells have the radiosensitivity of G$_1$ cells, and the standard GSM$^2$ without cell dynamics between fractions. (b) Phase composition immediately before the second fraction in the full model: G$_1$ (blue), S (green), G$_2$ (orange), M (red), and G$_0$ (dashed grey).}
    \label{fig:surv_phases_vs_time}
\end{figure}

Figure~\ref{fig:surv_phases_vs_time} isolates the contribution of cell-cycle dynamics and repopulation to the inter-fraction time dependence of survival probability for $^{1}$H ions at \SI{100}{\MeV}. In Figure~\ref{fig:surv_vs_time}, three model configurations are compared. The full model accounts for phase-dependent radiosensitivity, cell-cycle progression, and repopulation between fractions. The G$_1$-only case assigns the G$_1$ radiosensitivity to all cells, isolating the contribution of phase-selective damage. The standard GSM$^2$ model without cell dynamics serves as a reference, yielding a time-independent survival probability once the interval exceeds the repair time scale. Two distinct trends emerge in the range of 6--24 hours: a first increase in survival probability due to cells redistributing into the radioresistant S phase, followed by a decrease as the population re-enters the radiosensitive G$_2$/M phases. Figure~\ref{fig:phases_vs_time} provides a direct readout of the cell-cycle state immediately before the second fraction, showing how the phase composition evolves as a function of inter-fraction time in the full model. The oscillatory structure of the phase curves confirms the mechanistic origin of the survival trends described above.

\begin{figure*}[t]
  \centering
  \begin{subfigure}[b]{0.32\linewidth}
    \centering
    \includegraphics[width=\linewidth]{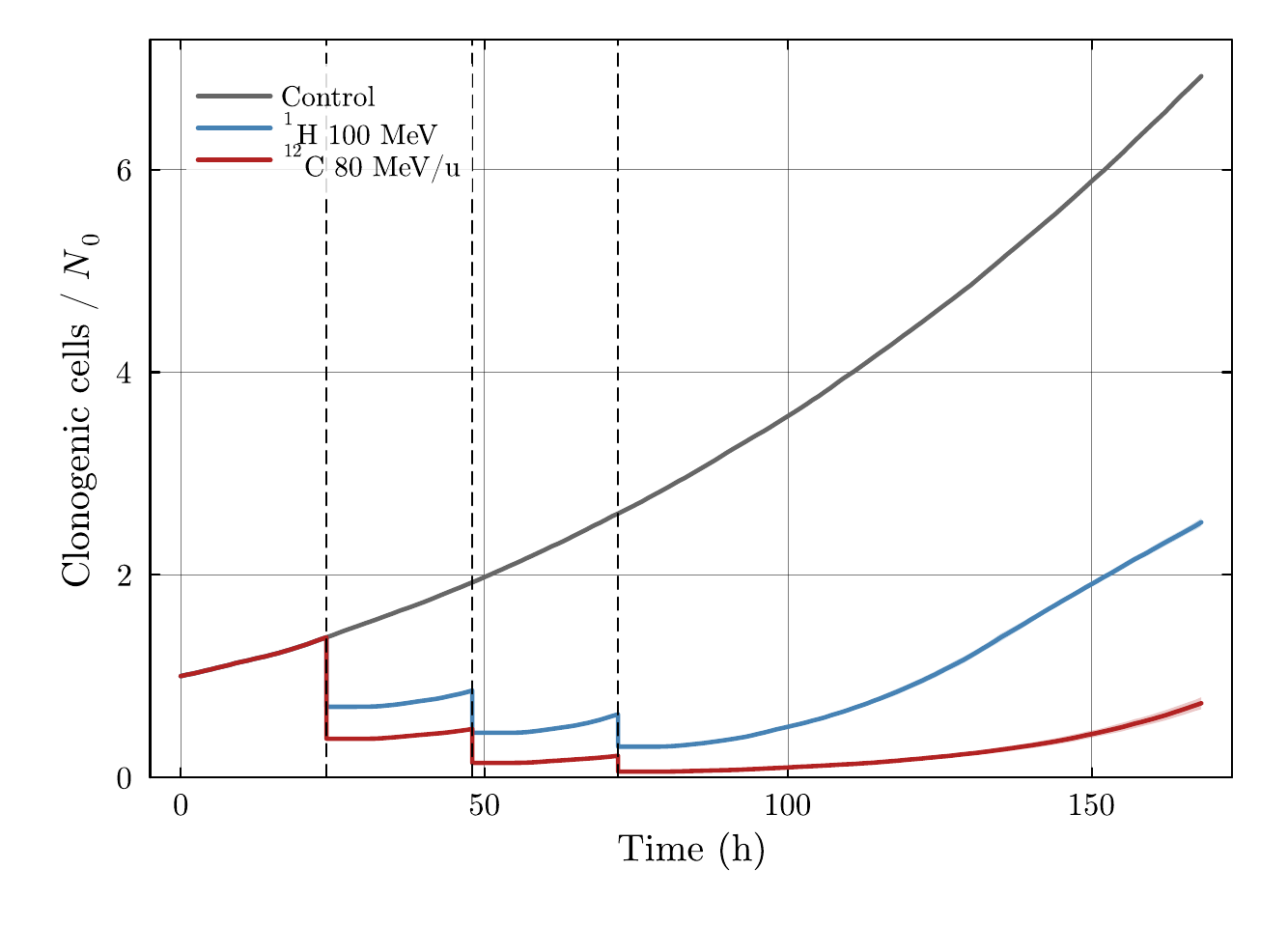}
    \caption{}
    \label{fig:fx_comparison}
  \end{subfigure}\hfill
  \begin{subfigure}[b]{0.32\linewidth}
    \centering
    \includegraphics[width=\linewidth]{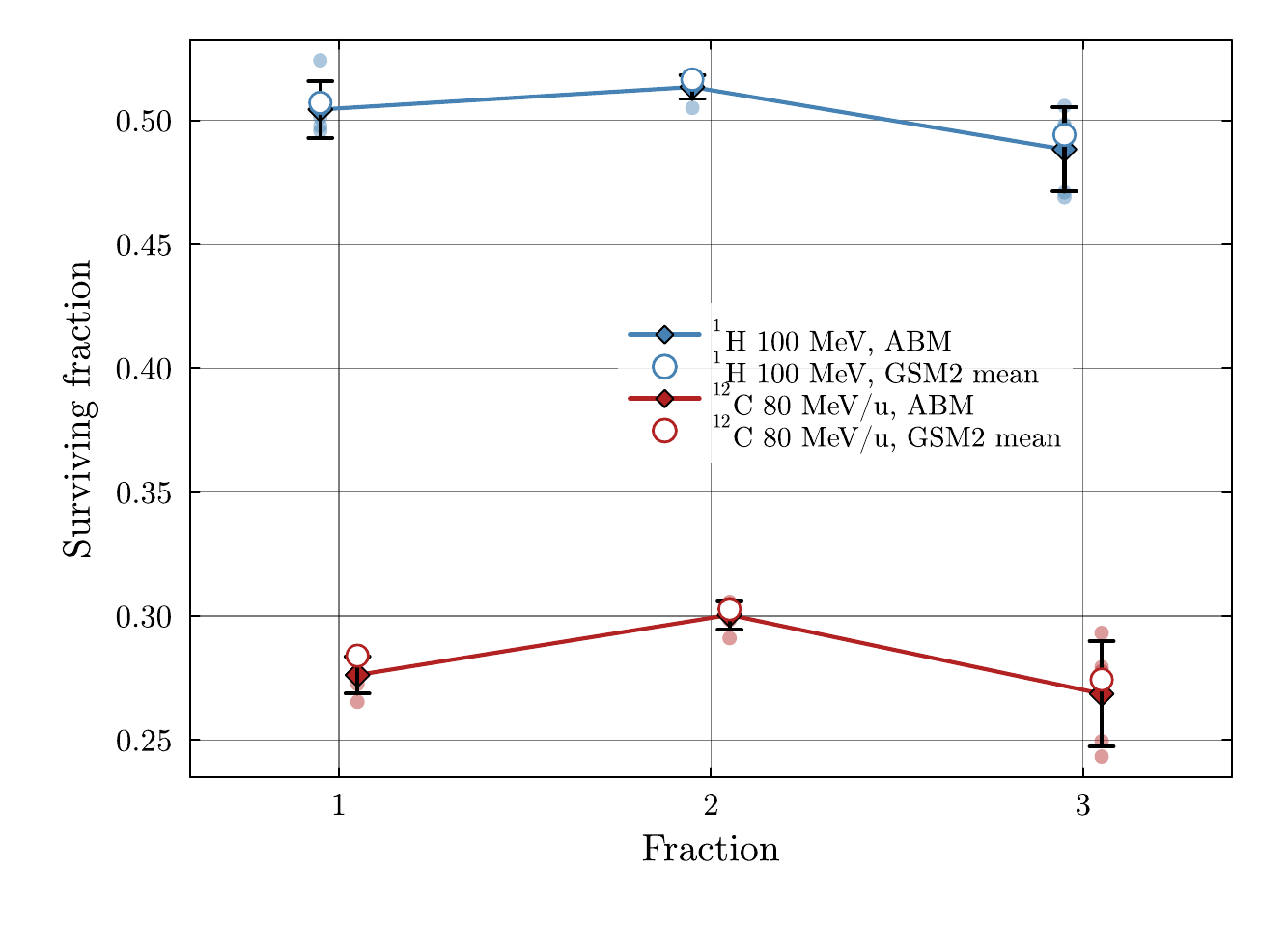}
    \caption{}
    \label{fig:fx_sf}
  \end{subfigure}\hfill
  \begin{subfigure}[b]{0.32\linewidth}
    \centering
    \includegraphics[width=\linewidth]{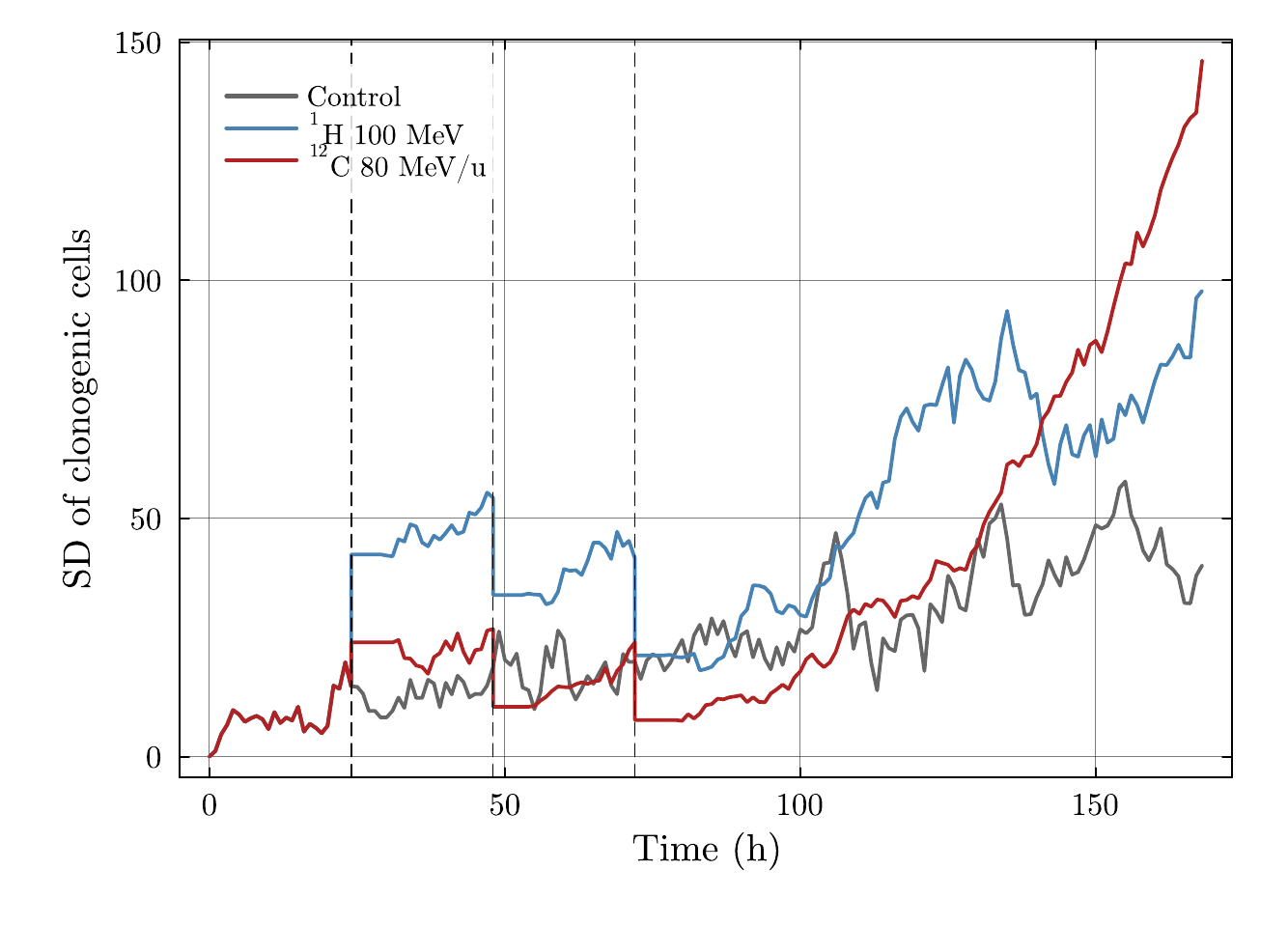}
    \caption{}
    \label{fig:fx_sd}
  \end{subfigure}
  \caption{Fractionated irradiation ($3 \times 2$~Gy, 24~h apart) of spheroids with $^{1}$H at 100~MeV and $^{12}$C at 80~MeV/u, and unirradiated control. (a)~Clonogenic cells normalized to the initial number, mean $\pm$ SD over realizations. (b)~Surviving fraction of each fraction: simulations (symbols, mean $\pm$ SD) and mean GSM$^2$ survival probability (open circles). (c)~Standard deviation of the clonogenic cell number across realizations. Dashed vertical lines: fractions.}
  \label{fig:fractionation}
\end{figure*}

\begin{figure}[t]
  \centering
  \begin{subfigure}[b]{0.49\linewidth}
    \centering
    \includegraphics[width=\linewidth]{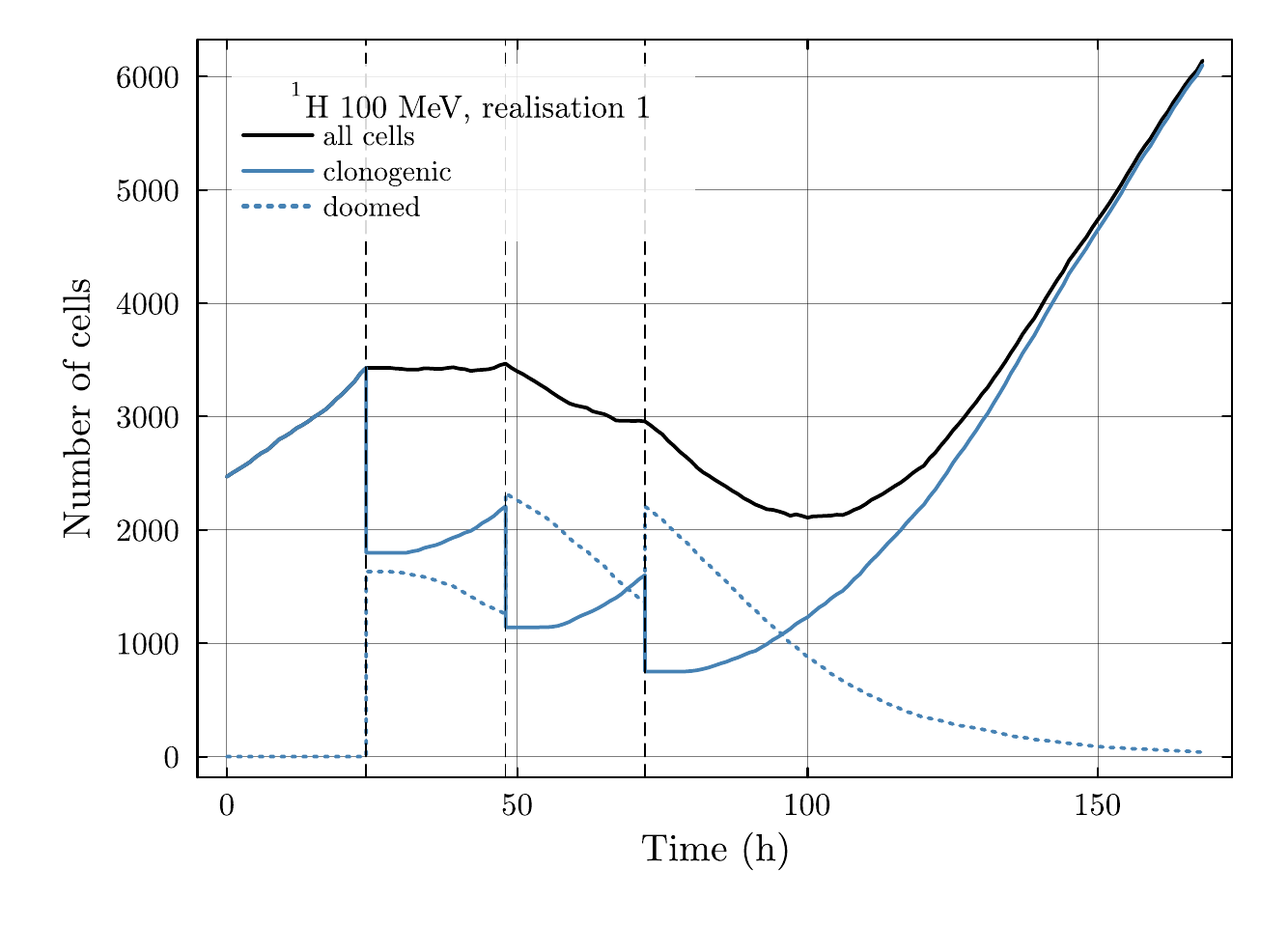}
    \caption{}
    \label{fig:fx_single_H}
  \end{subfigure}\hfill
  \begin{subfigure}[b]{0.49\linewidth}
    \centering
    \includegraphics[width=\linewidth]{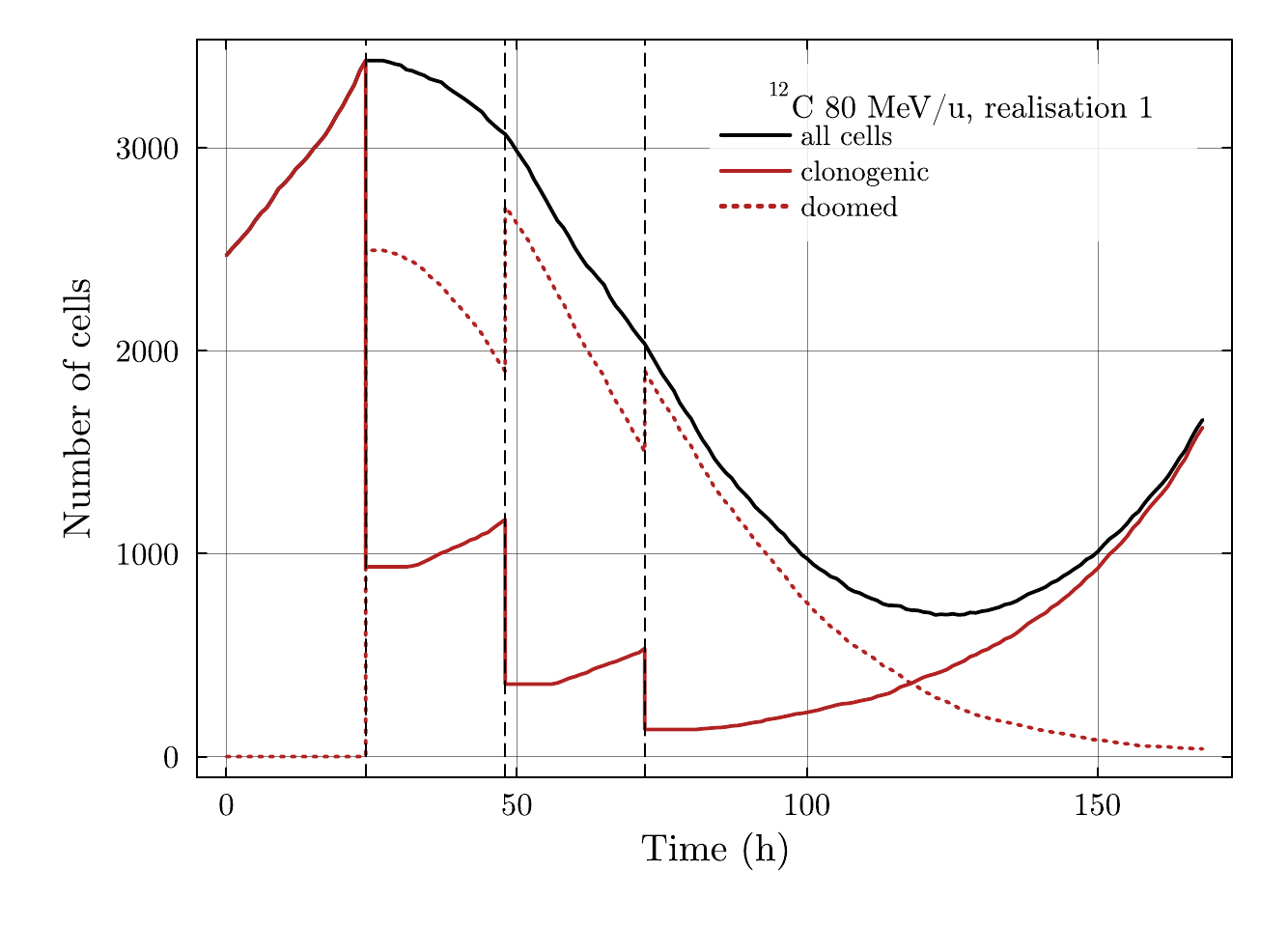}
    \caption{}
    \label{fig:fx_single_C}
  \end{subfigure}
  \caption{Single realization of the fractionated schedule for (a)~$^{1}$H at 100~MeV and (b)~$^{12}$C at 80~MeV/u: all cells, clonogenic cells, and doomed cells (committed to death, not yet removed). Dashed vertical lines: fractions.}
  \label{fig:fractionation_single}
\end{figure}

To assess the response to fractionated irradiation, spheroids were grown for 24~h, then irradiated with three 2~Gy fractions, 24~h apart, followed by 96~h of free evolution. All conditions started from the same spheroid, and each realization used the same random sequence up to the first fraction, so that differences in radiation quality reflect only the irradiation. Figure~\ref{fig:fractionation} summarizes the realizations: (a) the number of clonogenic cells, normalized to its value at the first fraction, as a function of time, for the control and the two radiation qualities (lines: mean; bands: standard deviation; dashed vertical lines: fractions); (b) the surviving fraction of each fraction, from the simulations (filled symbols) and from the mean GSM$^2$ survival probability of the irradiated cells (open circles); (c) the variability between realizations as a function of time. Figure~\ref{fig:fractionation_single} shows one realization for each radiation quality, with the total number of cells (black), the clonogenic cells (solid colored line) and the doomed cells (dotted colored line), i.e.\ cells committed to death but not yet removed.

At the population level (Figure~\ref{fig:fx_comparison}), each fraction produced an immediate drop in the number of clonogenic cells, followed by regrowth during the interval before the next fraction. The drop was larger for $^{12}$C than for $^{1}$H, and the clonogenic population after the third fraction was about five times smaller (clonogenic/$N_0 \approx 0.32$ for $^1$H vs.\ $\approx 0.06$ for $^{12}$C), which delayed regrowth relative to the control. The surviving fraction of individual fractions (Figure~\ref{fig:fx_sf}) was about 0.50 for $^{1}$H and 0.28 for $^{12}$C. It agreed with the mean GSM$^2$ survival probability of the irradiated cells, confirming that the population response is consistent with the single-cell model. The surviving fraction changed slightly from one fraction to the next, reflecting the redistribution of cells among cell-cycle phases and the recruitment of quiescent cells between fractions.

The relative spread between realizations (Figure~\ref{fig:fx_sd}) increased after each fraction and was larger for $^{12}$C than for $^{1}$H, since fewer cells survive each carbon-ion fraction and the dose is deposited by fewer, denser tracks, so that the fate of each realization depends on a smaller number of surviving clonogens; during regrowth after the last fraction, differences in the number and position of the surviving clonogens are amplified by proliferation.

Single realizations (Figure~\ref{fig:fractionation_single}) show how the spheroid responds within this schedule. At each fraction, the clonogenic count drops immediately, while the total number of cells barely changes. The lethally damaged cells remain in the spheroid as doomed cells for the duration of the death delay and are removed over the following one to two days. Total cell number, therefore, lags behind the clonogenic population, and the difference between the two curves is the number of doomed cells still present. For $^{12}$C, doomed cells make up a larger part of the spheroid after each fraction, and they accumulate across fractions, because a new fraction arrives before the cells doomed by the previous one have been cleared. After the last fraction, doomed cells are eliminated, and the total cell number converges to the clonogenic population as the spheroid regrows. This separation between committed and surviving cells, which is not captured by models that remove cells instantaneously, shows that the cell number or volume of an irradiated spheroid overestimates the surviving population in the days after irradiation.

\begin{figure*}[t]
  \centering
  \includegraphics[width=\linewidth]{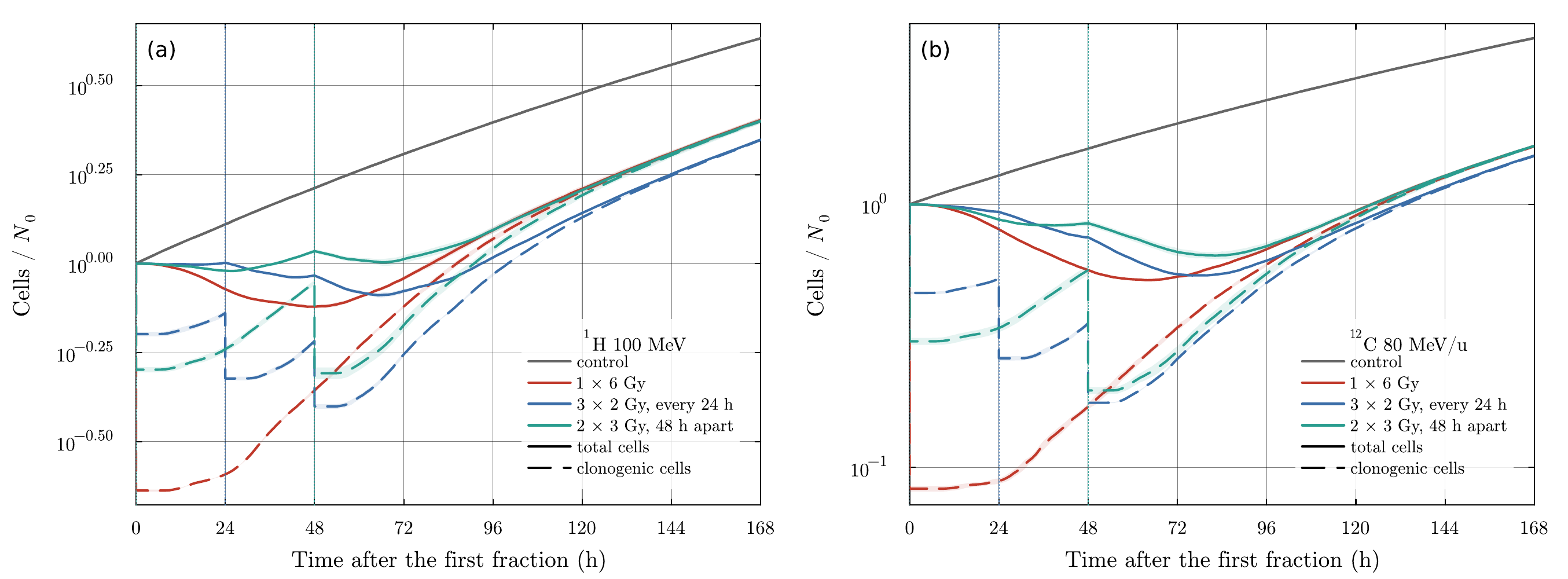}\\[0.5em]
  \includegraphics[width=0.85\linewidth]{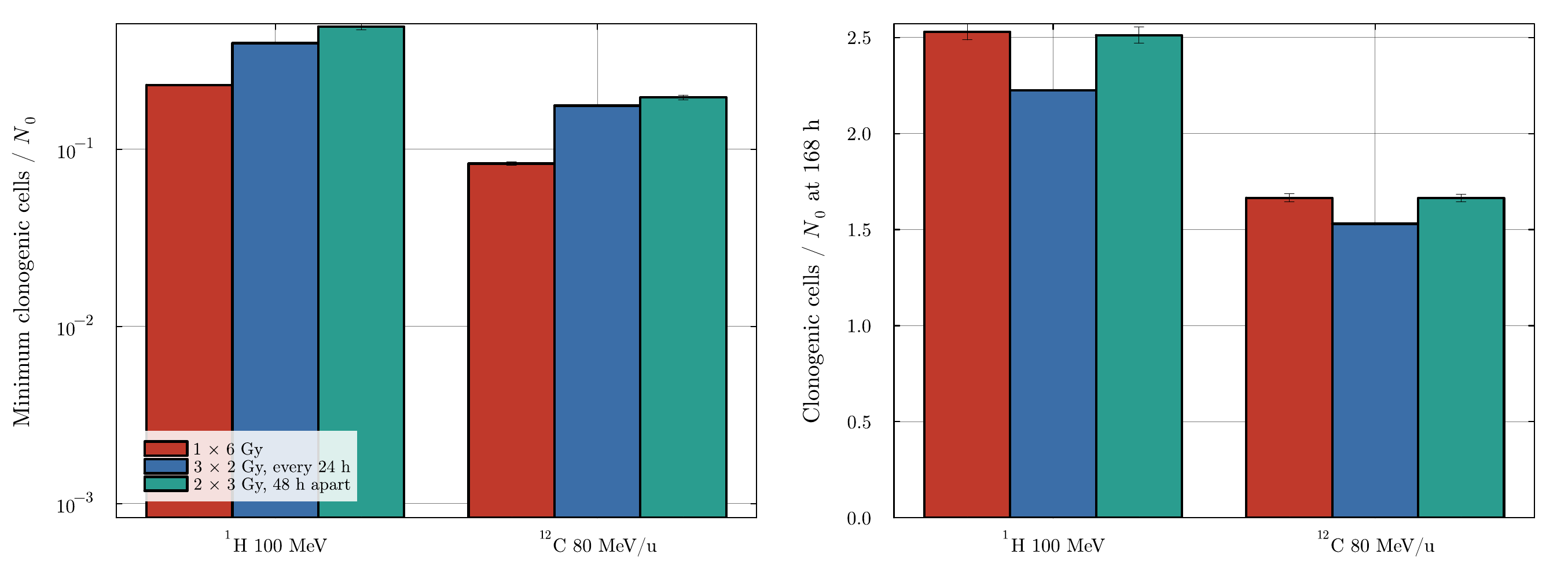}
  \caption{A total dose of 6~Gy delivered as $1\times6$~Gy, $3\times2$~Gy every 24~h, or $2\times3$~Gy 48~h apart, to spheroids of radius \SI{350}{\micro\metre} with a hypoxic core, with the oxygen profile recomputed before each fraction. Top: total (solid) and clonogenic (dashed) cells normalized to the initial number for (a) $^{1}$H at 100~MeV and (b) $^{12}$C at 80~MeV/u, with the unirradiated control; dotted lines: fractions. Bottom: (c) minimum and (d) final (168~h) clonogenic fraction for each schedule and radiation quality; mean $\pm$ SD over 3 realizations.}
  \label{fig:schedules}
\end{figure*}

Figure~\ref{fig:schedules} compares three ways of delivering the same total dose of 6~Gy to larger spheroids. Panels (a) and (b) show, for $^{1}$H and $^{12}$C respectively, the total number of cells (solid lines) and the clonogenic cells (dashed lines) normalized to their number at the first fraction, for the three schedules and the unirradiated control (grey); dotted vertical lines mark the fractions of each schedule. Panels (c) and (d) summarize the clonogenic population with two numbers per schedule: its minimum over the follow-up (c) and its value at 168~h (d). For both radiation qualities, the single fraction produced the deepest drop of the clonogenic population, to 0.23 of the initial value for $^{1}$H and 0.083 for $^{12}$C, whereas splitting the dose spared the population through repair between fractions, the recruitment of quiescent cells and repopulation: the lowest clonogenic number reached with the fractionated schedules was 1.7--2.1 times higher for $^{1}$H (0.40 and 0.49) and 2.1--2.4 times higher for $^{12}$C (0.18 and 0.20). At the end of the follow-up, however, this advantage had disappeared: the single fraction, delivered first, left the longest time for regrowth and ended with as many clonogenic cells as the $2\times3$~Gy schedule (2.53 vs.\ 2.51 times the initial number for $^{1}$H, 1.67 vs.\ 1.66 for $^{12}$C) and more than the $3\times2$~Gy schedule (2.23 and 1.53). The two fractionated schedules differ in the time available for regrowth and for the clearance of doomed cells: with 48~h between fractions, most cells doomed by the first fraction are removed before the second, and the surviving clonogens repopulate the space they leave, whereas with 24~h intervals, doomed cells accumulate across fractions. The $2\times3$~Gy schedule ended with 13\% ($^{1}$H) and 9\% ($^{12}$C) more clonogenic cells than the $3\times2$~Gy schedule, i.e.\ repopulation over the longer interval outweighed the lower number of fractions. 

\subsection{Computational performance}
\label{SEC:res_cost}

\begin{figure*}[t]
  \centering
  \includegraphics[width=.99\linewidth]{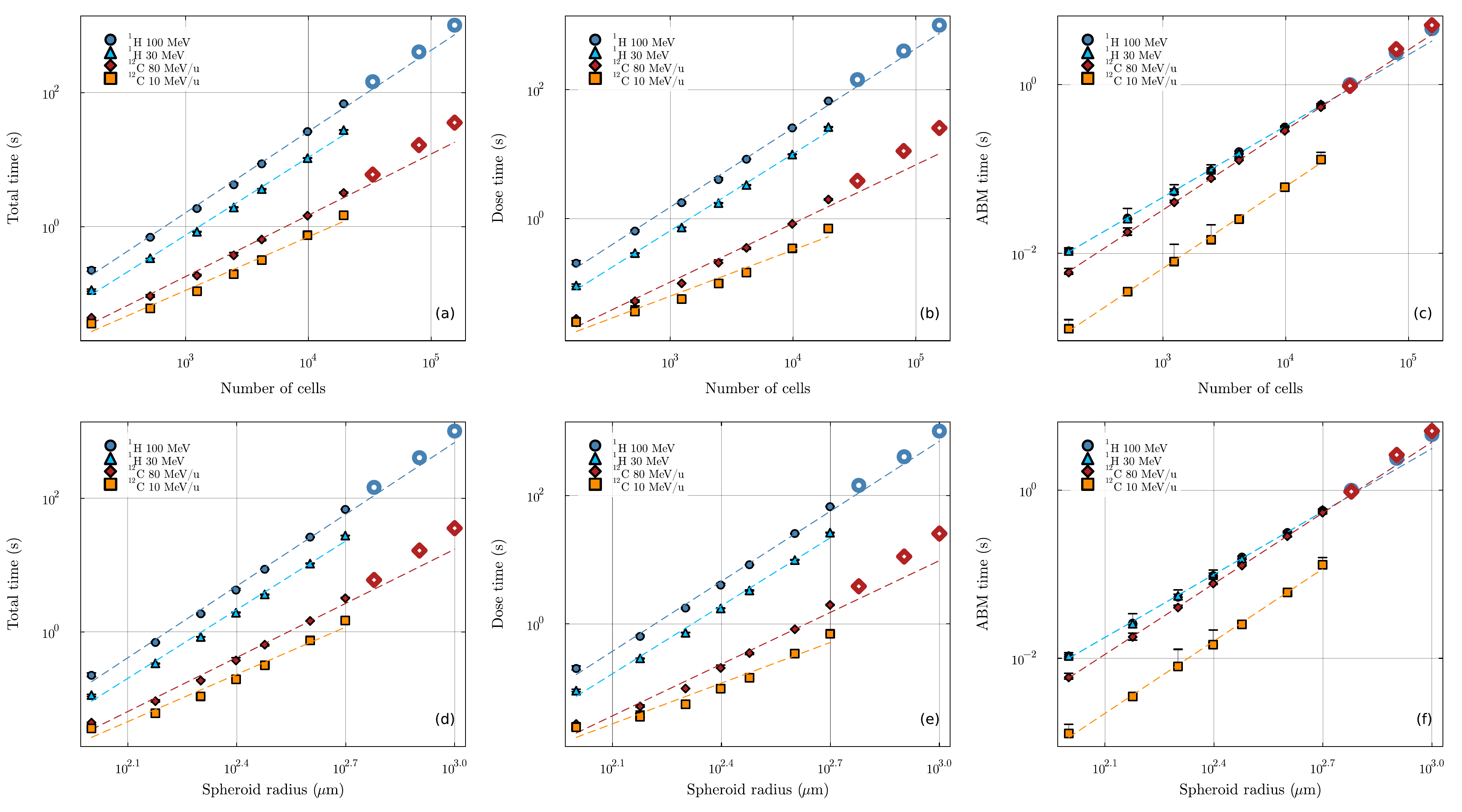}
  \caption{Computation time as a function of the number of cells (top row) and of the spheroid radius (bottom row) for a single 2~Gy irradiation followed by 72~h of agent-based simulation. (a,d)~Total time; (b,e)~dose deposition (amorphous-track Monte Carlo); (c,f)~agent-based simulation. Symbols: median over three repeats; error bars: range; open symbols: stress test (one repeat).}
  \label{fig:timing}
\end{figure*}

Figure~\ref{fig:timing} shows the computation time on logarithmic axes, as a function of the number of cells (top row) and of the spheroid radius (bottom row), for the total simulation (a, d), the dose deposition (b, e) and the agent-based simulation (c, f); each color corresponds to a radiation quality, filled symbols to the spheroid sizes used in this work and open symbols to the stress test. The computational cost of the framework was characterized for spheroid radii from 100 to 500~\textmu m, corresponding to $\approx 150$ to $\approx 19\,400$ cells, for four radiation qualities ($^{1}$H at 100 and 30~MeV, $^{12}$C at 80 and 10~MeV/u), a dose of 2~Gy and 72~h of post-irradiation dynamics (Figure~\ref{fig:timing}). Run times were measured after discarding a warm-up run to exclude compilation effects and are reported as the median of three repeats. For the spheroid sizes used in this work, a complete simulation (track-structure dose deposition, lesion sampling, fate assignment, and 72~h of agent-based dynamics) took from $\approx 0.04$~s ($\approx 150$ cells, $^{12}$C at 10~MeV/u) to $\approx 70$~s ($\approx 19\,400$ cells, $^{1}$H at 100~MeV) on a single workstation. The agent-based part required less than 1~s for all these sizes and scaled as $N^{0.9}$, close to linear as expected from the local, priority-queue update. The dose deposition dominated the cost; it scaled with the number of simulated particles, as $N^{1.2}$ for $^{1}$H at 100~MeV and $N^{0.9}$ for $^{12}$C at 80~MeV/u, because the number of particles grows with the irradiated area and, for protons, each track overlaps more domains as the spheroid grows; at $R = 500$~\textmu m it was about 35--40 times slower for $^{1}$H at 100~MeV than for $^{12}$C at 80~MeV/u. In the stress test, a spheroid of radius 1000~\textmu m ($\approx 1.5\times10^5$ cells) was simulated in $\approx 18$~min for $^{1}$H at 100~MeV and $\approx 40$~s for $^{12}$C at 80~MeV/u, of which only $\approx 5$~s for the agent-based dynamics, and the scaling observed for smaller spheroids extended to this size (open symbols).

\section{Discussion}
\label{SEC:discussion}

The agent-based framework presented in this work extends \GSM{} to include explicit single-cell dynamics, cell-cycle progression, and spatial organization within an irradiated cell population. While the tumor spheroid geometry is adopted here as a well-controlled and experimentally accessible model system \cite{staab2004response}, the underlying architecture imposes no geometric constraints: the lattice can, in principle, represent any tissue configuration, and the biological and physical modules are fully modular and interchangeable. Future work will investigate the response of highly heterogeneous tissues, with particular attention to skin, whose acute and late toxicity is among the most clinically relevant normal-tissue complications in particle therapy \cite{barnett2009normal, hall2006radiobiology}.

This single-cell resolution is physically motivated by the nature of particle interactions with tissue. Energy deposition by charged particles is inherently local, with track-structure effects producing heterogeneous dose distributions on the nanometric and micrometric scales \cite{cordoni2021generalized, battestini2025multiscale}. At these scales, cell-to-cell differences in energy deposition, repair capacity, cell-cycle state, and oxygenation are not small perturbations but can determine whether an individual cell survives or dies, as shown by the heterogeneous survival probability after carbon-ion irradiation (Figure~\ref{fig:spheroid3d_C}a). Resolving each cell individually allows this heterogeneity to propagate naturally into the biological response, rather than being averaged away at the population level. It also enables the direct incorporation of tissue-level heterogeneity in cell-cycle phase, proliferative capacity, and oxygenation, all of which influence the population-level response \cite{battestini2026ntcp}.

The present framework extends this landscape by grounding the single-cell biological response in \GSM{}, which has been validated across a broad range of LET values and particle species \cite{bordieri2024validation}, thereby making it applicable to any ion within a single, consistent formulation. By embedding \GSM{} within a next-event, continuous-time Markov architecture, the model resolves individual cells across all stages of the radiation response, from local energy release and DNA damage induction through repair, cell-cycle progression, and spatial reorganization, while remaining computationally tractable, as discussed below. The biological accuracy of the model is inherited from \GSM{}, whose predictions for cell survival have been validated against experimental data for multiple cell lines and irradiation conditions \cite{cordoni2021generalized, cordoni2022cell, missiaggia2024cell, bordieri2024validation}. Beyond this inherited validation, the benchmark against V79 spheroids (Section~\ref{app:staab}, Appendix~\ref{app:staab}) shows that two global parameters, adjusted on a single aerobic X-ray survival curve, suffice for the model to predict three independent conditions, a different radiation quality and a different oxygen level, with OER within 3\% and RBE within 12\% of the measured values. Because this endpoint was measured on cells dissociated immediately after irradiation, it tests the radiation-response component of the framework independently of the growth dynamics. The largest deviations occur for X-rays, where the two-factor rescaling underestimates the survival shoulder between 4 and 10~Gy and the predicted anoxic survival at intermediate doses, indicating that the phase-specific parameters of synchronized V79 cells \cite{Sinclair1966} describe the shape of the low-LET survival curve of this substrain only approximately.

The computational cost of the framework is summarised in Figure~\ref{fig:timing} for spheroid radii between 100 and \SI{500}{\micro\metre}, four radiation qualities, and a dose of \SI{2}{\gray}, with the irradiation and the agent-based stages timed separately. The two stages have different origins and scale differently. The cost of the irradiation step is set by the number of simulated particles, which increases with the prescribed dose and the irradiated area, and decreases with LET. At fixed dose, it therefore increases approximately as $N^{1.2}$ ($^1$H at 100~MeV) and $N^{0.9}$ ($^{12}$C at 80~MeV/u), i.e.\ as $R^{3.6}$ and $R^{2.7}$ (Figure~\ref{fig:timing}b,e). High-energy protons are a near-worst case: they need many more particles than heavier ions to deliver the same dose, and their wide penumbra spreads each track over more domains, so the dose deposition for $^{1}$H at \SI{100}{\MeV} is about 35--40 times slower than for $^{12}$C at \SI{80}{\MeV}/u. Energy deposition is scored in micrometer-scale domains with a radius of \SI{0.8}{\micro\metre}, which are smaller than the cell nucleus and enable spatially heterogeneous dose sampling within it. This provides a microdosimetric description of energy deposition in targets comparable in size to chromosomal and DNA-bearing structures, which are generally considered biologically relevant. The cost of the agent-based component follows from the next-event algorithm. Rather than advancing all cells through a fixed time step, the global scheduler draws the next biological or physical event from a priority queue. Each event updates only the cells within two lattice sites of the cell where it occurs, so the cost per event does not depend on population size, and computational effort is concentrated where activity occurs. As a result, the agent-based simulation scales approximately linearly with the number of cells (Figure~\ref{fig:timing}c,f) and depends only weakly on the radiation quality, through the number of surviving cells that continue to divide. This is particularly advantageous in large, partially quiescent spheroids, where only a fraction of the population is cycling or repairing damage at any given time, and makes it possible to simulate spheroids of $\sim 2\times10^4$ cells with track-structure resolution within about one minute, and of $\sim 1.5\times10^5$ cells within about 18 minutes, on a single workstation, with the agent-based dynamics of three days taking only a few seconds. These run times make the framework practical for the parameter scans, replicate simulations, and fractionation schedules considered here. Existing agent-based models of radiotherapy either describe the response of each cell with an LQ survival probability assigned at the time of irradiation \cite{kempf2015spatio, Kunz2024AMBER}, or couple Monte Carlo energy deposition to a cell population for a single irradiation \cite{garcia2024topas, Cogno2024LungFibrosis}. To our knowledge, no other framework for particle therapy combines track-structure energy deposition, time-resolved stochastic repair kinetics, and cell-cycle and population dynamics for each of $\sim 10^4$--$10^5$ cells at a comparable computational cost on a single workstation. A direct comparison of run times between frameworks is difficult because the models differ in their physical and biological content; the scaling reported here, close to linear in the number of cells for the agent-based dynamics and in the number of particles for the irradiation, does not depend on implementation details and indicates that larger populations or longer schedules remain within reach. The agent-based dynamics would remain tractable for considerably larger populations, since their cost grows nearly linearly with the number of cells; the limiting factors are the cost of the irradiation step, which grows with the number of simulated particles, and the memory required to store the lesion state of each domain, both of which could be reduced by parallelization on GPUs and by storing the lesion state only for damaged cells.

The explicit representation of the cell cycle is a key feature of the framework, motivated by the well-established dependence of radiosensitivity on cell-cycle phase: cells in late S phase are relatively radioresistant, while those in G$_2$/M are among the most sensitive \cite{hall2006radiobiology}. This phase-dependent heterogeneity has been identified as a relevant contributor to inter-tumor variability in treatment response \cite{inaniwa2024biological}, and capturing it at the single-cell level, rather than through averaged correction factors, is one of the primary motivations for the agent-based approach. As in previous agent-based spheroid models \cite{kempf2015spatio}, phase-specific radiosensitivity is taken from synchronized V79 cells \cite{Sinclair1966}; in contrast to those models, which assign an LQ survival probability at the time of irradiation, survival here results from the microdosimetric damage and the stochastic repair kinetics of each cell, which is what allows dose-rate effects, LET dependence, and the timing of death to emerge. In the present model, each cell progresses through the cycle independently, with phase durations sampled from Gamma distributions and radiosensitivity parameters assigned according to the current phase. The resulting phase-dependent dynamics, and their dependence on dose, are illustrated in Figures~\ref{fig:phase_breakdown}, \ref{fig:spheroid3d_H}, \ref{fig:spheroid3d_C} and \ref{fig:dose_phases}, without requiring any empirical adjustment.

Beyond the heterogeneity in cell-cycle state and oxygenation, this last being extensively studied in \cite{battestini2026ntcp}, the model extends the stochastic principles of \GSM{} to a higher level by coupling subcellular damage kinetics to population-level variability. At the subcellular level, \GSM{} generates stochastic lesion counts and repair trajectories for each cell independently, reflecting the randomness of energy deposition and biochemical repair \cite{cordoni2021generalized, cordoni2022cell}. The resulting cell-to-cell variability in damage evolution is illustrated in Figure~\ref{fig:damage_evolution}, where the spread of individual trajectories around the ensemble mean is largest immediately after irradiation and narrows over time as repair proceeds. At the population level, the agent-based architecture produces stochastic realizations of the spheroid's evolution, capturing variability in division, death, and spatial reorganization (Figure~\ref{fig:fractionation}c). This dual stochasticity allows the model to generate distributions of possible outcomes rather than single deterministic predictions, providing a basis for uncertainty quantification in treatment response. The spread of these distributions reflects both the biological variability of the tumor and the stochasticity of the radiation field, and could, in principle, inform estimates of inter-patient variability in treatment outcomes, complementing the approach of \cite{battestini2026ntcp}.

In the limit of a static population without cell-cycle progression, the dose-rate dependence predicted by the model agrees with the Lea--Catcheside factor evaluated with the effective repair rate of \GSM{}, for all radiation qualities and over four orders of magnitude in dose rate. In most radiobiological models, dose-rate effects are introduced through this factor \cite{LeaCatcheside1942}, which follows from first-order repair of sublethal damage and pairwise interaction of lesions from different tracks, but is applied as an analytical correction to population-averaged quantities and cannot by itself account for concurrent cell-cycle progression, proliferation, or spatial effects. In the present model, the protraction factor is never used: each particle arrival adds lesions to the current state of each cell, repair proceeds continuously between arrivals, and residual damage is carried forward through the shared event queue. The agreement is therefore a quantitative verification of the event-driven coupling between irradiation and repair, on which all time-dependent results of this work rely. The LET dependence of the dose-rate effect follows directly from the quadratic share of the cell kill (Figure~\ref{fig:quadratic_share}). For high-energy protons, the inter-track term accounts for about half of the logarithmic cell kill at 6~Gy and is almost entirely spared by protraction. For low-energy carbon ions, cell killing is dominated by single-track events; the quadratic contribution is already small under acute irradiation, and survival is essentially independent of dose rate (Figure~\ref{fig:sparing}). The quadratic share thus provides a compact upper bound on the dose-rate sparing achievable for a given radiation quality and dose. These trends emerge without any ion-specific tuning of the dose-rate response; comparable results over this range of particles and dose rates have not, to our knowledge, been reported from a single mechanistic framework, and dedicated experimental validation remains an important next step.

The static limit is only the case in which an analytical reference exists. The Lea--Catcheside factor applies to a homogeneous population in which all cells share the same dose-rate history, receive many tracks, and neither cycle nor proliferate during irradiation. The event-driven formulation developed in this work requires none of these assumptions, meaning that each cell receives its own sequence of particle arrivals, so that spatially non-uniform dose rates, mixed radiation qualities, and the stochastic delivery of a few tracks per cell are explicitly described, while repair, cell-cycle progression, redistribution, and repopulation proceed concurrently with the irradiation. This makes the framework suited to regimes in which irradiation and cell dynamics occur on comparable time scales. Examples include low-dose-rate brachytherapy, where continuous irradiation over hours to days in steep spatial dose-rate gradients overlaps with cell-cycle redistribution and repopulation, and the chronic low-dose-rate exposure of space radiation, in which individual cells are traversed by rare high-LET particles at intervals long compared with repair, so that the response is determined by the stochastic arrival of single tracks rather than by a mean dose rate. The study of these regimes, which requires cell-line-specific cycle and checkpoint behavior, is left to future work.

The split-dose results of Figures~\ref{fig:surv_prob_vs_time} and \ref{fig:surv_phases_vs_time} illustrate the inter-fraction time dependence of survival probability for a \SI{1.5}{\gray} $+$ \SI{1.5}{\gray} fractionation scheme. As shown in Figure~\ref{fig:surv_vs_time} for $^{1}$H ions at \SI{100}{\MeV}, the survival probability exhibits a non-monotonic dependence on inter-fraction time: an initial increase reflecting sublethal damage repair, a second increase as cells redistribute into the radioresistant S phase, and a subsequent decrease as the population re-enters the radiosensitive G$_2$/M phases. This oscillatory behavior reproduces the classical split-dose recovery pattern reported experimentally for low-LET radiation \cite{hall2006radiobiology} and is directly supported by the phase-distribution curves of Figure~\ref{fig:phases_vs_time}.

The comparison with the G$_1$-only and no-repopulation reference cases in Figure~\ref{fig:surv_vs_time} helps to separate the contributions of individual mechanisms. The second survival increase is absent in the G$_1$-only case, confirming its association with redistribution into S phase. In the standard GSM$^2$ reference, only damage repair is accounted for, while repopulation and cell-cycle reassortment between fractions are neglected. As a result, survival initially increases with split time due to repair, but then reaches a plateau and becomes time-independent once split times exceed the repair timescale. This repair-only description is the one adopted in particle therapy, where the effect of the time structure of dose delivery on biological effectiveness is modeled through sublethal damage repair between irradiations, for example in the microdosimetric kinetic model extended to carbon-ion treatment planning \cite{inaniwa2013effects}, while redistribution and repopulation between fractions are not described. The full model diverges from this reference at intermediate and late inter-fraction times, a difference that would not be captured by population-level approaches. Taken together, these results show that the model reproduces three of the classical Rs of radiobiology \cite{hall2006radiobiology}: \textit{repair}, reflected in the initial survival increase at short inter-fraction intervals; \textit{reassortment}, in the phase-driven oscillatory modulation at intermediate times; and \textit{repopulation}, in the increase of the surviving cell number at late intervals. The fourth R, \textit{reoxygenation}, enters only through the recomputation of the quasi-static oxygen profile before each fraction and is not explored dynamically in the present work.

Cells committed to radiation-induced death are not removed instantaneously but remain in the population for a stochastic, phase-dependent delay, after which they are eliminated. This reflects the experimental observation that in most solid tumor cell lines, lethally irradiated cells die predominantly through mitotic catastrophe after one or more abortive divisions over a period of one to several days \cite{Chu2002, ErikssonStigbrand2010}. In the model, doomed cells keep their lattice sites, do not divide, and block neighboring sites until they are removed. This has two consequences that are visible in the fractionation results (Figure~\ref{fig:fractionation_single}) and in the spatial reconstructions (Figures~\ref{fig:spheroid3d_H} and~\ref{fig:spheroid3d_C}). First, the clonogenic and total cell populations separate after each irradiation: cell number or spheroid volume overestimates the surviving population for days after exposure, which is relevant when spheroid growth curves are used as a surrogate for cell kill. Second, recruitment of quiescent cells into the cycle is delayed because the space that allows them to re-enter the cycle becomes available only when doomed cells are removed, not when they are committed to death. The delay is currently described by a phenomenological Gamma distribution whose mean scales with the cycle length. It should be calibrated against time-lapse or growth-delay data for specific cell lines, and refined to distinguish death pathways (apoptosis, mitotic catastrophe, senescence), which are currently merged into a single binary fate.

Under fractionated irradiation (Figures~\ref{fig:fractionation} and~\ref{fig:schedules}), the model combines, within a single simulation, the processes that, in population-level models, are treated through separate corrections: sublethal damage repair between fractions, redistribution of surviving cells among cell-cycle phases, recruitment of quiescent cells, and repopulation. The surviving fraction of each fraction agrees with the mean \GSM{} survival probability of the irradiated cells (Figure~\ref{fig:fx_sf}), showing that the population-level response remains consistent with the underlying single-cell model, while its variation from one fraction to the next reflects the changing phase composition of the population. For $^{12}$C ions, the larger fraction of cells killed by each fraction, combined with the death delay, leads to an accumulation of doomed cells across fractions when the interval between fractions is shorter than the time needed to clear them. The comparison of schedules with the same total dose (Figure~\ref{fig:schedules}) shows that this interplay among the irradiation schedule, death kinetics, and repopulation changes the outcome of a given dose, an effect that cannot be captured by models that remove cells at the time of irradiation. Three limitations should be noted. Oxygenation is quasi-static, recomputed before each fraction but without transport dynamics between fractions. In addition, quiescence is governed by contact inhibition on the lattice rather than by nutrient and oxygen gradients, which influences how quickly the core is recruited after irradiation. Finally, fates are resolved at each fraction, so that residual sublethal damage from one fraction does not interact with the damage of the next; this is a good approximation for the 24 and 48~h intervals considered here.

Because all realizations start from the same spheroid and share the same random sequence up to the first fraction, the spread between realizations isolates the variability generated by irradiation and subsequent cell dynamics (Figure~\ref{fig:fractionation}c). This spread, measured relative to the mean, increases after each fraction and is consistently larger for $^{12}$C ions, where fewer cells survive and energy is deposited by fewer, denser tracks, so that the outcome of each realization depends on a smaller number of surviving clonogens. This is precisely the regime relevant to tumor control, which is determined by the last few surviving cells rather than by the mean surviving fraction. A natural extension is to separate the contribution of physical stochasticity (track positions and lesion yields) from that of biological stochasticity (repair, cycle progression, division); the common random number design used here makes this possible by fixing one source while varying the other.

The framework is intended to show how a cell population's response to particle irradiation emerges from stochastic damage, repair, and cell dynamics, rather than to reproduce a specific tumor. Its biological description is therefore deliberately a simplification: cell fate is binary, with death occurring after a phenomenological, phase-dependent delay, and the population is described by generic parameters for a representative tumor cell, whose radiation response has been tested against independent data but whose growth and death kinetics have not been calibrated against a specific cell line. For this reason, the results on growth, regrowth and fractionation schedules should be read as mechanistic trends rather than quantitative predictions. The model also contains no intercellular signaling, so non-targeted effects such as bystander responses are not described, and the low-dose hyper-radiosensitivity observed below about 0.5~Gy lies outside the dose range considered here. The modular structure of the framework allows these aspects to be addressed without changing its event-driven architecture; the inclusion of distinct death and arrest pathways, such as apoptosis and senescence, and a quantitative comparison with time-resolved spheroid growth measurements are currently in progress.

\section{Conclusions}

We have presented a stochastic agent-based extension of \GSM{} in which every cell of a three-dimensional spheroid evolves as a continuous-time Markov process, with particle arrivals, lesion repair and misrepair, cell-cycle progression, division, and delayed death treated as competing events in a single-event queue. Because each event updates only the cells in its neighborhood, spheroids of $\sim 1.5\times10^5$ cells are irradiated with track-structure resolution and followed for three days in about 18 minutes on a single workstation.

Three results support the framework. In the limit of a static population, the simulated dose-rate dependence of survival agrees with the Lea--Catcheside factor for protons and carbon ions over four orders of magnitude in dose rate, although no protraction factor is imposed, and the loss of dose-rate sparing at high LET follows from the small contribution of inter-track lesion interactions. With two global factors adjusted on a single X-ray survival curve of V79 spheroids, the model predicts three independent conditions, with different radiation quality and oxygenation, with OER within 3\% and RBE within 12\% of the measured values. Applied to spheroids, it reproduces repair, reassortment and repopulation within a single description and shows that cells committed to death persist for days, so that cell number or spheroid volume overestimates the surviving population after irradiation and the outcome of a given dose depends on how it is fractionated.

The framework provides a mechanistic, single-cell basis for studying time-dependent effects and their variability in particle therapy, and a natural route towards tumor control and normal tissue complication probabilities derived from single-cell outcomes.

\subsubsection*{Acknowledgments}

This work has been partially supported by the INFN CSN5 project GAP, by the CSN5 young investigator grant (Bando Giovani) PROBE, and by the Pianoforte grant PRESTO.

\cleardoublepage
\bibliographystyle{apalike}
\bibliography{bib}

\appendix
\section{Calibration of GSM$^2$ parameters to phase-specific linear-quadratic data}
\label{app:Fit}

The GSM$^2$ rate parameters $(r, a, b)$ for each cell-cycle phase are calibrated by fitting the GSM$^2$ survival curve to the phase-specific LQ response characterized by the $(\alpha, \beta)$ values reported in Section~\ref{SEC:params}. The fitting procedure minimizes the deviation between the GSM$^2$ survival curve and the LQ reference curve $S(D) = \exp(-\alpha D - \beta D^2)$ over a clinically relevant dose range. The calibration is performed independently for each phase, with the domain radius $r_d = \SI{0.8}{\micro\metre}$ and nuclear radius $R_N = \SI{7.2}{\micro\metre}$ held fixed throughout.

\begin{figure}[htbp]
    \centering
    \begin{subfigure}[t]{0.32\linewidth}
        \centering
        \includegraphics[width=\linewidth]{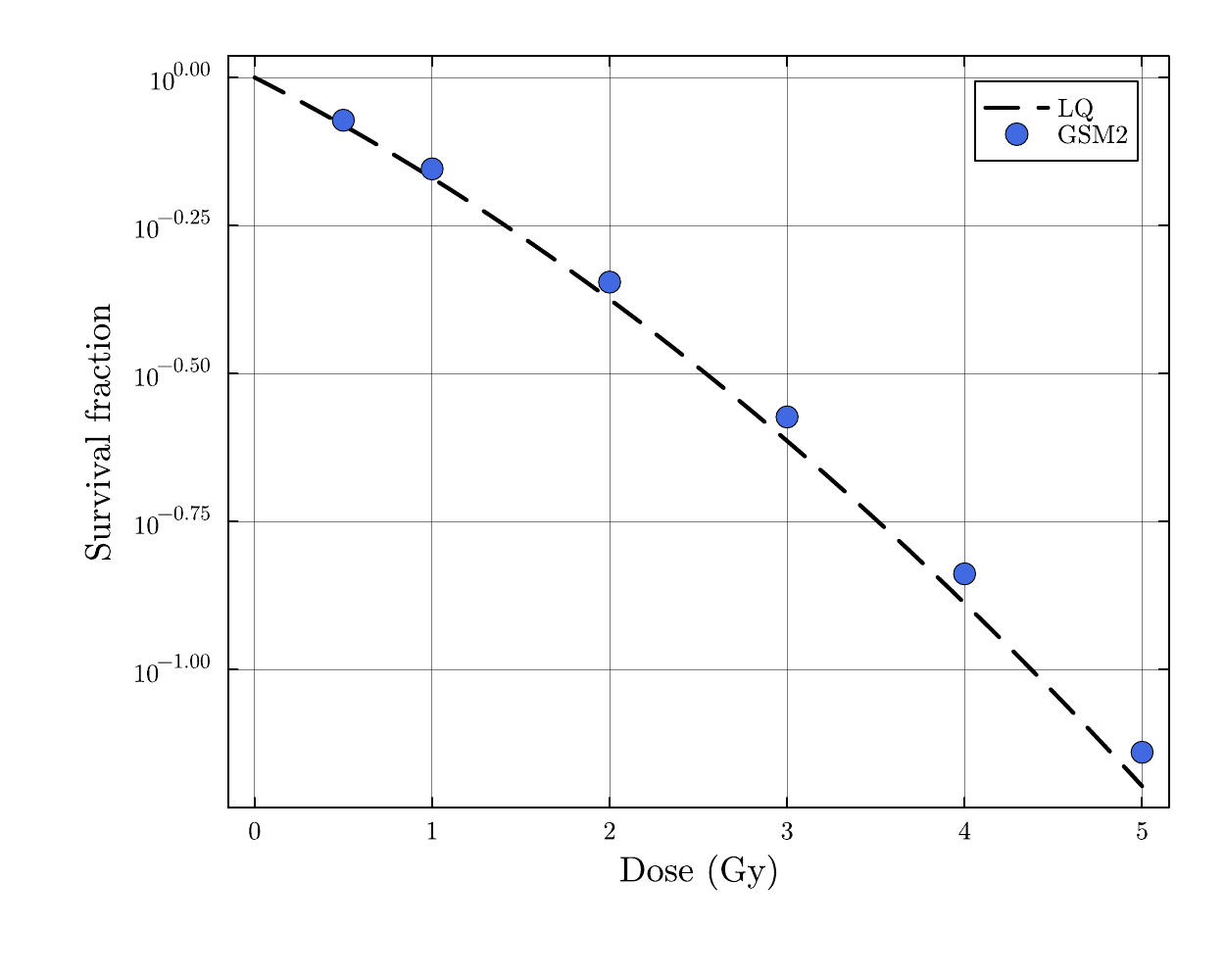}
        \caption{$\mathrm{G_1}$ phase: $\alpha = \SI{0.351}{\per\gray}$, $\beta = \SI{0.040}{\per\gray\squared}$.}
        \label{fig:fit_G1}
    \end{subfigure}
    \hfill
    \begin{subfigure}[t]{0.32\linewidth}
        \centering
        \includegraphics[width=\linewidth]{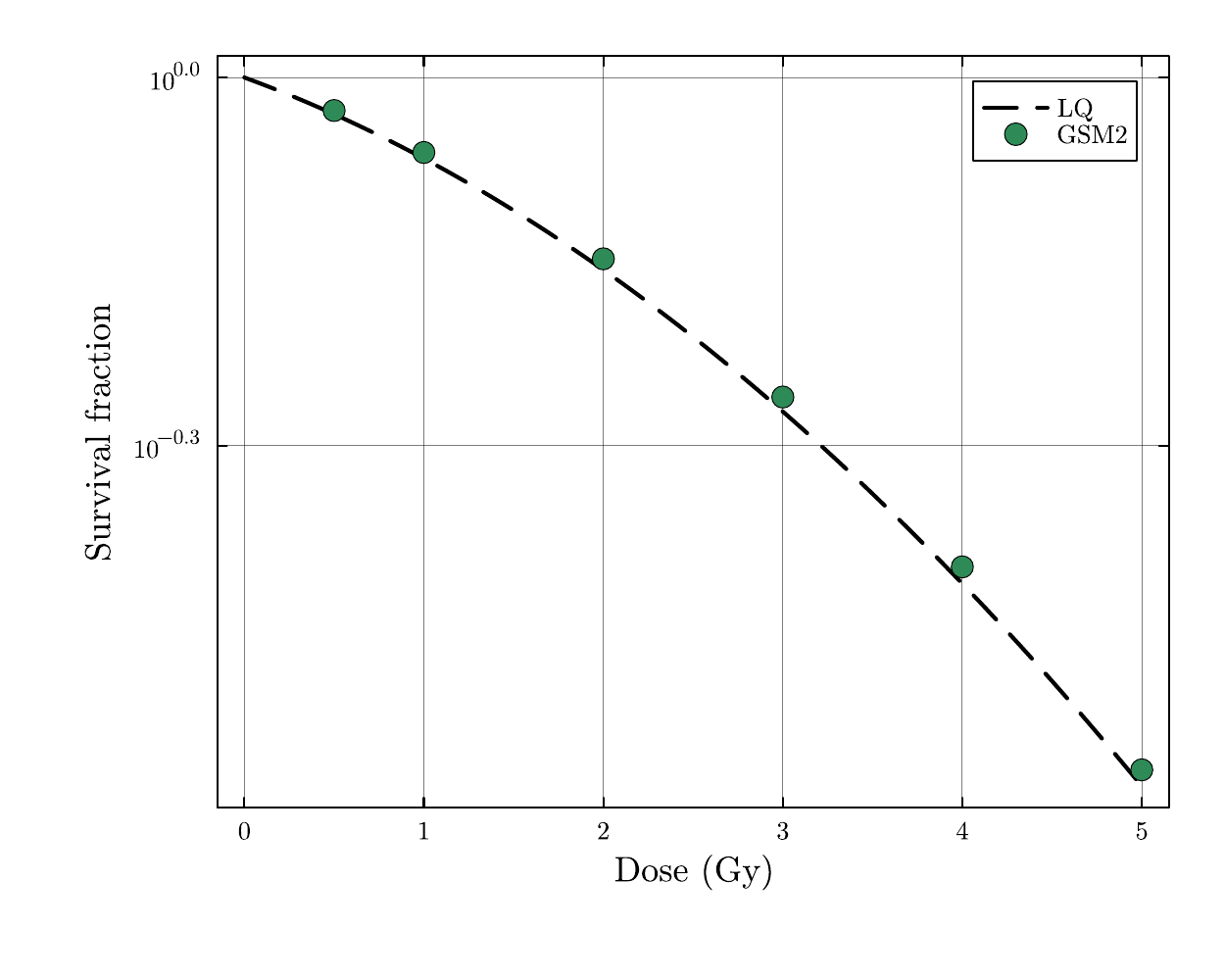}
        \caption{S phase: $\alpha = \SI{0.124}{\per\gray}$, $\beta = \SI{0.029}{\per\gray\squared}$.}
        \label{fig:fit_S}
    \end{subfigure}
    \hfill
    \begin{subfigure}[t]{0.32\linewidth}
        \centering
        \includegraphics[width=\linewidth]{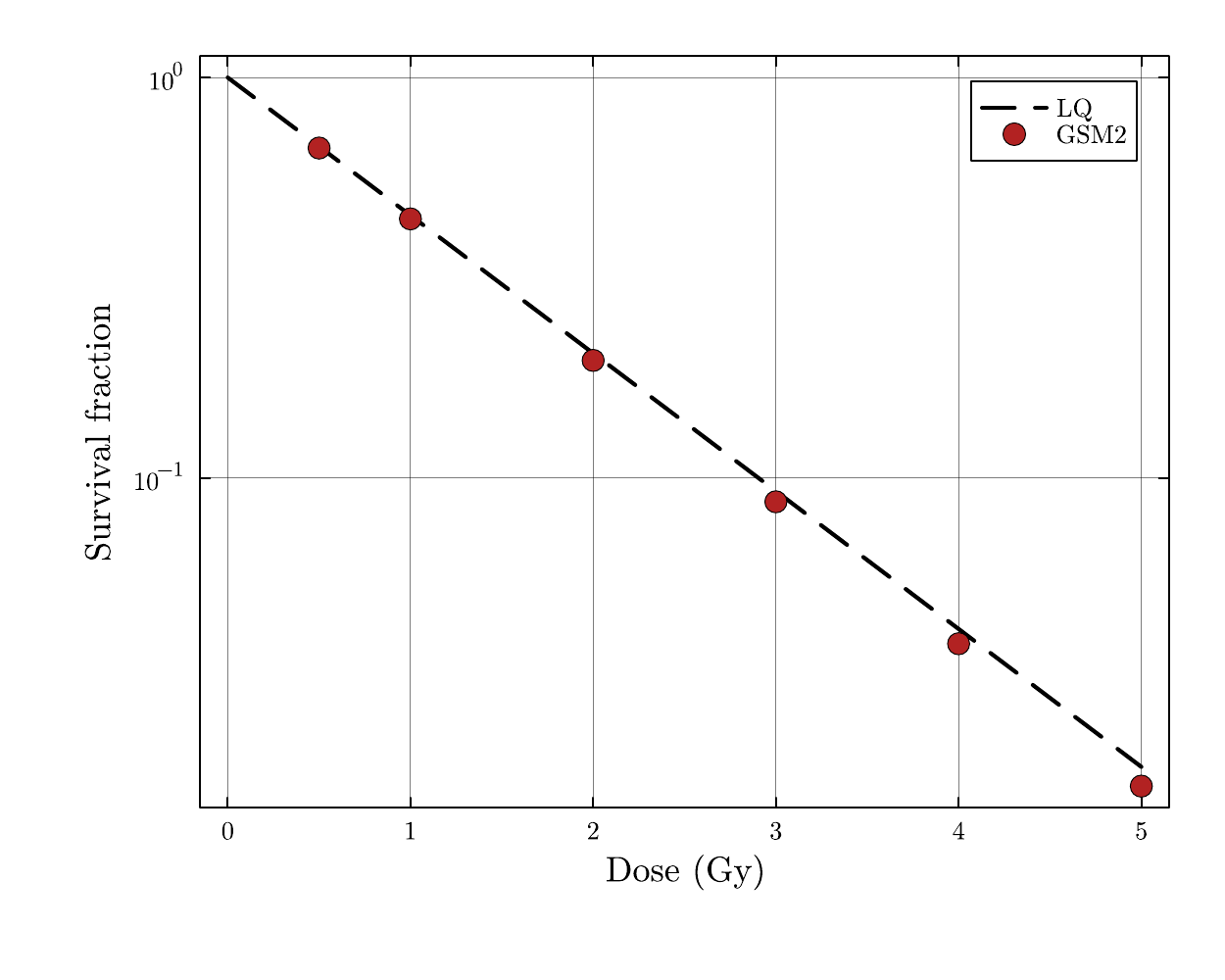}
        \caption{$\mathrm{G_2}$/M phase: $\alpha = \SI{0.793}{\per\gray}$, $\beta = \SI{0.000}{\per\gray\squared}$.}
        \label{fig:fit_G2M}
    \end{subfigure}
    \caption{Calibration of GSM$^2$ parameters to phase-specific linear-quadratic survival data \cite{hall2006radiobiology, Sinclair1966}. In each panel, the dashed line shows the LQ reference curve, and the points show the calibrated GSM$^2$ survival curve, plotted on a logarithmic scale versus dose.}
    \label{fig:LQ_calibration}
\end{figure}

The resulting fits are shown in Figures~\ref{fig:fit_G1}--\ref{fig:fit_G2M}. In each panel, the LQ reference experimental curve is shown as a dashed line and the calibrated GSM$^2$ survival curve as points, plotted on a logarithmic scale as a function of dose.

\begin{table}[htbp]
\centering
\caption{GSM$^2$ rates after calibration on the synchronized clonogenic cell survival of the V79 cell line.}
\label{tab:LQ_cellcycle}
\begin{tabular}{lccc}
\hline
Phase & $r$ & $a$ & $b$ \\
\hline
$\mathrm{G_1}$, $\mathrm{G_0}$ & 2.780 & 0.01287 & 0.04030 \\
S & 5.840 & 0.00589& 0.05794 \\
$\mathrm{G_2}$, M & 1.772 & 0.02431 & $5.70\times10^{-5}$ \\
\hline
\end{tabular}
\end{table}

Table \ref{tab:LQ_cellcycle} reports the calibrated \GSM{} parameters for the different cell cycles. 

The agreement between the GSM$^2$ and LQ curves is good across the dose range considered for all three cell-cycle phases, indicating that the calibrated GSM$^2$ parameters consistently reproduce the dominant radiobiological features encoded in the LQ fits.

For the $\mathrm{G_1}$ phase (Figure~\ref{fig:fit_G1}), the survival curve exhibits the characteristic shouldered shape associated with a nonzero $\beta$ coefficient, reflecting a significant quadratic contribution from pairwise sublethal lesion interactions. The calibrated GSM$^2$ reproduces this behavior through a moderate repair rate, $r = \SI{2.780}{\per\hour}$, together with a nonzero interaction rate $b$. This parameter combination is consistent with the intermediate radiosensitivity of $\mathrm{G_1}$ cells, in which non-homologous end joining predominates, and misrepair of interacting lesions contributes appreciably to cell inactivation.

For the S phase (Figure~\ref{fig:fit_S}), the reference survival curve is noticeably shallower than in $\mathrm{G_1}$, reflecting the well-established radioresistance of cells undergoing DNA synthesis. This behavior manifests as a lower $\alpha$ value and a reduced initial slope, which the GSM$^2$ reproduces via a substantially higher repair rate, $r = \SI{5.840}{\per\hour}$. The increased repair rate implies more rapid clearance of sublethal damage, consistent with the enhanced repair capacity characteristic of S-phase cells, in which homologous recombination pathways predominate.

For the $\mathrm{G_2}/\mathrm{M}$ phase (Figure~\ref{fig:fit_G2M}), the reference curve is purely exponential, with $\beta = 0$, corresponding to a linear survival response with the steepest initial slope among the three phases. This regime represents the most radiosensitive configuration and is captured within GSM$^2$ by a high conversion rate $a$ relative to the repair rate $r$ and a near-zero interaction rate, $b \approx 0$. In this case, pairwise lesion interactions play a negligible role, and cell killing is dominated by single-track lethal events and the conversion of sublethal into lethal lesions, consistent with the limited opportunity for effective repair during late cell-cycle progression and mitosis.

Across all three phases, the GSM$^2$ framework provides a mechanistic account of the phase-specific LQ parameters without requiring separate empirical fits for each radiation quality: once the rate constants $(r, a, b)$ are fixed, the dose-rate and LET dependence of the survival response follows from the model equations together with the LET dependence of the lesion yield, Equation~\eqref{eq:kappa}. This is the key advantage of the GSM$^2$ parameterization over the direct use of the LQ model for particle therapy and variable-dose-rate irradiation. Appendix~\ref{app:staab} tests this property against experimental data.

\section{Benchmark against V79 spheroid survival}
\label{app:staab}

The phase-specific parameters of Appendix~\ref{app:Fit} describe synchronized V79 cells and are used in the main text as representative values for a generic tumor cell. To test whether the framework, once calibrated to a single condition, predicts responses to other radiation qualities and oxygen tensions, we used the clonogenic survival of V79-379A multicellular spheroids reported by Staab et al.~\cite{staab2004response}. Spheroids of $200 \pm 30$~\textmu m diameter were irradiated in stirred medium with 250~kV X-rays and with 227~MeV/u $^{12}$C ions in the plateau region (LET 18~keV\,\textmu m$^{-1}$), at external oxygen tensions of 145~mmHg (aerobic) and 0~mmHg (anoxic), and dissociated and plated immediately after irradiation. Because cells were plated immediately after exposure, this endpoint tests the radiation-response component of the model, from energy deposition and lesion induction to repair and the oxygen effect, independently of the spheroid's cell-cycle, growth, and death kinetics. The simulation protocol is described in Section~\ref{SEC:staab_methods}: all cells were assigned the oxygen level of the gas phase (19\% and 0.1\% O$_2$), and X-rays were represented by a low-LET field of 1.7~keV\,\textmu m$^{-1}$. Cell-cycle phases were assigned as in all spheroid simulations: cells start asynchronously distributed over the cycle, and cells with no free neighboring site are quiescent, giving a composition dominated by G$_0$/G$_1$ cells, as in the untreated spheroids of the experiment (about 65\% G$_0$/G$_1$, 25\% S and 8\% G$_2$/M \cite{staab2004response}). The phase composition was not adjusted to the data, so that the two calibration factors describe the radiosensitivity of this mixture rather than of a single phase.

The phase-specific rates derived in Appendix \ref{app:Fit} were multiplied by two global factors, one for the conversion rate $a$ and one for the interaction rate $b$, which were fitted to aerobic X-ray survival only. The fit gave $s_a = 0.506$ and $s_b = 0.631$, identical for all phases by construction, so that the relative radiosensitivity of the cell-cycle phases was preserved; the resulting rates are listed in Table~\ref{tab:staab_params}. The reduction of both rates relative to the values of Appendix~\ref{app:Fit} reflects the higher radioresistance of the V79-379A substrain used in the spheroid experiments, whose aerobic X-ray survival corresponds to $\alpha = \SI{0.092}{\per\gray}$ and $\beta = \SI{0.030}{\per\gray\squared}$~\cite{staab2004response}, compared with an asynchronous $\alpha$ of about $\SI{0.3}{\per\gray}$ for the phase-specific values. The repair rates $r$ were not modified.

\begin{table}[htbp]
\centering
\caption{GSM$^2$ rates after calibration on the aerobic X-ray survival of V79-379A spheroids.}
\label{tab:staab_params}
\begin{tabular}{lccc}
\hline
Phase & $r$ & $a$ & $b$ \\
\hline
$\mathrm{G_1}$, $\mathrm{G_0}$ & 2.780 & 0.00652 & 0.0254 \\
S & 5.840 & 0.00298 & 0.0366 \\
$\mathrm{G_2}$, M & 1.772 & 0.0123 & $3.60\times10^{-5}$ \\
\hline
\end{tabular}
\end{table}

\begin{figure}[htbp]
    \centering
    \includegraphics[width=\linewidth]{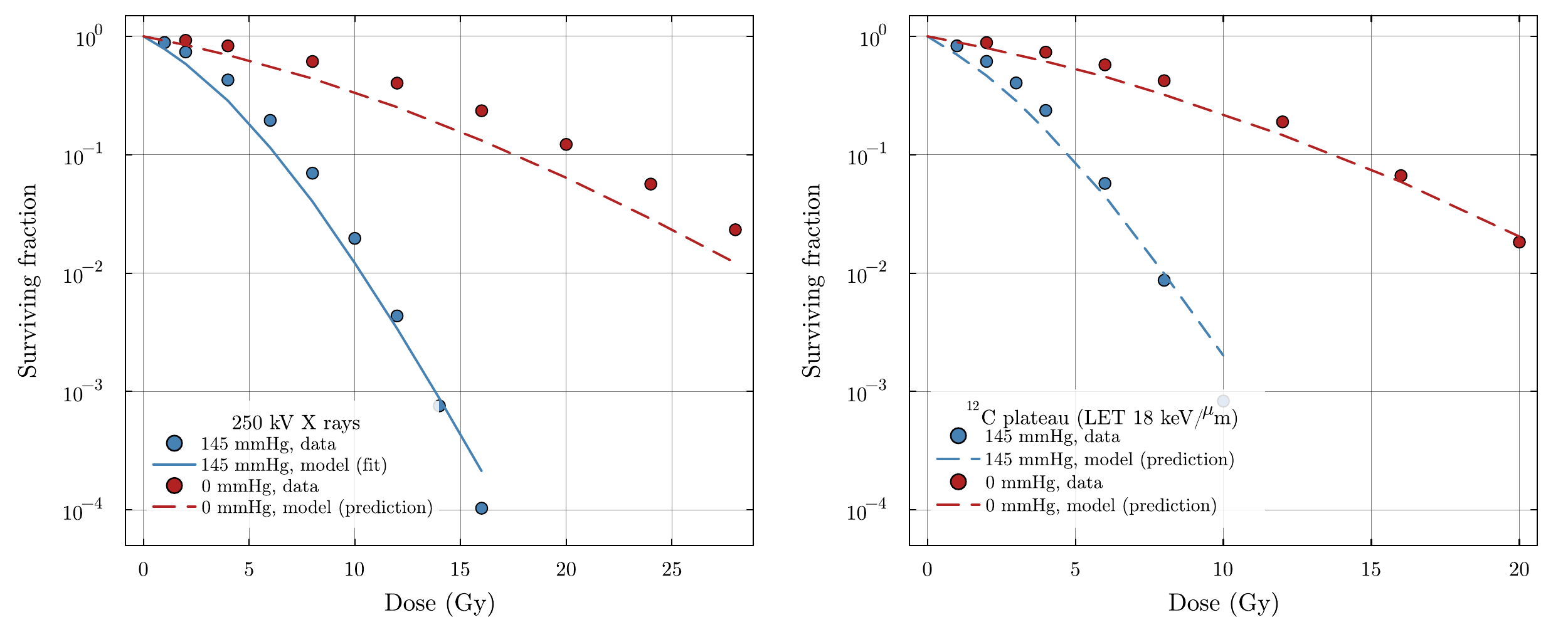}
    \caption{Clonogenic survival of V79-379A spheroids after (a) 250~kV X rays and (b) 227~MeV/u $^{12}$C ions in the plateau region (LET 18~keV\,\textmu m$^{-1}$), at 145~mmHg (blue) and 0~mmHg (red). Symbols: measurements~\cite{staab2004response}; solid line: GSM$^2$ calibrated on the aerobic X-ray survival; dashed lines: predictions without further adjustment.}
    \label{fig:staab}
\end{figure}

Survival under anoxia and after carbon-ion irradiation was then predicted without further adjustment (Figure~\ref{fig:staab}). OER and RBE were computed from LQ fits to the simulated curves as dose ratios averaged over survival levels of 37\%, 10\%, 1\%, and 0.1\%, following the procedure of the experimental study~\cite{staab2004response}, and are compared with the measured values in Table~\ref{tab:staab}. The predicted OER agrees with the measurements within 2\% for X-rays (2.82 vs.\ 2.87) and within 3\% for carbon ions (2.85 vs.\ 2.76). The predicted RBE of plateau carbon ions is 1.29 under aerobic conditions and 1.28 under anoxia, 7\% and 12\% below the measured values of 1.39 and 1.45.

\begin{table}[htbp]
\centering
\caption{Oxygen enhancement ratio (OER) and relative biological effectiveness (RBE) of V79-379A spheroids: model predictions and measurements~\cite{staab2004response}. The model was calibrated only on aerobic X-ray survival.}
\label{tab:staab}
\begin{tabular}{lccc}
\hline
Quantity & Model & Measured & Difference \\
\hline
OER, X rays & 2.82 & 2.87 & $-2\%$ \\
OER, $^{12}$C plateau & 2.85 & 2.76 & $+3\%$ \\
RBE, $^{12}$C plateau, aerobic (145~mmHg) & 1.29 & 1.39 & $-7\%$ \\
RBE, $^{12}$C plateau, anoxic (0~mmHg) & 1.28 & 1.45 & $-12\%$ \\
\hline
\end{tabular}
\end{table}

The agreement of the oxygen effect follows from the oxygen parameterization of Equation~\eqref{eq:OER}, which acts as a dose-modifying factor on the lesion yields and, for the oxygen levels of the experiment, gives a reduction of about 2.9 in both cases; at this LET range the model OER is therefore nearly independent of LET, whereas the measured OER decreases slightly from X rays to plateau carbon ions. The RBE of plateau carbon ions is determined mainly by the increase of the sublethal-lesion yield with LET, Equation~\eqref{eq:kappa}, whose ratio between 18 and 1.7~keV\,\textmu m$^{-1}$ (about 1.3) sets the predicted RBE almost entirely. A further deviation is visible in the calibrated curve itself: the two-factor rescaling reproduces the aerobic X-ray survival at high dose but underestimates the survival shoulder between about 4 and 10~Gy, by up to a factor of 2, indicating that the phase-specific parameters of synchronized V79 cells describe the shape of the low-LET survival curve of this substrain only approximately. The predicted carbon-ion survival, in contrast, agrees with the data over the whole dose range under both oxygen conditions.

Irradiation in the extended Bragg peak of the same study was not simulated because the LET spectrum of the mixed field was not reported; representing it by a single effective LET would require fitting that LET to the data. The model OER reaches the measured Bragg-peak value of about 1.4 only above about 150~keV\,\textmu m$^{-1}$, consistent with the dominant contribution of the high-LET components of a carbon-ion spread-out Bragg peak to the oxygen effect. Overall, with two parameters fitted to a single survival curve, the framework predicts the oxygen effect within 3\% and the RBE of carbon ions within 12\% for an independent experimental dataset.

\section{Simulation parameters}
\label{app:params}

This appendix collects the parameters used in the simulations: the GSM$^2$ repair kinetics (Table~\ref{tab:app_gsm2}), and the geometry, irradiation and statistics of each simulation protocol (Tables~\ref{tab:app_geometry} and~\ref{tab:app_irradiation}). Parameters not listed here (lesion yields, OER and amorphous-track parameters) are given in Section~\ref{SEC:DNA} and Table~\ref{tab:params}, and the oxygenation parameters in Table~\ref{tab:O2params}.

\paragraph{GSM$^2$ repair kinetics.} Two parameter sets are used (Table~\ref{tab:app_gsm2}). The representative set, calibrated on the phase-specific LQ parameters of synchronized V79 cells (Appendix~\ref{app:Fit}), is used in all simulations of the main text. The V79 spheroid set is used only for the benchmark of Appendix~\ref{app:staab}; it is obtained from the representative set by multiplying the conversion rate $a$ by $s_a = 0.506$ and the interaction rate $b$ by $s_b = 0.631$ in all phases, with the repair rates $r$ unchanged. In both sets, G$_0$ cells use the G$_1$ parameters and M cells the G$_2$ parameters. 

\begin{table}[htbp]
\centering
\caption{GSM$^2$ repair kinetics (rates in h$^{-1}$). Representative set: all simulations of the main text; V79 spheroid set: benchmark of Appendix~\ref{app:staab} only ($s_a = 0.506$, $s_b = 0.631$).}
\label{tab:app_gsm2}
\begin{tabular}{llccc}
\hline
Set & Phase & $r$ & $a$ & $b$ \\
\hline
Representative & G$_1$, G$_0$ & 2.780 & 0.01287 & 0.04030  \\
 & S & 5.840 & 0.00589 & 0.05794  \\
 & G$_2$, M & 1.772 & 0.02431 & $5.70\times10^{-5}$ \\
\hline
V79 spheroid & G$_1$, G$_0$ & 2.780 & 0.00652 & 0.0254  \\
 & S & 5.840 & 0.00298 & 0.0366  \\
 & G$_2$, M & 1.772 & 0.0123 & $3.60\times10^{-5}$ \\
\hline
\end{tabular}
\end{table}

\paragraph{Spheroids and simulation protocols.} Tables~\ref{tab:app_geometry} and~\ref{tab:app_irradiation} list, for each protocol, the geometry, the irradiation, the follow-up and the number of independent realizations. Because the lattice spacing exceeds the size of typical tumor cells, a spheroid of given radius contains fewer cells than its experimental counterpart; spheroid sizes are therefore also given as cell numbers.

\begin{table}[htbp]
\centering
\caption{Geometry and oxygenation of the simulation protocols. $R_{\mathrm{sph}}$: spheroid radius; $X_{\mathrm{box}}$: half-size of the simulation box; $N_c$: initial number of cells; uniform: all cells at the stated oxygen level; profile: oxygen profile of Section~\ref{SEC:oxygen}.}
\label{tab:app_geometry}
\begin{tabular}{p{0.24\linewidth}p{0.17\linewidth}p{0.15\linewidth}p{0.15\linewidth}p{0.16\linewidth}}
\hline
Protocol & $R_{\mathrm{sph}}$ (\textmu m) & $X_{\mathrm{box}}$ (\textmu m) & $N_c$ & Oxygen \\
\hline
Continuous dose rate & 250; 450$^{a}$ & 300; 560$^{a}$ & $\approx 2420$; $14\,100^{a}$ & uniform, 21\% \\
Split dose & 350 & 550 & $\approx 6650$ & profile \\
Single irradiation & 350 & 550 & $\approx 6650$ & profile \\
Single-dose response & 250 & 550 & $\approx 2420$ & profile \\
Regrowth & 150 & 400 & $\approx 520$ & profile \\
Fractionation & 250 & 450 & $\approx 2420$ & profile, recomputed at each fraction \\
Equal-dose schedules & 350 & 700 & $\approx 6650$ & profile, recomputed at each fraction \\
V79 benchmark & 200 & 350 & $\approx 1240$ & uniform, 19\% and 0.1\% \\
Computational cost & 100--500 & $R_{\mathrm{sph}} + 150$ & $\approx 150$--$19\,400$ & profile \\
Stress test & 600--1000 & $R_{\mathrm{sph}} + 150$ & up to $\approx 155\,000$ & profile \\
\hline
\multicolumn{5}{l}{\footnotesize $^{a}$ $^{12}$C at 10~MeV/u; the other radiation qualities use the first value.}
\end{tabular}
\end{table}

\begin{table}[htbp]
\centering
\caption{Irradiation, follow-up and statistics of the simulation protocols. Within each protocol, all conditions start from the same initial spheroid and share the same random sequence up to the first irradiation.}
\label{tab:app_irradiation}
\begin{tabular}{p{0.19\linewidth}p{0.22\linewidth}p{0.19\linewidth}p{0.15\linewidth}p{0.15\linewidth}}
\hline
Protocol & Radiation & Dose & Follow-up & Realizations \\
\hline
Continuous dose rate & $^{1}$H 100~MeV; $^{12}$C 80 and 10~MeV/u & up to 6, 3 and 1.5~Gy; $10^{-5}$--$10^{-2}$~Gy\,s$^{-1}$ and acute & end of irradiation & until 600 expected survivors, at most 30 \\
Split dose & $^{1}$H 2, 10, 100~MeV; $^{12}$C 15, 20, 80~MeV/u & $1.5 + 1.5$~Gy, 0--30~h apart & end of second fraction & 3 \\
Single irradiation & $^{1}$H 100~MeV; $^{12}$C 80~MeV/u; control & 2~Gy & 72~h & 1 \\
Single-dose response & $^{1}$H 100~MeV; $^{12}$C 80~MeV/u & 0, 2, 4, 6~Gy & 168~h & 3 \\
Regrowth & $^{1}$H 100~MeV; $^{12}$C 80~MeV/u & 0--12~Gy ($^{1}$H); 0--8~Gy ($^{12}$C) & up to 14~d & 20 (4 dose fields) \\
Fractionation & $^{1}$H 100~MeV; $^{12}$C 80~MeV/u; control & $3 \times 2$~Gy, 24~h apart & 24~h before, 96~h after & 5 \\
Equal-dose schedules & $^{1}$H 100~MeV; $^{12}$C 80~MeV/u; control & $1\times6$~Gy; $3\times2$~Gy, 24~h; $2\times3$~Gy, 48~h & 24~h before, 168~h after the first fraction & 3 \\
V79 benchmark & 1.7 and 18~keV\,\textmu m$^{-1}$ & 1--28~Gy, acute & end of irradiation & 3 lesion samples \\
Computational cost & $^{1}$H 100, 30~MeV; $^{12}$C 80, 10~MeV/u & 2~Gy & 72~h & 3 \\
Stress test & $^{1}$H 100~MeV; $^{12}$C 80~MeV/u & 2~Gy & 72~h & 1 \\
\hline
\end{tabular}
\end{table}

The continuous dose-rate simulations use a static G$_1$ population without cell-cycle progression, division or migration, and evaluate survival as the exact GSM$^2$ probability of completing repair at the end of irradiation; the V79 benchmark evaluates clonogenic survival immediately after acute irradiation, as in the experiment. All other protocols follow the full agent-based dynamics with delayed death. Within each protocol, all conditions start from the same initial spheroid, and realization $k$ uses the same random sequence for all radiation qualities up to the first irradiation, so that differences between conditions reflect only the irradiation.

\end{document}